\documentclass[journal]{IEEEtran}

\usepackage[table,xcdraw]{xcolor}
\usepackage{amsmath,amsfonts}
\usepackage{algorithmic}
\usepackage{algorithm}
\usepackage{textcomp}
\usepackage{stfloats}
\usepackage{url}
\usepackage{verbatim}
\usepackage{graphicx}
\usepackage{cite}
\usepackage[normalem]{ulem}
\usepackage{multirow,booktabs,bigstrut,longtable,array}

\author{Mehdi Bennis,
Sumudu Samarakoon,
Tamara Alshammari,
Chathuranga Weeraddana, \\
Zhoujun Tian,
and
Chaouki Ben Issaid
\thanks{
All authors are from Centre for Wireless Communications, University of Oulu, Finland (emails: \{firstname.lastname\}@oulu.fi).
}
}

\usepackage{setspace}
\usepackage[most]{tcolorbox}    
\tcbuselibrary{breakable} 
\usepackage[acronym]{glossaries}

\usepackage{subcaption}
\usepackage{bm}
\usepackage{soul}
\definecolor{myblue}{rgb}{0.18, 0.015, 0.588}

\newcommand{\set}[1]{\mathcal{#1}}
\newcommand{\vect}[1]{\boldsymbol{#1}}
\newcommand{\dom}[1]{\mathbb{#1}}
\newcommand{\timeDomain}{\dom{T}}
\newcommand{\realDomain}{\dom{R}}

\newcommand{\zero}{\mathbf{0}}
\newcommand{\identity}{\vect{I}}
\newcommand{\expect}{\mathbb{E}}
\newcommand{\prob}{\mathbb{P}}
\newcommand{\indict}{\mathbb{I}}
\newcommand{\normal}{\mathcal{N}}

\newcommand{\signal}{\vect{\xi}}
\newcommand{\signalMapping}{\mu}
\newcommand{\formula}{\varphi}
\newcommand{\interval}{I}

\newcommand{\until}[1]{\text{U}_{#1}}
\newcommand{\eventually}[1]{\text{F}_{#1}}
\newcommand{\always}[1]{\text{G}_{#1}}
\newcommand{\stlRobustness}[1]{\rho(#1,\signal,t)}
\newcommand{\stlResilience}[2]{\text{R}_{#1,#2}}
\newcommand{\tRecover}{t_{\text{r}}}
\newcommand{\tDurable}{t_{\text{d}}}

\newcommand{\statePlant}[1]{\boldsymbol{x}^{(t)}_{#1}}
\newcommand{\stateControl}[1]{\hat{\boldsymbol{x}}^{(t)}_{#1}}
\newcommand{\stateVec}{\hat{\boldsymbol{x}}^{(t)}}

\newcommand{\stateTarget}{\boldsymbol{x}'}
\newcommand{\stateThreshold}{\delta}

\newcommand{\comPolicySingle}{\zeta^{(t)}}
\newcommand{\comPolicy}{\boldsymbol{\zeta}^{(t)}}
\newcommand{\controlAction}[1]{\boldsymbol{u}^{(t)}_{#1}}
\newcommand{\controlVec}{\boldsymbol{u}^{(t)}}
\newcommand{\plantNoise}{\boldsymbol{\epsilon}^{(t)}}
\newcommand{\transpose}{^\top}
\newcommand{\noiseStressor}{\sigma^{(t)}}
\newcommand{\stressorThreshold}{s}

\newcommand{\txpower}{p}
\newcommand{\aoi}{\eta^{(t)}}
\newcommand{\timeHorizon}{\set{K}}

\newcommand{\timeRecovery}{\alpha}
\newcommand{\timeSustain}{\beta}
\newcommand{\stlThreshold}{\epsilon}

\newcommand{\var}{\text{Var}}

\newcommand{\agent}{n}
\newcommand{\AGENT}{N}
\newcommand{\agentSet}{\set{\AGENT}}
\newcommand{\knowledge}[1]{{K}_{#1}}
\newcommand{\knowledgeCollective}[1]{E_{#1}}
\newcommand{\languageKripke}{L_{\AGENT}}
\newcommand{\kripke}{M_t}
\newcommand{\WORLD}{W}
\newcommand{\worldSet}{\set{\WORLD}}
\newcommand{\world}{w}
\newcommand{\relation}{\set{R}}
\newcommand{\valuationFunc}{v}
\newcommand{\prop}{p}
\newcommand{\propSet}{\set{P}}

\newcommand{\graph}{G}
\newcommand{\subgraph}{M}

\newcommand{\EDGE}{E}
\newcommand{\edgeSet}{\set{\EDGE}}
\newcommand{\node}{l}
\newcommand{\NODE}{L}
\newcommand{\nodeSet}{\set{\NODE}}
\newcommand{\motif}{\pi}
\newcommand{\motifDistribution}{\boldsymbol{\motif}}

\newcommand{\configuration}{x}
\newcommand{\configMat}{\boldsymbol{x}}
\newcommand{\permittedEdges}{\Lambda}

\newcommand{\attackProb}{p}
\newcommand{\attackDistribution}{\boldsymbol{\attackProb}}
\newcommand{\targetNodeCount}{K}

\newcommand{\atk}[2]{#1^{(#2)}}
\newcommand{\init}[1]{\atk{#1}{0}}

\newcommand{\minDegree}{\alpha}
\newcommand{\kldiv}{d}
\newcommand{\targetNodeSet}{\set{\targetNodeCount}}

\newcommand{\vertex}[1]{\mathbf{v}_{#1}}
\newcommand{\PCvertex}[1]{\mathbf{x}_{#1}}

\newcommand{\droneSet}{D}
\newcommand{\activeSet}{\droneSet}
\newcommand{\droneCount}{N}
\newcommand{\droneState}{s}
\newcommand{\dronePos}{p}
\newcommand{\areaSize}{S}
\newcommand{\area}{A}
\newcommand{\coverage}{C}

\newcommand{\controlActions}{u}
\newcommand{\failureProb}{\lambda}
\newcommand{\transTime}{T_c}
\newcommand{\redundancy}{\alpha}
\newcommand{\riskAreas}{R}

\newacronym{kpi}{KPI}{key 6G performance indicator}
\newacronym{ai}{AI}{artificial intelligence}
\newacronym{ood}{OOD}{out-of-distribution}
\newacronym{ma}{MA}{multi-agent}
\newacronym{pd}{PD}{persistence diagram}
\newacronym{pc}{PC}{point cloud}
\newacronym{bsc}{BSC}{binary symmetric channel}
\newacronym{Tx}{Tx}{transmitter node}
\newacronym{Rx}{Rx}{receiver node}
\newacronym{bch}{BCH}{Bose, Chaudhuri, and Hocquenghem}
\newacronym{ml}{ML}{machine learning}
\newacronym{iid}{IID}{independent identically
distributed}
\newacronym{p2p}{P2P}{point-to-point}
\newacronym{fep}{FEP}{free energy principle}
\newacronym{jepa}{JEPA}{joint embedding  predictive architecture}
\newacronym{nn}{NN}{neural network}
\newacronym{its}{ITS}{information transition system}
\newacronym{stl}{STL}{signal temporal logic}
\newacronym{rsv}{RSV}{robustness satisfaction value}
\newacronym{tda}{TDA}{topological data analysis}
\newacronym{ph}{PH}{persistent homology}
\newacronym{cdf}{CDF}{cumulative distribution function}
\newacronym{aos}{AOS}{age of semantics}
\newacronym{itur}{ITU-R}{international telecommunication union radio communication sector}
\newacronym{uav}{UAV}{unmanned aerial vehicle}
\newacronym{kl}{KL}{Kullback–Leibler}
\newacronym{lidar}{LiDAR}{Light Detection and Ranging}
\newacronym{vr}{VR}{Vietoris-Rips}
\newacronym{wncs}{WNCS}{wireless network control system}
\newacronym{mpc}{MPC}{model predictive control}
\newacronym{aoi}{AoI}{age of information}
\newacronym{bs}{BS}{base station}

\title{Resilient-Native and Intelligent Next-Generation Wireless Systems:  Key Enablers, Foundations, \\ and Applications}
\begin{document}
\maketitle

\begin{abstract}
Just like power, water, and transportation systems, wireless networks are a crucial societal infrastructure. 
 As natural and human-induced disruptions continue to grow, wireless networks must be resilient. 
 This requires them to withstand and recover from unexpected adverse conditions, shocks, unmodeled disturbances and cascading failures. 
 Unlike robustness and  reliability, resilience is based on the understanding that disruptions will inevitably happen. 
 Resilience, as \textit{elasticity}, focuses on the ability to bounce back to favorable states, while resilience as \textit{plasticity} involves  agents and networks that can flexibly expand their states and hypotheses through real-time adaptation and reconfiguration. 
 This situational awareness and active preparedness, adapting world models and counterfactually reasoning about potential system failures and the best responses, is a core aspect of resilience. 
 This article will first  disambiguate  resilience from reliability and robustness, before delving into key mathematical foundations of resilience  grounded in abstraction, compositionality and emergence. Subsequently, we focus our attention on a plethora of techniques and methodologies pertaining to the unique characteristics of resilience, as well as their applications through a comprehensive set of use cases.  Ultimately, the goal of this  paper is to establish a unified foundation for understanding, modeling, and engineering resilience in wireless communication systems, while laying a roadmap for the next-generation of resilient-native and intelligent wireless systems.
\end{abstract}


\section{Resilience: Motivation}

\begin{figure*}
	  \begin{center}
      \includegraphics[width=.8\linewidth]{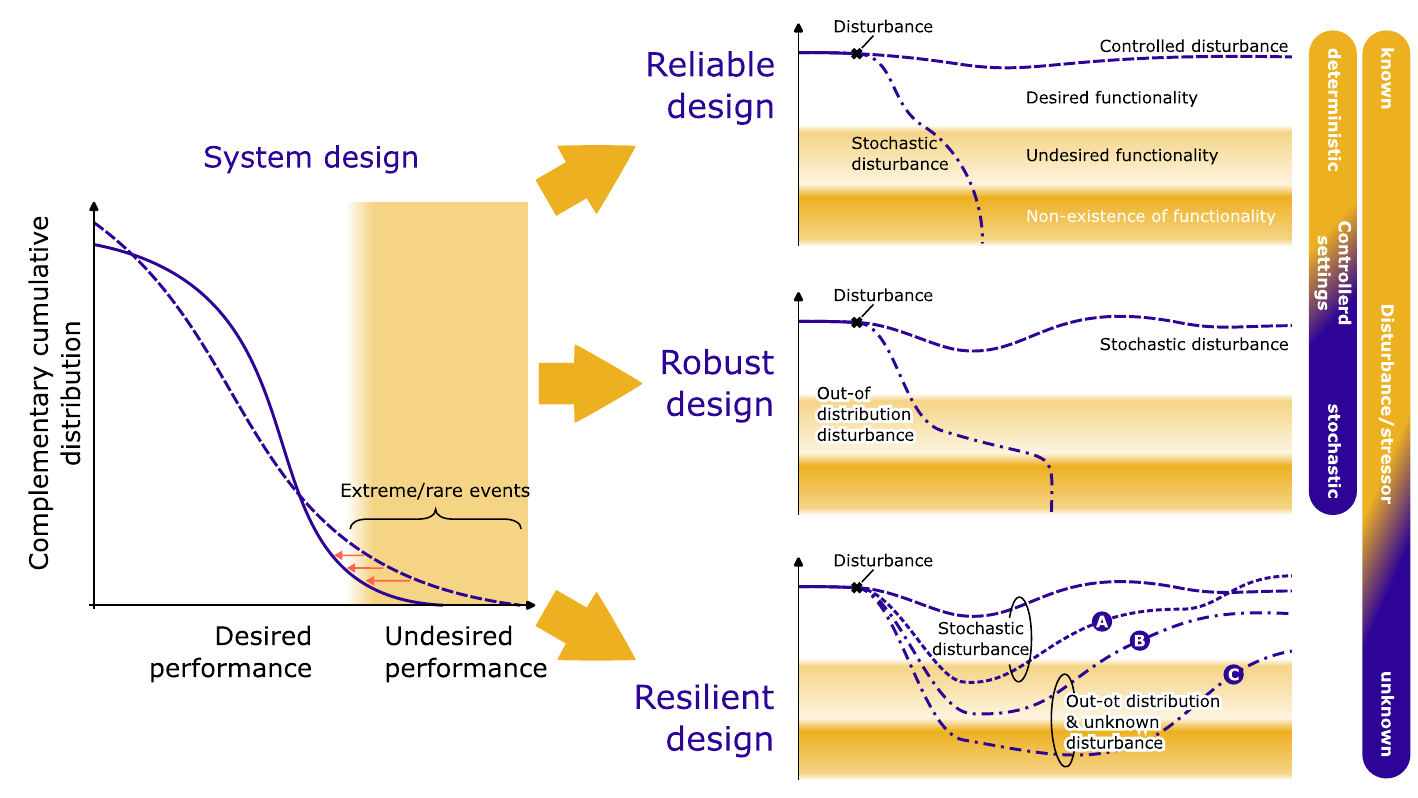}
	  \end{center} 
      \caption{The three R's of reliability, robustness and resilience.
      }
	\label{dino}
\end{figure*}

Resilience, arguably one of the \glspl{kpi} has been discussed in various fora, panels, and
articles \cite{ khaloopour2024resilience, vesterby_2022,9963527,weissberger2023imt2030,cioschina2023miit}.  
Yet to this day, resilience remains an elusive and ill-defined concept allowing for broad and varied interpretations in general.
Despite its crucial and far-reaching importance, the  mathematical foundations of resilience are sorely lacking \cite{ieeespectrum}. 
Resilience is often equated with reliability or robustness causing confusion. 
It is noteworthy that robustness refers to
the ability to withstand adverse conditions and accommodate known uncertainties, such as measurement errors and sensory/actuators' faults. 
Robustness can be modeled through various frameworks, including min-max optimization, stochastic optimization, or zero-sum game formulations incorporating risk-sensitive objective functions. 
On the other hand, reliability, a distinguishing feature of 5G, concerns the statistics of extreme/rare events, aiming to characterize  and tame the tail distribution of performance metrics, such as transmission rate and latency~\cite{8472907}. 
For instance, a link-level reliability of $99\%$ implies that, under known uncertainty (i.e., known unknowns), at most one packet loss occurs out of every $100$. 
However, these definitions fall short in scenarios  requiring performance guarantees during prolonged periods of consecutive losses. Furthermore, and more crucially, they provide no assurance of desired functionality under unexpected or unknown stressors (unknown unknowns). 
In sharp contrast to both reliability that suffers from stochastic disturbances and robustness which is affected by \gls{ood} disturbances, resilience assumes unforeseen events will occur, warranting the ability to  withstand  stressors, recover  and adapt in real-time as depicted in Fig. \ref{dino}. 
Stressors  can be known or unknown, ranging from node/network/component failure, misinformation, compromised sensor/node, adversarial  attacks, jamming, link/network outages,    environmental factors, and other unforeseen disruptions.

While both robustness and resilience address the impact
of stressors on system functionality, they differ in their objectives. 
In particular, robustness entails optimizing system performance under worst-case realizations of uncertainties or in expectation over their known distributions. 
As such, robustness focuses on maintaining performance levels within a predefined range of stressors or based on statistical characterizations of those stressors.
In contrast to robustness,  resilience does not aim to entirely prevent failures, rather, it is rooted in quick
recovery from disruptions and adaptation (i.e., plasticity) in response to unknown stressors or unforeseen variations in known stressors. 
This fundamental distinction arises from the capacity for online adaptation or configuration inherent in resilience, in contrast to offline design principles associated with robust approaches. Consequently, robust system designs, while effective against known uncertainties, often require significant resource over-provisioning, such as redundancy, to maintain performance under unknown stressors. This over-provisioning accounts for unknown variations of known stressors or anticipated types of stressors whose specific parameters are uncertain.
In contrast, resilient system designs prioritize adaptability and rapid recovery. This is achieved, for example, by allocating resources toward system identification (for detecting unknown stressors or changes in known ones) and dynamic reconfiguration, which may involve structural transformation of the network  rather than relying solely on static over-provisioning.
For instance, a network can be engineered to be robust against failures of up to $N$ nodes. 
However, if failures exceed this threshold, a purely robust design may become vulnerable. 
More importantly, designing for robustness against the worst conceivable scenario (e.g., failure of a very large number of nodes) becomes impractical and necessitates excessive over-provisioning. 
Therefore, resilience becomes essential, enabling the network to maintain continued functionality through recovery and adaptation under unforeseen events, obviating the need for impractical levels of static over-provisioning. 

The above discussion highlights a critical implication: the design choice between  robustness and  resilience inherently involves a tradeoff, particularly concerning resource efficiency. This robustness-resilience-efficiency tradeoff is a concept of crucial importance for future wireless network design, wherein robustness, as a precursor to resilience,  contributes to resilience while resilience can mitigate  events that cannot be handled through robustness. Furthermore, when unforeseen events  lie beyond the system's prior knowledge, resilience is the only viable solution.

Resilience is found at various microscopic and macroscopic scales, ranging from biological brains, animals (e.g., bird flocks, ants colony) to human-made power grids, cities and  other infrastructure systems.   Taking an inspiration from the brain, (loosely) seen as  networked, heterogeneous, hierarchical and modular control-loops that  sense their environments, perceive, abstract,  predict/infer, communicate  and plan accordingly, resilience can be seen through the lens of elasticity and plasticity. 
While elasticity  simply means returning to preferred states after disruption (case A in Fig. \ref{dino}), plasticity requires planning, structural reorganization and reconfigurability/transformation (cases B and C in Fig. \ref{dino}).

\subsection{State-of-the-Art}

Over the years, numerous articles, surveys, and books have been written on resilience across various fields such as ecology, climate, social networks, engineering, robotics, and cyber-physical systems (e.g., \cite{Royce, mitre, scheffer2009early,dai2012generic, barker2013resilience,fang2016resilience,liao2018resilience,STERBENZ20101245, gao2016universal,hulse2022understanding,wu2024towards}, to name just a few).  
More recently, resilience has gained significant attention in the context of future 6G networks \cite{khaloopour2024resilience, resil, brothersarms}, as evidenced by dedicated research programs like the NSF-funded Resilient \& Intelligent NextG Systems (RINGS I-II) \cite{WinNT}. This increasing focus is driven by the crucial role wireless networks play as a societal infrastructure and the growing impact of natural and human-induced disruptions \cite{brothersarms}. 

The existing literature discusses three conceptual interpretations of resilience: inertia (resistance), elasticity, and plasticity \cite{miller2022resilience,cottam2019defining}. 
Inertia enables a network to endure perturbations with minimal deviation from intended operations, relying on passive and preemptive mechanisms. Elasticity permits transient deviations but ensures a rapid return to the original state, characterized by preemptive and reactive responses. Neither entails structural adaptation and instead depends on static provisioning. In contrast, plasticity embodies a transformative form of resilience by actively detecting, reorganizing, and adapting internal and external structures to maintain functionality under previously unseen or evolving~conditions.

Despite the significant recent research interest in resilience \cite{khaloopour2024resilience, vesterby_2022, 9963527, ieeespectrum, resil, brothersarms}, it remains an elusive and ill-defined concept, allowing for broad and varied interpretations. 
Resilience is still frequently conflated with related concepts such as robustness and reliability, leading to conceptual confusion. While some prior works have discussed resilience in specific wireless contexts, such as security frameworks or as an add-on feature \cite{arfaoui2017security, chorti2022, gil2023physicality}, a holistic and unified definition tailored to the complex and dynamic nature of future 6G networks, which must handle mixed-criticality services and unpredictable stressors, is still lacking. Moreover, despite its crucial importance, a rigorous and much-needed mathematical formalism for defining and quantifying resiliency in a principled and consistent manner is sorely missing in the literature \cite{ieeespectrum, resil, brothersarms}. Existing metrics often focus on individual aspects like absorption, adaptation, or recovery time, or tend to be heavily biased towards one perspective, failing to capture the multi-faceted nature of resilience across different time scales and resource considerations \cite{najarian2019design,eldosouky2021resilient,shui2024resilience}.

Given the increasing severity and frequency of unforeseen events and attacks, coupled with the disaggregated, distributed, and virtualized nature of future networks, resilience has become a necessity rather than an option across diverse application domains~\cite{9963527}. Resilience offers the potential to redefine the current vision and requirements for 6G, which to date has been an incremental evolution of 5G  \cite{ieeespectrum}, \cite{8869705}. Worth highlighting is that the resilience of \gls{ai}/\gls{ml} models to unforeseen uncertainties  is  essential not only for maintaining \gls{ai} model performance, but also for upholding the overall resilience of future networks, given the increasing integration of \gls{ai}-native wireless systems~\cite{letaief2019roadmap,NextG-AI-Native-Wireless-Networks,hoydis2021toward,iovene2023defining}. In general, resilience in ML/AI plays a crucial role in addressing challenges such as out of distribution generalization, ensuring safety, and facilitating formal verification of models, especially for mission-critical applications, where current statistical models often fall short. 
As previously alluded to, unlike robustness and reliability, resilience does not focus on preventing failures based on static models of known unknowns with predefined stressors' characteristics, inducing   over-provisioned system designs that are still vulnerable to unforeseen uncertainties. Instead, resilient systems must dynamically adapt to unforeseen and unmodeled uncertainties and reconfigure or transform effectively to maintain the desired system functionality. Towards this goal, the formulation of fundamental questions and  development of mathematical frameworks tailored to the unique characteristics of resilience are critical, as we shall detail next.

\subsection{Contributions}

The main contribution of this paper is to establish a foundational framework for designing resilient next-generation wireless systems by systematically comparing and contrasting existing approaches. 
To this end, we raise four fundamental questions that address resilience from statistical, dynamical, topological, and logical standpoints, highlighting the importance of abstraction, anticipation, adaptation, compositionality, and  higher-order reasoning. 
We then discuss relevant resilience metrics along with critical tradeoffs between temporal characteristics, resource requirements, and stressor handling capabilities.
Building on these insights, we present key mathematical tools that support resilient system design suited for link- and network-level operations with both centralized and distributed decision-making capabilities. 
Finally, we demonstrate the applicability of these tools through a set of diverse use cases, supported by detailed numerical validations.

\begin{figure}[!t]
\centering
	\includegraphics[width=\linewidth]{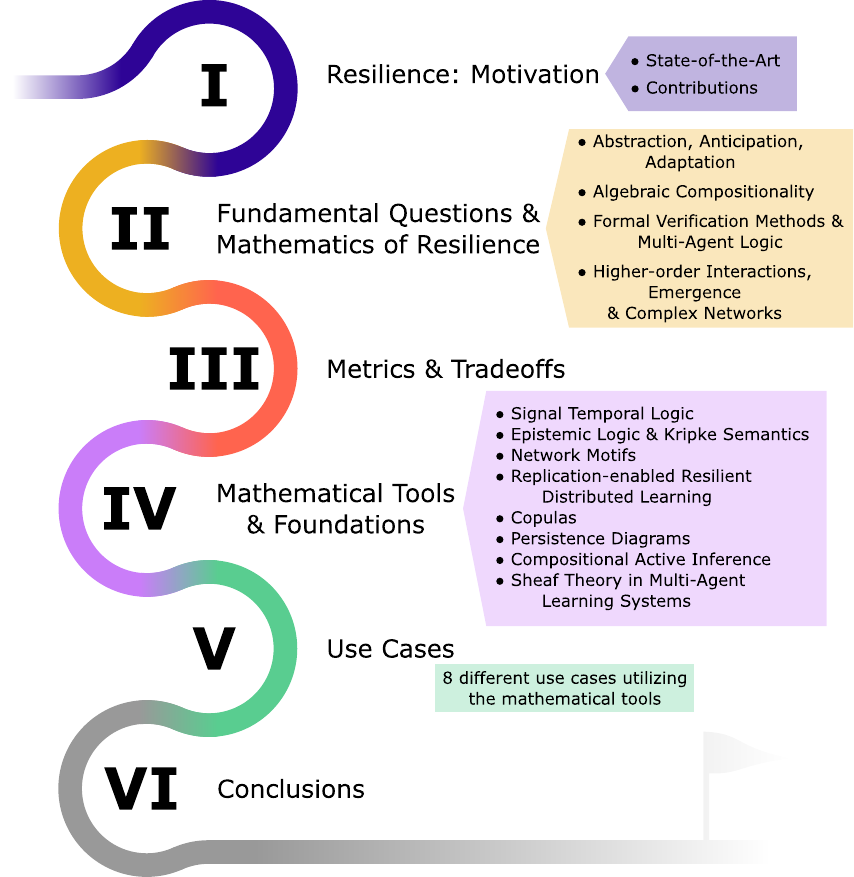} 
	\caption{Organization and roadmap of the article.}
	\label{fig:ToC}
\end{figure}

The remainder of the article is organized as follows:
Section \ref{sec:Fundamental-Questions and-Mathematics-of-Resilience} presents the four fundamental questions that guide the design of resilient systems, supported by relevant mathematical foundations.
Section \ref{sec:metrics_tradeoffs} introduces several qualitative and quantitative metrics for assessing resilience and discusses key trade-offs. 
In section \ref{sec:tools}, we detail a plethora of foundational mathematical tools applicable to resilient system design. 
Section \ref{sec:use_cases} illustrates the implementation of such designs through selected use cases, including validation results.
Finally, the conclusions are drawn in section \ref{sec:conclusions}.
For ease of reading, the structure of the paper is summarized in Fig.~\ref{fig:ToC}.

\section{Fundamental Questions and Mathematics of Resilience}\label{sec:Fundamental-Questions and-Mathematics-of-Resilience}

Studying resilience through first principles hinges on addressing several fundamental questions, sitting at the intersection of many mathematical disciplines. In what follows, we  formulate foundational questions that catalyze the development of resilience-oriented design methodologies within the context of networks of agents (hereafter referred to as \textit{networks}). In this respect, we outline prospective mathematical frameworks for  the characterization, design, and optimization of resilient systems. We assume that agents are endowed with key primitives, namely sensing (perception), computation and storage (local processing and memory), reasoning/planning and communication (grounded interactions). These primitives constitute the minimal desiderata for the  foundations of resilience. 

\subsection{Abstraction, Anticipation, Adaptation (A$^\textrm{3}$)}

\begin{tcolorbox}[colback=yellow!10, colframe=white!20!black, boxrule=1.9pt, boxsep=2pt,left=2pt,right=2pt,top=2pt,bottom=2pt,enhanced,breakable]
    \noindent \textsf{Q1}: For situational awareness and active preparedness, how do agents learn abstractions and concepts, referred to as statistical/geometrical/topological internal structures, from their (multimodal) sensorimotor signals? How can  agents detect time instances when their existing internal structures become inadequate,  necessitating structural modifications to preserve their functionality?
How to minimize the timelapses during which the existence of network functionality  is
compromised? 
    \end{tcolorbox}

\subsubsection{Learning Internal Structures (Abstractions)}

\begin{figure}[!t]
\centering 
    \includegraphics[width=.9\linewidth]{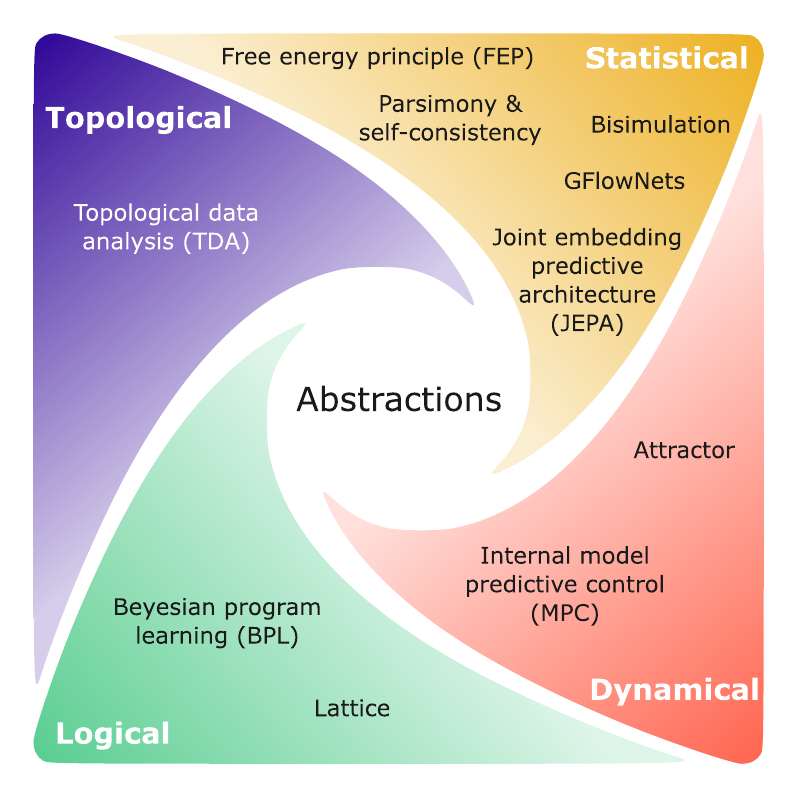} 
	\caption{Abstractions along the statistical, logical, dynamical and topological continuum.}
	\label{fig:Stat-Log-Topo-dyn-Continuum}
\end{figure}

Addressing the first part of \textsf{Q1} requires agents (or networks of agents) that sense, track and continuously update their internal structures about the external environment. Key methodologies for computing these structures can be explored through mathematical frameworks formulated within statistical, logical, topological, and dynamical paradigms, as illustrated in Fig.~\ref{fig:Stat-Log-Topo-dyn-Continuum}. \Gls{fep} \cite{FRISTON200670,friston2023path,WAFR} is a statistical inference formulation, which models the action-perception loop in terms of variational (approximate) Bayesian inference and self-consistent compressive control loops \cite{ma2022principles, dai2021closed}. Central to these statistical-based learning approaches is the ability to actively infer an internal compressed representation (i.e., an internal structure)  from the external environment via internal feedback. 
To model hierarchical abstractions such as preferences or multi-resolution sensory data, ordered algebraic structures like \textit{lattices} offer a rigorous and expressive framework. These structures enable the formalization of equivalence relations and refinement operations on sensory data and their representations, thereby facilitating efficient inference and providing significant structural compression gains~\cite{Hans_lattices,Lav_lattices}.
Moreover, from a formal logic standpoint, abstractions/concepts pertain to probabilistic programs, where concepts are represented as simple probabilistic programs and richer concepts are built compositionally from simpler primitives \cite{doi:10.1126/science.aab3050}. Unlike statistical inference (e.g., \gls{fep}), in program inference, agents aim to generate compressed mathematical theories (programs) of their external world, which entails searching for the minimal program from the space of all programs that explain their sensorimotor data.  From a dynamical systems perspective, abstractions/concepts  relate to low-dimensional attractors~\cite{Bengio_attractor}, which are learned (for e.g., using continuous Gflownets). In essence, Gflownets learn  a diverse distribution of  attractor states (discrete basins) by sampling trajectories of  states such that the distribution of final states is proportional to an unnormalized  target density \cite{bengio2023gflownet}. 

\subsubsection{Anticipation and Proactive/Look-ahead Planning}
Situational awareness and active preparedness as described in the first part of \textsf{Q1} necessitates mathematical formulations  for characterizing anticipation, which pertains  to agents' internal structures. Quite simply, anticipation relies on the internal structures of agents, which encode representations of past experiences and environmental models to enable planning, followed by decision making. The internal structures facilitate such proactive decisions by enabling agents to incorporate not only their own state and objectives but also inferred behaviors and potential intentions of other agents in the network. Anticipation and planning based on statistically grounded internal structures, often referred to as \textit{world models}, have a well-established body of literature, primarily developed within the \gls{ai}/\gls{ml} community, particularly in the context of single-agent settings~\cite[\S~8]{sutton-reinforcement-second-edition}, \cite{ha2018world}.
Predictive-decision making based on world models has also been extensively studied in model-predictive control, where \textit{model} refers to the plant's state dynamics, and in model-based\footnote{In contrast,  model-free RL, akin to perception-action loop, lacks planning whereby agents fail to adapt to local changes without re-experiencing the entire trajectory after changes.} RL, where a (world) model refers to  the state transition probabilities used for look-ahead planning with works going back to Schmidhuber et. al. 2018  \cite{DBLP:journals/corr/abs-1803-10122}. Very recently,  planning based on world models have witnessed a resurgence in the \gls{ai}/\gls{ml} community, as evidenced by Lecun's \gls{jepa}~\cite{lecun2022path, assran2023self, assran2025v}, latent state space models and many others. Beyond these statistical approaches, from a formal logic standpoint, planning is akin to automatic theorem-proving, where the agent's inference proposes solutions to prove theorems through a sequence of steps, and the world model sequentially checks the proof's correctness~\cite{Young_Theorem}. Moreover, agents can leverage counterfactual reasoning by simulating what-if scenarios, thereby improving agents' proactive decision-making.

\subsubsection{Fast Recovery with Adaptive Topologies and Dynamics}

The latter aspects of \textsf{Q1} (fast recovery) require computational frameworks capable of not only operationalizing adaptation but also optimizing resilience characteristics. Operationalizing adaptation requires that agents dynamically detect the need for change and respond accordingly, either by reconfiguring their internal structures or undergoing structural transformation. For example, in the case of \gls{nn}-based internal structures, this can be achieved through backpropagation, where appropriate error signals  update the parameters of \glspl{nn} maintained by each agent. On the other hand, the minimization of time intervals during which the system's functionality is compromised necessitates effective mathematical modeling that offers tractable formulations for performing the required optimizations. Alternatively, systematic procedures for tuning  parameters of the underlying optimization-based solutions must be investigated, e.g., changes to the stepsize rules for backpropagation in the preceding example.

Furthermore, owing to the diversity of real-world applications and heterogeneous resource requirements, current approaches, even those focused on statistically based internal structures or world models (e.g., \gls{jepa}), for learning  statistically-based internal structures or world models remain fundamentally limited in their scope. On the other hand, internal structures based on logical, topological, and dynamical primitives  offer a more expressive class of world models than their statistical counterparts,  enabling more effective representation of both known unknowns and unknown unknowns.  
However, the development of effective and scalable computational frameworks for enabling agents to learn such non-statistical structures remains unexplored. Similarly, computational methods that enable agents to learn a collection/hierarchy/category of sensorimotor world models through observations and interactions are challenging and  remain as open problems. Moreover, since world models may employ different syntaxes (e.g., probabilistic, geometric, mathematical theories), a rigorous methodology is needed for their integration and cross-modal knowledge transfer (e.g., combining vision-based and language-based models). In addition, due to epistemic uncertainty stemming from lack of knowledge, agents should have an innate curiosity by continuously foraging for information based on their current models and uncertainty, generating hypotheses based on their world models. Akin to prompt engineering,  besides world model and inference, agents need an attention policy to internally optimize/learn which information or agent to attend to (generate a prompt or query) for uncertainty reduction and maximum compression. 
It is important to emphasize that  adaptation  within resilience is  an online operation, triggered and guided by carefully engineered active detection mechanisms during network operation.  This warrants  efficient computational frameworks for modeling internal structures, regardless of whether they are statistical, logical, topological, or dynamical in nature.

The form of adaptation considered in \textsf{Q1} pertains to the concept of resilience as plasticity~\cite{miller2022resilience}, which emphasizes transformative capabilities through active detection, reorganization, and adaptation of internal structures or world models to sustain functionality in the face of novel or evolving conditions. Although resistance and elasticity may not inherently capture the adaptive dynamics central to A$^\textrm{3}$, they represent complementary dimensions of resilience that deserve further exploration.

\begin{tcolorbox}[colback=yellow!10, colframe=white!20!black, boxrule=1.9pt, boxsep=2pt,left=2pt,right=2pt,top=2pt,bottom=2pt,enhanced,breakable]
    \noindent \textsf{Q2}:    How to algebraically compose concepts/nodes/subsystems and their semantics to emerge a resilient system? Under which conditions does network resilience emerge out of its individual (and possibly non-resilient) components?   How do distributed algebraic structures with different syntax, language and models,  communicate and coordinate  to achieve a goal?   
    \end{tcolorbox}

\subsubsection{Algebraic Compositionality}

\begin{figure}[t]
\centering
	\includegraphics[width=\linewidth]{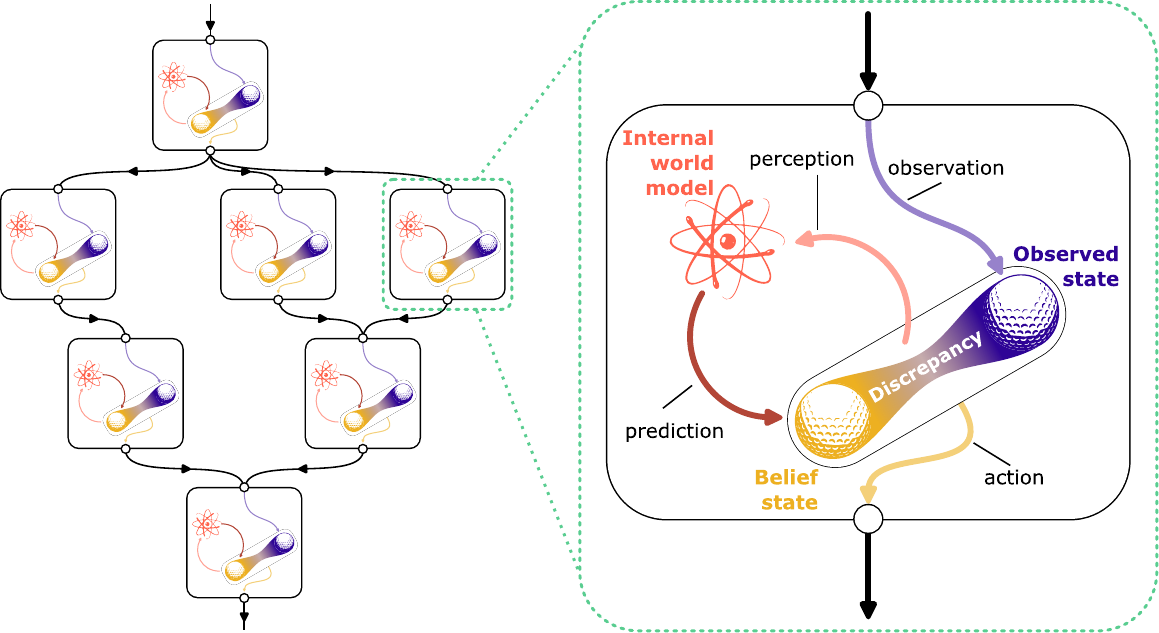} 
	\caption{Algebraic Composition of  Active Inference Loops. 
    }
	\label{compose}
\end{figure}

Humans possess an extraordinary cognitive ability to create high-level relational representations (abstractions) and combine them for quick adaptation and generalization in dynamic environments. 
This skill of algebraically composing concepts, subsystems, and bidirectional control loops (see Fig. \ref{compose}) plays a crucial role in solving a wide range of tasks, including reasoning, planning and communication. Broadly speaking, the design of resilient systems can be modeled as the composition of multiple interacting components, via carefully chosen algebraic operators, with the goal of analyzing and controlling the behavior of the resulting composite system. 
Moreover, given the distributed, interconnected, and multiscale nature of these inherently complex networks, understanding how resilience emerges and scales with the number of nodes, connectivity, topological structures, and their dynamics is essential. 
In fact, the resilience of a single agent does not guarantee the resilience of the network of agents. It is worth pointing out that the resilience of a network is linked to the network's global ability to maintain desirable functionality in the presence of unprecedented disruptions.
In contrast, the resilience of an individual agent is its ability to maintain or recover its own operational role (e.g., sensing, computing, storage, communicating) in response to unprecedented disruptions.
In this context, algebraic compositionality is a promising mathematical tool \cite{brogi1995fully,clarke2010semantic} to model how the resilience of individual components influences the overall resilience of the composite network, governed by well-defined  rules. As such, in addressing the second part of \textsf{Q2}, it is important to study and characterize the compositional resilience of agents to emerge a (globally) resilient network based on (locally) resilient agents or possibly non-resilient ones. Such a characterization  is key for defining suitable ordering primitives, for systematically comparing networks based on their global resilience, or agents based on  their local resilience.

\subsubsection{Integration of Heterogeneous Algebraic Structures}

Addressing the latter part of \textsf{Q2} requires mechanisms for the integration of heterogeneous algebraic structures for subsequent higher-order decision-making tasks. 
It is important to highlight that distributed algebraic structures, such as those arising from multimodal sensory inputs with heterogeneous syntactic and structural forms, as well as internal representations or world models, must interact/communicate across diverse operational contexts and stages (e.g., \textrm{A}$^\textrm{3}$) to ensure a desirable network functionality. 
In such settings, sheaf theory offers a promising and generalized mathematical framework for coherently integrating these structurally diverse algebraic entities~\cite{bredon_sheaf_1997}.
Such designs can be posed as consensus problems and solved using sheaf-theoretic frameworks that enable the composition of local structures into global ones~\cite{bredon_sheaf_1997}.

\begin{figure}\centering 
    \includegraphics[width=.9\linewidth]{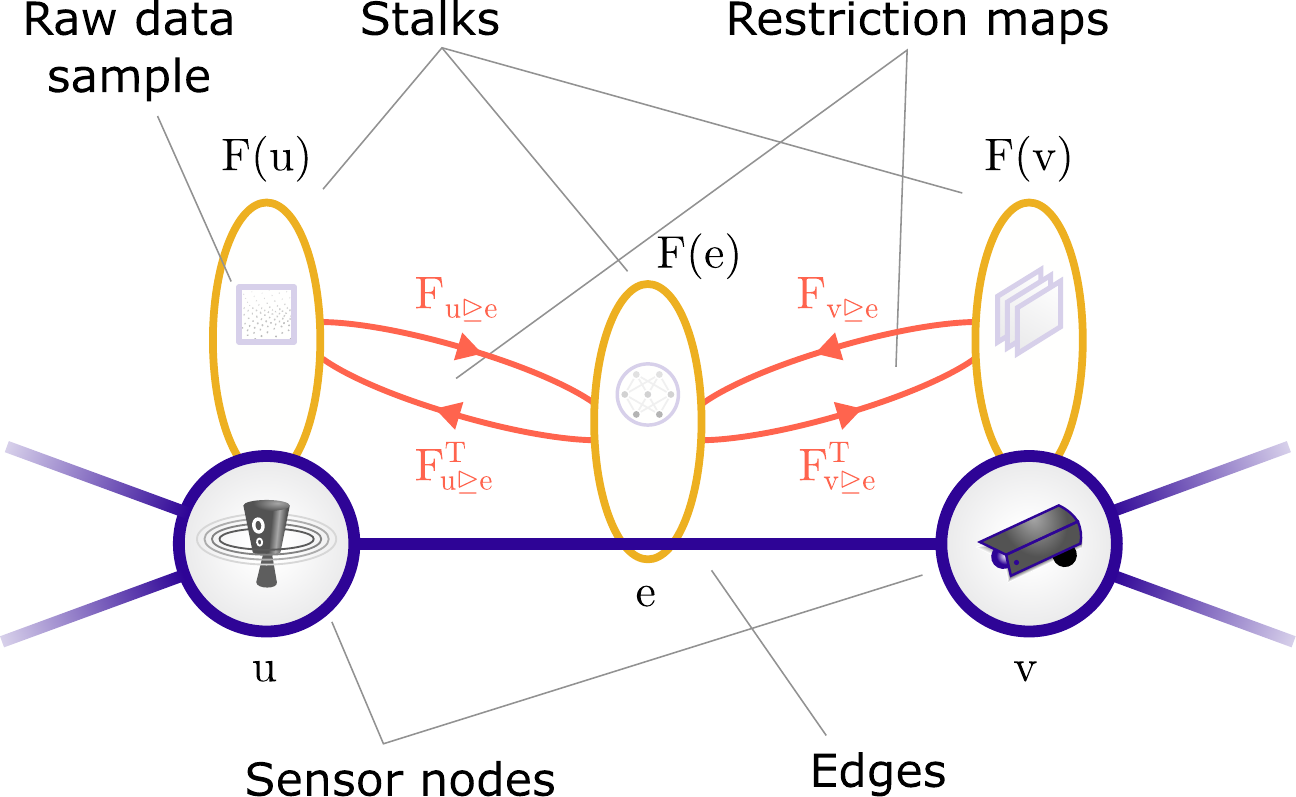} 
	\caption{Sheaf-theoretic modeling to communication  between a camera node and a wireless RF node.  Vector space-valued Sheaves (data structures) on edge ($e$) and vertices ($u,v$) with learnable (linear) restriction maps.}
	\label{sheaf1}
\end{figure}

For example, for different internal structures or world models among network agents, sheaf theory can resolve disagreements  enabling agents to align their beliefs through interaction to predict each other's actions and intentions  \cite{issaid2025tacklingfeaturesampleheterogeneity} (see Fig. \ref{sheaf1} for an illustration). 
Furthermore, for adaptability, which is an imperative for resilience, reasoning through the composition of knowledge or concepts to test hypothetical actions is crucial. In this respect, communication among agents is tantamount to agents \textit{composing} their internal information spaces (represented as a \textit{sheaf of world models})  for mutual predictability, a prerequisite for understanding. It is worth noting that a similar mechanism, inspired by enactivist approaches such as \gls{fep} and dynamical systems theory, is the compositionality of coupled transition or dynamical systems~\cite{Mahault2024}. 
Therein, each agent is modeled as a pair of coupled dynamical transition systems: an external system that interacts with the physical environment, and an internal system that evolves over a belief space. 
Belief states transition based on incoming observations, while communication is conceptualized as the composition or alignment of agents' internal belief spaces, effectively coordinating their world models for mutual predictability~\cite{Mahault2024}.

\subsection{Formal Verification Methods and Multi-Agent Logic}

\begin{tcolorbox}[colback=yellow!10, colframe=white!20!black, boxrule=1.9pt, boxsep=2pt,left=2pt,right=2pt,top=2pt,bottom=2pt,enhanced,breakable]
    \noindent \textsf{Q3}:  As networks are becoming disaggregated, distributed, and virtualized, how to certify and verify the  functioning, interoperability and explainability of models, protocols and networks? Beyond statistical methods, how do formal methods and   logical reasoning provide a rigorous modeling, analysis and optimization framework for resilience?
    \end{tcolorbox}

\subsubsection{Formal Verification \& Temporal Logic}

For mission-critical applications, resilience must be achieved in real-time with certificates on performance guarantees. This requirement precludes the flexibility of extensive offline computations and excludes reliance on heavy and empirical validation methods that are commonplace in most contemporary \gls{ai}/\gls{ml} systems and frameworks. Accordingly, reliance solely on contemporary \gls{ai}/\gls{ml} frameworks is inadequate for the design of resilient systems, underscoring the need for novel methodologies that enable resilience with formal performance guarantees, In this context, formal verification techniques and logical systems, such as temporal, epistemic, and modal logic, offer promising tools for certifying and explaining the proper functioning of models, communication protocols and networks, as well as ascertaining resilience under belief manipulation and  misleading~information. 

In particular, \gls{stl} formally provides high-level temporal  specifications that can be systematically translated into hard optimization constraints (i.e., safety-critical constraints) or soft objectives (best-effort criteria), providing a mathematical formulation for resilient designs with well-defined safety and performance limits~\cite{chen2023stl}.
Therein, a formal specification language is used to capture temporal dynamics (e.g., of a real-valued sensory signal) that can be encoded into constructs called logical formulas. 
An \gls{stl} formula is defined recursively according to some grammar and admits a quantitative semantics given by a corresponding real-valued function. 
The level of this function quantifies the extent to which the signal satisfies the formula, and it encodes hard constraints or an objective function of an underlying optimization problem formulated for resilience design.
In other words, the quantitative semantics of the logical specifications enable the incorporation of these requirements into an optimization framework for solving specific tasks \cite{chen2023stl}, \cite{GirgisSTL}. 
\Gls{stl} specifications have been employed as a rich language to formalize resilience in terms of quantifiers known as \textit{recoverability} and \textit{durability}. 
In terms of formal guarantees, \gls{stl}-based resilience specifications allow proving the soundness and completeness of their semantics. 
Moreover, in the context of networks of agents, the resulting \gls{stl}-based optimization problems inherently involve coupling in both constraints and objective functions. 
Consequently, appropriate decentralized and scalable optimization techniques are needed. 
Nevertheless,  standard \gls{stl} formulations typically assume the stressor is known and described by distributions or bounded uncertainty. 
Therefore, under unprecedented disturbances such as shifts in known unknowns or emergence of unknown unknowns, related frameworks must be  extended to address A$^\textrm{3}$.

\subsubsection{Multi-Agent Reasoning with Modal and Epistemic Logic}

In a similar vein, in multi-agent Kripke systems (modal logic), the meaning of a logical formula (Kripke semantics) is defined based on a set of possible worlds and a truth assignment to each formula at each possible world. This allows agents to reason about the truth of a formula in different possible worlds and modeling other agents' knowledge to refine their understanding of the (real) world for collaborative problem solving. Although certain approaches do exist, such as those based on hard-coded and/or external human verifiers, more principled approaches remain largely unexplored. Recent efforts include methods based on internal verifiers, as advocated in~\cite{Kambhampati_2024} and category-theoretic constructs that embed logical constraints into  structures~\cite{jia2024category}, allowing  to formally verify properties such as consistency, correctness, and safety of communication protocols and models. 
However, unlike \gls{stl},  modal logic frameworks generally lack quantifiable representations to encode the related logical reasoning into optimization frameworks to produce a practically feasible computational platform. This highlights a critical need for  bridging this gap and enabling the practical integration of modal logic primitives into computational frameworks. A similar research effort is also essential in the case of epistemic logic, where the constructs for formal reasoning about knowledge and beliefs must be encoded in computationally tractable representations to support effective resilience designs.

\subsection{Higher-order Interactions, Emergence  and Complex Networks} 

\begin{tcolorbox}[colback=yellow!10, colframe=white!20!black, boxrule=1.9pt, boxsep=2pt,left=2pt,right=2pt,top=2pt,bottom=2pt,enhanced,breakable]
    \noindent \textsf{Q4}: What is the impact of network topologies  (scale-free, small-world, grids, multiscale-nested, random) and their higher-order interactions and dynamics on the emergence of resilience?   What are the scaling laws of resilience in terms of node density, connectivity/topological structures? 
    \end{tcolorbox}

\subsubsection{Statistical and Topological Analysis Frameworks}

Complex networks, ranging from the Internet and transportation systems to power grids, biological networks,  and social structures, are dynamic structures woven into patterns of heterogeneity, modularity, and hierarchy. The diversity and connectivity of these structures have deep implications for how information is stored, processed, composed, and communicated. However, the extent to which these characteristics can be harnessed to enable \textit{emergent resilience}, as posed in \textsf{Q4}, remains underexplored and requires systematic investigations.

\begin{figure}[!t]
    \centering 
    \includegraphics[width=.9\linewidth]{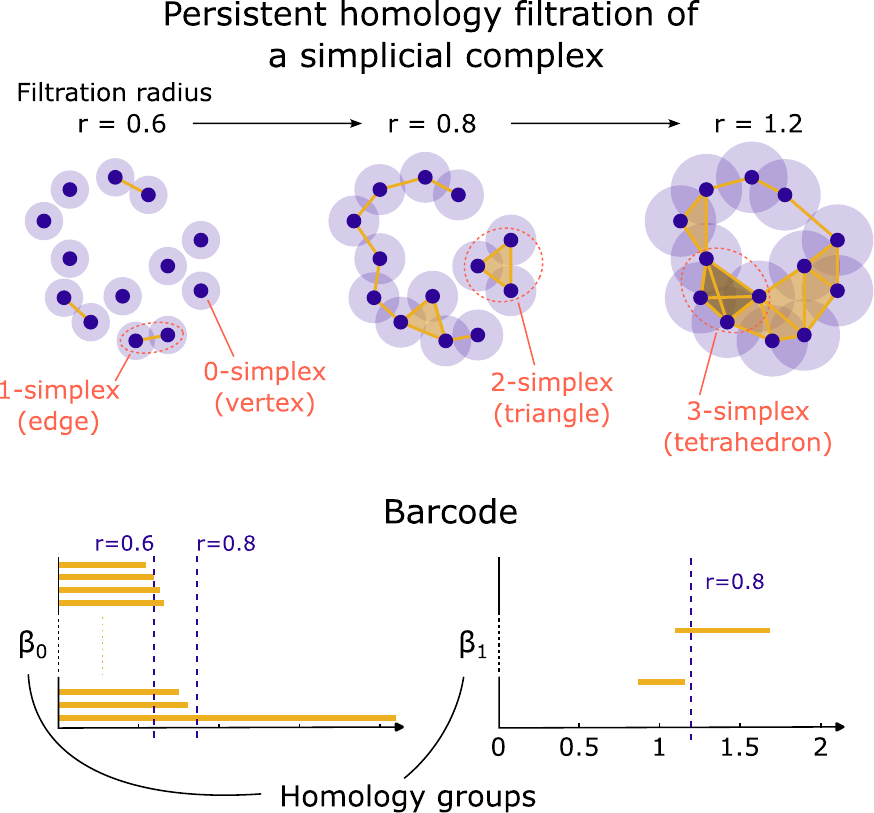}
	\caption{Simplicial complexes, persistent homological filtration and barcodes. } 
	\label{Fig:PD-Beeta}
\end{figure}

Owing to highly heterogeneous connectivity, most real-world networks of agents (or nodes) exhibit robustness against random internal failures caused by a malfunction or removal of a node or an edge. Unlike regular networks (e.g., grids/lattices), nodes in complex networks are not structurally or functionally identical. For example, the intrinsic heterogeneity of complex networks gives rise to diverse vulnerability patterns under failures or attacks. In homogeneous networks, where node degrees are relatively uniform, the impact of random failures or targeted attacks is generally consistent across nodes. However, in heterogeneous networks, such as scale-free networks, node degrees vary significantly, with a few high-degree hubs playing a critical role in connectivity. As a result, while random failures typically affect low-degree nodes and have limited impact, targeted attacks on hubs can severely disrupt the network's functionality.
Therefore, a systematic
understanding of how different network structures, including scale-free, small-world, and hierarchical, among others, respond to disruptions is essential for identifying network topologies that  enable more resilient network designs against a range of perturbations.  
In this respect, it is important to scrutinize not only lower-order global network connectivity features, such as node degree distribution, mean degree, small world properties, and betweenness centrality, but also higher-order local topological structures, such as \textit{motifs} \cite{dey2019network,milo2002network}. 
This motivates a motif-based analysis of resilience,  to examine how higher-order structures, potentially quantified through multivariate motif concentrations and motif lifetime distributions, influence both resilience properties and the global network topology. It is important to investigate diverse forms of local higher-order structures and their relations to the global network structure, as set forth in the latter part of \textsf{Q4}. Such analyses may employ methodologies from statistics, algebraic topology, dynamical systems theory, and related mathematical frameworks, as discussed in the sequel.

Seen through the lens of compositionality, a motif-based analysis of resilience  boils down to composing individual motif characteristics
to emerge a resilient network based on motifs as its primitives. As such, the necessary algebraic compositionality operations pertaining to individual motifs and their characteristics underscores the need for statistical frameworks for modeling beyond typical \gls{iid} assumptions and other mean-field type approximations towards characterizing inter-dependencies among different system elements, to assess how failures or disturbances in one component or system element affect others. A relevant and promising statistical framework for analyzing structure dependencies between random variables is copula theory \cite{copula}. Unlike traditional parametric distributions (e.g., multivariate Gaussian or $t$-distributions), Copulas link the joint distribution function with the marginal distributions, enabling the separation of the marginal distributions from the dependence structure. In particular, modeling the dependencies between a set of random variables (such as subsystems or components) first requires selecting appropriate marginal distributions and a copula. 
Then, 
based on data, parameters for both the marginals and the copula are estimated.
For instance, in collaborative multi-agent systems, mean-field approximations often fail to capture the complex dependency structure and coordination of joint actions. However, interestingly, the framework of copulas can explicitly model correlation and coordination by learning marginal distributions that capture individual agent behaviors, along with a copula function that models the dependencies among agents to accomplish a task. 

On the other hand, \gls{tda} provides a deep understanding of the characteristics of underlying network topologies, higher-order relational structures, and interactions in the functionality of a complex network~\cite{chazal2021introduction,lum2013extracting}.  Methods available in \gls{tda} extend well beyond  pairwise relationships typically characterized by traditional graph-based approaches and related analyses. One such tool, \gls{ph}, provides a unique view of data by capturing higher-order topological features~\cite{aktas2019persistence,wiseman2025persistent}. In particular, \gls{ph} uncovers topological invariants (shapes) of any sensory
data through a multi-resolution view of nested simplicial complexes. Such invariant characteristics include topological signatures, such as \glspl{pd} and Betti numbers, among others. Loosely speaking, these topological invariants capture the evolution of geometric properties of the underlying data, in terms of the number of connected components, loops, voids, and so on, at different scales specified by a scalar parameter, see Fig.~\ref{Fig:PD-Beeta}.
It is important to investigate the  potential of these intrinsic topological invariants, not only for the modeling and design of resilience but also for quantifying it. Their versatility stems from their ability to serve as sparse abstractions, encode structural semantics~\cite{asirimath2024rawdatastructuralsemantics},   capturing information about the persistence of underlying features.

\subsubsection{Dynamical Systems Perspective}

From a dynamical systems theory standpoint, understanding how higher-order interactions and their dynamics impact the emergence of resilience, how a system responds to perturbations and how this affects its structural and dynamical stability, is of utmost importance. In this context, complex networks are hierarchical or modular structures consisting of coupled non-linear high-dimensional dynamical systems. In this context, characterizing, inferring system resilience, and  predicting critical phase transitions and cascade failures, is a daunting task. To address it,   \textit{bifurcation theory} in dynamical systems  provides a systematic analysis of resilience in term of analysing changes in the behavior of a dynamical system as the parameters underlying the system vary,  defining regions of multi-stability~\cite{ananthkrishnan2009computational}. For example, the system's behavior can change from a stable operating point (equilibrium) to a new equilibrium, from an equilibrium to an unstable setting, or from an unstable setting to an equilibrium due to  changes of the underlying parameters. These critical parameter values  capturing system behavioral changes are called \textit{bifurcation points}. As such, from a resilience design point of view, canonical forms under which transitions to such system's changes happen are of crucial importance to facilitate $\textrm{A}^{\textrm{3}}$. 
Moreover, analyzing the bifurcation structure helps  anticipate how the system responds to changes (e.g., disturbances) and  provides adaptive mechanisms to steer the system away from undesirable operating regimes, potentially through optimization over the parameter space~\cite{melot2024multi}. Furthermore, the bifurcation characteristics within the parameter space can offer valuable guidance for developing quantifiable measures of resilience. The associated research remains largely unexplored and requires  further investigation.

\section{Metrics and Tradeoffs} \label{sec:metrics_tradeoffs}

Quantitative measures of resilience play a crucial role when evaluating how resilient a node or network is, as as well how can a system  withstand disruptions by quickly recovering and maintaining a desirable level of performance.  
In addition, these metrics are inherently connected to other competing system design objectives, such as energy consumption, computational load, latency and so forth thereby entailing a careful evaluation of such trade-offs among these objectives during the system design process.

\subsection{Metrics}\label{subsec:metrics}
Some of the primary metrics for resilience include resistance, recoverability (time it takes to recover after performance degradation), and durability (the extent to which functionality is preserved, whether elastic or plastic). Relevant metrics depend on the mathematical framework used, whether statistical, logical, dynamical, or topological. 
Recall  the three types of resilience designs, namely inertia, elasticity, and plasticity. Irrespective of the type, `resistance' (to change) is a primary metric that is to be quantified. On the other hand, in the case of elasticity and plasticity, a key metric is `recoverability', which can be quantitatively evaluated  based on the time it takes to recover after an unexpected performance degradation. Moreover, `durability' serves as another primary metric or an indicator of long-term system performance. The semantics of durability may change depending on the type of  resilience. For example, in the case of inertia-based resilience,  durability is similar to resistance   indicating how long/how much a system endures  unknown stressors or known stressor changes without structural failures. In contrast, durability in the case of elasticity-based resilience needs to capture the extent to which the system can repeatedly recover over time. For example, one can consider the number of unexpected intensifying stressor changes, the system can sustain by bouncing back to the original performance without  structural failures. On the other hand, in the case of plasticity-based resilience designs, durability  indicates the system’s ability to maintain performance in its newly adapted, post-disturbance configuration over time. In addition, in a plasticity-based resilience design,  when an unknown stressor change can no longer be overcome, resistance still delays the onset of degradation, while durability governs the rate at which the system degrades over time. However, a systematic investigation of metric definitions relevant to resilient-native intelligent next generation systems remains largely underexplored. In what follows, we show how the analytical tools, either statistical, topological, dynamical, or logical, considered as response to the fundamental questions presented in section~\ref{sec:Fundamental-Questions and-Mathematics-of-Resilience}  can help identify and define meaningful resilience metrics.

\subsubsection{Statistical Metrics}

From a statistical standpoint, resilience metrics focus on characterizing the probability distributions of system performance indicators, quantifying deviations from nominal behavior, and analyzing the system's ability to adapt its statistical models in response to disruptions. 
For instance, resistance can be quantified by measuring the statistical divergence (e.g., \gls{kl} divergence \cite{cover1999elements}) between the system's performance distribution under stress and its nominal distribution, or by tracking the probability of key performance indicators exceeding predefined critical thresholds. 
Recoverability can be assessed by the time required for key statistical moments (e.g., mean, variance) or the entire performance distribution to converge back to the desired baseline after a disruption, often identified using statistical change-point detection techniques \cite{basseville1993detection}. 
Furthermore, a copula-based analysis offers powerful metrics by characterizing the dependence structure between multiple system variables or performance indicators \cite{nelsen2006introduction}. 
Changes in the estimated copula function under stress can quantify alterations in system interdependencies, providing a nuanced measure of resistance beyond marginal distributions. 
The rate at which the copula converges back to a baseline or adapts to a new stable structure serves as a metric for recoverability and plasticity, respectively. 
Moreover, durability relates to the long-term stability of statistical properties post-adaptation or the rate at which prediction errors (e.g., from internal statistical world models discussed in \textsf{Q1}) decrease and stabilize in the new operating regime. 
Information-theoretic measures like predictive information or entropy rates can also quantify the system's sustained predictability or information processing capacity under continued or evolving stress \cite{shannon1948mathematical,bialek2001predictability}.

\subsubsection{Topological Metrics}

From a topological standpoint, resilience metrics can be derived from the structural properties of underlying data representations, such as geometric configurations of network agents modeled as point clouds, and temporal data including inputs, outputs, and dynamic states, together with their time-delay embedding structures. For example,
an agent's or network's ability to maintain its intrinsic geometrical or topological properties can be captured by topological signatures, 
such as \glspl{pd} or Betti numbers \cite{carlsson2009topology}. 
\Glspl{pd} can identify stable geometrical/topological features (e.g. loops, voids) that persist across scales \cite{ghrist2008barcodes}, thereby enabling a quantitative assessment of the sustainability (resistance) of features under varying stressor perturbations or system configurations. 
\Glspl{pd} can also be employed to quantify discrepancies in system behavior using metrics such as bottleneck distances between segments of time-series data \cite{chazal2016structure}, which define recoverability and durability. 
Alternatively, measures, such as  persistence entropy based on \glspl{pd}, quantify how evenly distributed the lifetimes of features are. 
As such, possible links of these definitions for building relevant resilience metrics need further investigations.
Moreover, higher-order geometric features, such as motifs and their lifetime and concentration are key measures that can be used at both the system and network levels \cite{milo2002network}. 
Finally, there exist other topological resilience indicators, such as  the \gls{aos}, which measures the persistence of internal structures. On the other hand, in the area of distributed learning, $(r,s)$-robustness is a measure that characterizes how a network  resists misinformation, isolate faults, and continue functioning despite node-level adversarial behaviors \cite{sundaram2010distributed}.

\begin{figure}[!t]
\centering 
    \includegraphics[width=.9\linewidth]{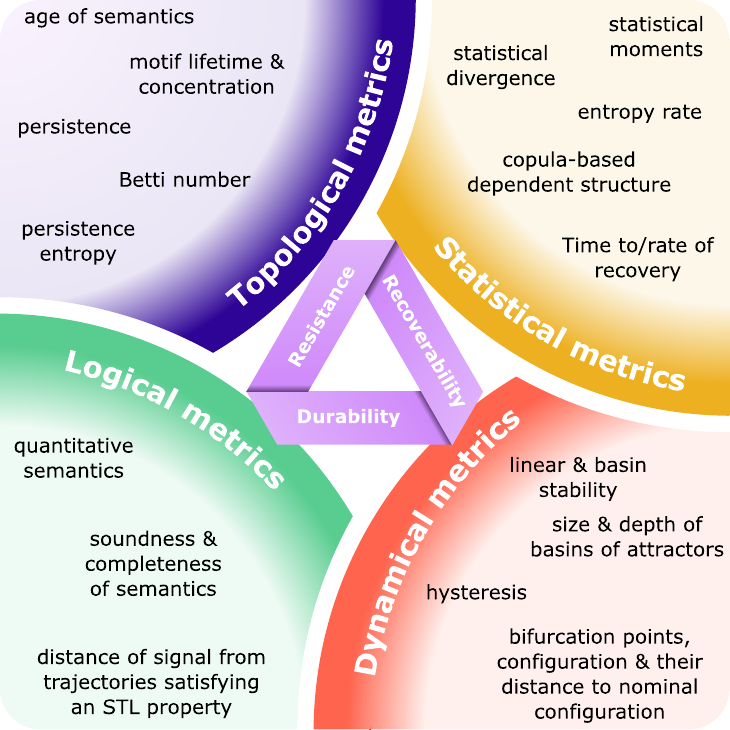} 
    \caption{Resilience metrics along the statistical, topological and logical continuum.}
	\label{fig:metrics}
\end{figure}

\subsubsection{Dynamical Systems Metrics}

From the perspective of dynamical systems theory, bifurcations  play a critical role in providing  important resilience metrics, as they mark qualitative changes in system behavior \cite{strogatz2024nonlinear}. 
These changes are intimately linked to the structure and stability of invariant spaces (e.g., attractors), which often undergo transformation at bifurcation points.  
For instance, the system’s response to perturbations, whether it remains near an equilibrium under small deviations in its state, shifts to a different equilibrium under large deviations in its state, or undergoes abrupt transitions due to slight parameter variations of the system (e.g., system's configurations), is governed by the geometry of equilibrium curves in bifurcation diagrams \cite{scheffer2009early}. 
These characteristics are inherently related to the system's resistance to change and its capacity to maintain its functionality under varying conditions. Moreover, it is worth exploring the utility of potential landscapes, representations of stability via high and low energy over system states, in gaining insights into system dynamics, particularly in low-dimensional dynamical systems \cite{truhlar2013potential}, which can be  leveraged for defining resilience metrics. 
For instance, when the system’s configuration or parameters are far from a bifurcation point, the basin of attraction in the potential landscape tends to be large, which corresponds to a faster rate of recovery from state perturbations. 
Conversely, as the system approaches a bifurcation point, the basin of attraction contracts, leading to a slower recovery from even small perturbations \cite{gao2016universal}. 
This highlights the relevance of basin size as a basis for defining resilience metrics, particularly those related to recoverability. 
Additionally, a shift in the system’s configuration can induce a transition from the current equilibrium to a different one as    pointed out earlier.
However, restoring the original equilibrium  cannot be achieved by merely reverting the configuration to its initial state. 
Thus, it is worth highlighting that this phenomenon, known as hysteresis, is closely associated with the system’s recoverability, and thus with resilience metrics capturing recoverability \cite{levins1968evolution}. 
Finally, from the potential landscape perspective, the depth of a basin of attraction quantifies a form of resistance, which defines resilience metrics associated with the system's resistance to perturbations \cite{guckenheimer2013nonlinear}. 
However, under complex dynamical systems as those expected in future next generation systems, leveraging the aforementioned characteristics to inform the development of relevant resilience definitions, appealing from both theoretical and practical standpoints, remains an open~challenge.

\subsubsection{Formal Methods based Metrics}

For mission-critical and safety-critical applications, purely statistical, topological, or dynamical measures alone are insufficient, necessitating  definitions for resilience metrics that ensure formal guarantees \cite{baier2008principles}. 
Therefore, it is essential to investigate formal methods and logical frameworks, such as temporal, modal, or epistemic logic that support the rigorous specification of resilient behaviors, enabling such properties to be intrinsically embedded within the definitions of resilience metrics \cite{maler2004monitoring,van1998common}.
In this context, for modeling resilience, using formal syntax, such as \gls{stl} formulas representing a specific temporal specification and corresponding semantics, such as sets of recoverability and durability pairs that satisfy the \gls{stl} formulas is important.

In summary, the following characteristic features should be taken into account, though not exhaustively, when defining resilience through the lens of resistance, recoverability, and durability.
These potential metrics are illustrated in Fig. \ref{fig:metrics}.
\begin{itemize}
    \item \textbf{Statistical}: statistical divergence, probability of key performance indicators exceeding predefined critical thresholds, classic statistical moments, copula-based dependence structure between system variables or performance indicators, predictive information/entropy rates. 
    \item \textbf{Topological}: Motifs concentration and lifetime, topological invariants (Betti number), recoverability and durability in terms of topological structures (persistence from \glspl{pd}), persistence entropy, age of semantics.
    \item \textbf{Dynamical}: Bifurcation points and their configuration, the size of basins of attractors, the depth of the basins of attractors (e.g., persistence of topology of the basins), distance to bifurcation points form nominal system's configuration, hysteresis.
    \item \textbf{Logical}:  Quantitative semantics using resilience satisfaction value functions.
\end{itemize}

\subsection{Tradeoffs}

Within the \gls{itur} framework IMT-2030, resilience is recognized as an important capability~\cite{NextG-ITU-R-Framework-IMT-2030,ITU-T-WTSA2024,ITU-R-M2160,ITU-R-M2516}.
However, achieving resilience  involves trade-offs with other performance indicators envisioned for future communication systems, spanning several key dimensions, as listed in the sequel.

\subsubsection{Robustness versus Resilience}\label{subsubsec:Robustness-versus-Resilience}

\begin{figure}[!t]
    \centering
    \includegraphics[width=.95\linewidth]{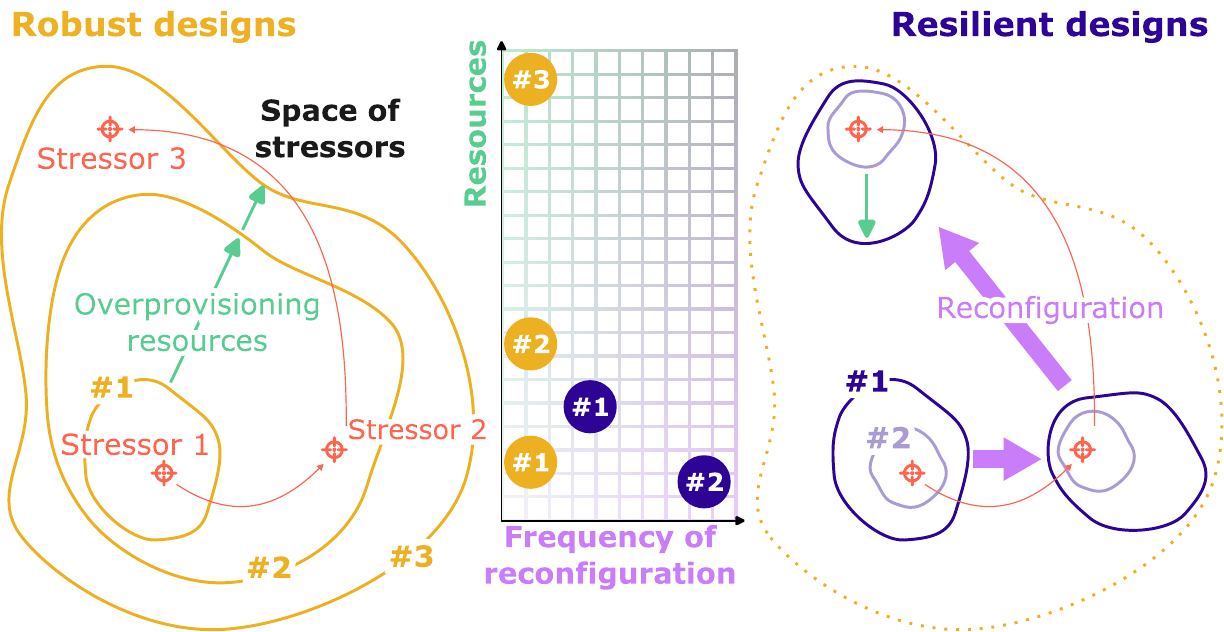}
    \caption{Illustration of how robust and resilient designs handle stressors.}
    \label{fig:robustness-resilience-with-uncertainty-sets}
\end{figure}

As alluded to earlier, robust systems are designed to withstand a predefined set of known disturbances, and consequently maintain the system's ability to meet its \glspl{kpi}. Most formulations typically assume worst-case scenarios for disturbances and are  handled based on a static one-shot resource configuration. As such,  handling unknown disturbances lead to  resource over-provisioning since a robust formulation is not designed to reconfigure resources in response to unknown disruptions. In contrast, resilient systems  reconfigure under unknown disturbances to meet the intended \glspl{kpi}.  In this context, a system should strike a balance between maintaining desired operations and preparing for unforeseen disruptions. 
A conceptual diagram is depicted in Fig.~\ref{fig:robustness-resilience-with-uncertainty-sets} highlighting the  interplay between  resources needed for over-provisioning for robustness and resources for designing a reconfiguration mechanism for resilience. If appropriately quantified (see Section~\ref{subsec:metrics}),  the pair (\texttt{robustness}, \texttt{resilience})  defines a system operating point  in the robustness–resilience space,   revealing a trade-off between the two. As such, depending on the application domain, given a fixed set of resources (e.g.,  computational capacity, memory, storage, power, and bandwidth),  the most relevant operating point is found. However,  a computationally tractable mathematical formulation for determining such a trade-off frontier between robustness and resilience is lacking, highlighting the need for a systematic study.

Note that the trade-off and  associated system operating points are inherently linked to the system's capability to satisfy a given \gls{kpi} either through robustness, resilience, or a hybrid combination thereof. From this perspective, robustness and resilience act as operational qualities rather than primary system capabilities. In particular, they govern the characteristics of the behavior of the system’s capability-related \glspl{kpi} in the presence of disruptions or uncertainties. Table~\ref{table:resilience_vs_robustness} lists some \gls{itur}, IMT-2030 capabilities and implications of robust-oriented and resilient-oriented designs on their behavior.

\renewcommand{\arraystretch}{1.25}

\begin{table*}[t]
\rowcolors{2}{myblue!10}{white}
\centering
\caption{Comparison of Resilience-Oriented vs. Robustness-Oriented Design for IMT-2030 Capabilities \cite{ITU-R-M2160}.}
\begin{tabular}{p{.22\linewidth} p{.35\linewidth} p{.35\linewidth}} 
\multicolumn{1}{c}{\cellcolor{myblue}\color{white}\textbf{IMT-2030 Capability}} & 
    \multicolumn{1}{c}{\cellcolor{myblue}\color{white}\textbf{Resilience-Oriented Design}} & 
    \multicolumn{1}{c}{\cellcolor{myblue}\color{white}\textbf{Robustness-Oriented Design}} \\
Peak Data Rate & Temporarily degrade but recover from disruption & Maintain peak rate under over-provisioned resources \\
User Experienced Data Rate & Focus on maintaining acceptable levels over time with slight disruptions & Focus on consistent performance with over-provisioned resources\\
Area Traffic Capacity & Dynamically redistributes load in response to failures & Over-provisioned for worst-case scenarios \\
Connection Density & Reconfigures resource allocation dynamically & Pre-allocated resources \\
Mobility & Allowed maximum speeds may drop momentarily  & Designed for maximum speeds with over-provisioning \\
Latency & Accepts temporary increase with fast recovery & Over-provisioned to ensure bounded latency  \\
Coverage & Recovers coverage gaps via reconfiguration (e.g., \gls{uav} repositioning) & Coverage is ensured via redundant infrastructure \\
Sensing-related Capabilities & Related accuracy, resolution, detection rate, and false alarm rate momentarily degrade & Ensures minimum degradation via redundancy techniques \\
Applicable \gls{ai} Capabilities & \gls{ai} models adapt, retrain, or switch input form in response to changing environments & Multiple \gls{ai} models trained under different environment conditions \\
Positioning & May tolerate temporary degradation and correct over time & Designed for high-precision in all possible conditions \\
\end{tabular}
\label{table:resilience_vs_robustness}
\end{table*}
\renewcommand{\arraystretch}{1}

\subsubsection{Sustainability versus Resilience}\label{subsubsec:Sustainability-versus-Resilience}

Sustainability, particularly environmental sustainability, refers to the system’s capacity to minimize greenhouse gas emissions and mitigate environmental impacts across its entire life cycle. Key considerations include enhancing energy efficiency, optimizing resource utilization, and enabling reuse and recycling. In this context, robustness-oriented designs are very likely less favorable compared to resilience-oriented designs, primarily because robustness often relies on over-provisioning, such as redundant infrastructure or excessive resources (e.g., power and bandwidth), which naturally leads to higher energy consumption and increased material usage. In contrast, resilient designs emphasize dynamic reconfiguration and adaptive resource management, enabling systems to respond to disruptions with minimal excess capacity, thereby promoting long-term environmental~sustainability. 

However, a clear trade-off can be identified between the energy expenditure and the reconfiguration capacity of a resilient system. Broadly speaking, reconfiguration capacity characterizes the system’s ability to autonomously transition to a stable operational state following a disruption, both in terms of response speed and the degree to which pre-disruption \gls{kpi} levels are restored. Higher reconfiguration capacity often requires increased computational and communication overhead, leading to higher energy~consumption. Specifically, increased energy consumption during reconfiguration may arise from processes such as real-time sensing and monitoring, state estimation, detection, control signaling for  decision making and parameter updates.

Note that sustainability, as defined by \gls{itur} in the context of IMT-2030, exhibits distinct characteristics compared to  the other system capability \glspl{kpi} listed in Table~\ref{table:resilience_vs_robustness}. In particular, when resilience mechanisms are employed to maintain or restore these capabilities under disruptions, the associated energy expenditure of the reconfiguration processes  influences sustainability outcomes. For example, consider the case of restoring coverage after an unforeseen disruption. A system equipped with greater computational resources and a more aggressive reconfiguration mechanism may achieve faster recovery of coverage, but this improvement comes at the cost of increased energy consumption, i.e., poor sustainability. Thus, the trade-off lies in balancing resilience (reconfiguration capacity) against the energy budget.

\subsubsection{Recoverability versus Durability}
\label{subsubsec:Recovery-versus-durability}

Unlike the trade-offs discussed in Sections \ref{subsubsec:Robustness-versus-Resilience} and \ref{subsubsec:Sustainability-versus-Resilience}, the  recoverability-durability trade-off  gives rise to conflicting demands  within the concept of resilience. Recoverability focuses on the system's ability to return to a functional state following a disruption, prioritizing responsiveness at the onset of a degradation and speed of reconfiguration to recover after the degradation. In contrast, durability emphasizes the system's capacity to maintain its functionality over extended time periods, even under prolonged or repeated stress, aiming to minimize degradation or rate of degradation over time. 

A system designed for high recoverability may  tigger fast resource optimization strategies  to accelerate recovery, at the expense of repetitive component reconfigurations and excessive energy consumption. Conversely, high durability leads to conservative system behaviors, limiting the aggressiveness of reconfiguration actions to preserve long-term system performance under repeated disturbances, probably with a slight degradation, resulting in slower recovery times. For example, consider a solar-powered swarm of \gls{uav} transceivers tasked with providing wireless coverage over a designated area. In response to unexpected and repeated disruptions, aggressively reconfiguring the \glspl{uav}' positions to immediately restore coverage can accelerate battery depletion, thereby compromising the system's overall durability. In such scenarios, a more conservative reconfiguration strategy that accepts a temporary reduction in performance and a slower recovery, may be preferable, as it allows sufficient time for the \glspl{uav} to recharge, ultimately enabling more persistent coverage over an extended period. 

Therefore, striking the right balance between recoverability and durability requires careful consideration of the system's operational context, together with the criticality of fast performance restoration versus longevity. Similar to the system's operating points discussed in section~\ref{subsubsec:Robustness-versus-Resilience}, the (\texttt{recoverability}, \texttt{durability}) operating points discussed in this section are   directly linked to the \gls{itur}, IMT-2030 system's capabilities listed in Table~\ref{table:resilience_vs_robustness}.

\section{Mathematical Tools and Foundations of Resilience} \label{sec:tools}

\begin{figure}[!t] 
\centering 
	\includegraphics[width=.9\linewidth]{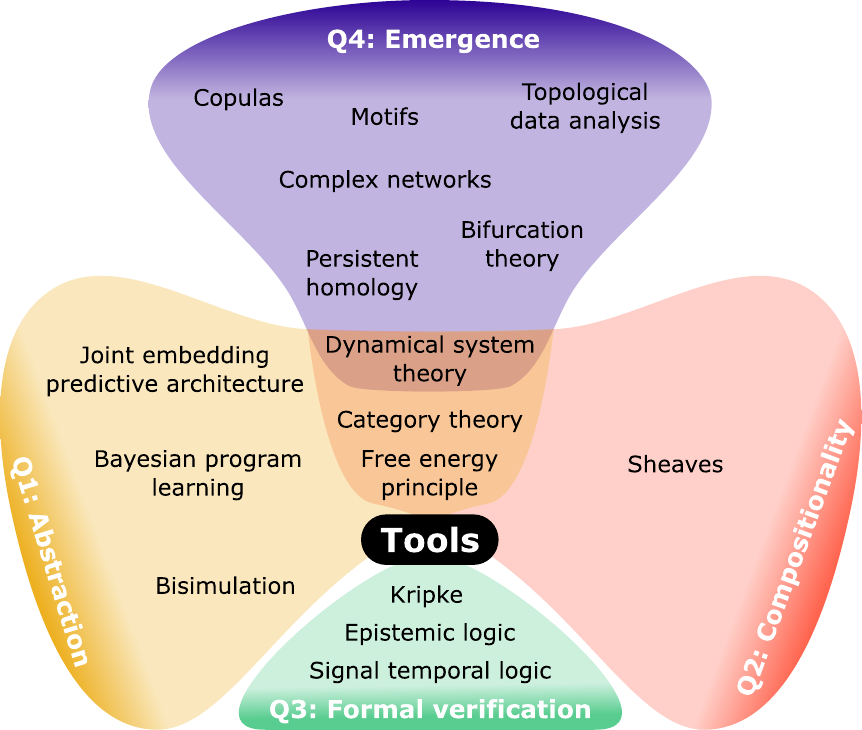} 
	\caption{Mathematical tools tailored to the unique characteristics of resilience.   }
	\label{fig:Raw-Material-Fundamental-Question-Association}
\end{figure}

In the endeavor to design resilient-native and intelligent NextG wireless systems, this section discusses in detail several potential mathematical tools addressing the fundamental questions posed in Section~\ref{sec:Fundamental-Questions and-Mathematics-of-Resilience}. The link between each mathematical tool or technique and the fundamental questions \textsf{Q1}-\textsf{Q4} is concisely illustrated in Fig.~\ref{fig:Raw-Material-Fundamental-Question-Association}.

\subsection{Signal Temporal Logic (STL)}

In classical logic, the propositions are evaluated as either true or false disregarding the notions of their temporal dependencies. 
Towards describing time-dependent behaviors of systems and reasoning about sequences of discrete events over time, \emph{temporal logic} is used. 
\Gls{stl} is an extension of the temporal logic framework that generalizes the analysis of discrete time events with Boolean propositions to continuous time real-valued signals~\cite{maler2004monitoring}. 

Formally speaking, we consider an $n$-dimensional signal $\signal:\timeDomain\to\realDomain^n$ defined over the time domain $\timeDomain$, the set of non-negative real numbers.
Then, an \gls{stl} formula $\formula$ is defined over $\signal$ as the equivalence $\formula \equiv \signalMapping\left(\signal(t)\right)\geq c$ with the mapping $\signalMapping:\realDomain^n\to\realDomain$ known as a predicate and a constant $c\in\realDomain$.
In this view, the notion of $\signal$ satisfying $\formula$ starting from time $t$ is denoted by $(\signal,t) \models \formula$.
The operators of negation ($\neg$) and conjunction ($\wedge$) for formulae $\formula$ and $\formula'$ are defined as follows:
\begin{align}
    (\signal, t) \models \neg\formula &\iff \neg\left((\signal, t) \models \formula \right), \\
    (\signal, t) \models \formula\wedge\formula' &\iff (\signal, t) \models \formula \wedge (\signal, t) \models \formula'.
\end{align}
To analyze the temporal conditions, three additional operators, \emph{eventually}, \emph{always}, and \emph{until}, are mainly used in \gls{stl}.
For a given time interval $\interval\in\timeDomain$, they are formalized as follows {\cite{maler2004monitoring}}:
\begin{align*}
	\textit{Eventually:} &&
    \!\!\!\!\!(\signal,t) \models \eventually{\interval} \formula
	& \!\!\iff
	\!\!\exists t' \in t + \interval, (\signal,t') \models \formula, \\
	\textit{Always:} &&
    \!\!\!\!\!(\signal,t) \models \always{\interval} \formula
	& \!\!\iff
	\!\!\forall t' \in t + \interval, (\signal,t') \models \formula, \\
	\textit{Until:} &&
    \!\!\!\!\!(\signal,t) \models \formula \until{\interval} \formula'
	& \!\!\iff
	\!\!\left( \exists t' \in t + \interval, (\signal,t') \models \formula' \right) \\
    &&&\phantom{000}\wedge \left( \forall t'' \in [t,t'], (\signal,t'') \models \formula \right).
\end{align*}
The until operator describes one formula must be held until the other becomes true. 
As such, the eventually operator that presents the notion of satisfying a formula at any time can be denoted by $\eventually{\interval}\formula = \top \until{\interval} \formula$ where $\top$ denotes the always true proposition.
Similarly, the always operator describes the formula is held true for all times followed by the interval is equivalent to $\always{\interval}\formula = \lnot \eventually{\interval} \lnot \formula$. 
Under these semantics, the \gls{stl} syntax is formally represented by the grammar \cite{maler2004monitoring}, 
\begin{equation*}
    \formula ::=  \signalMapping \,\vert\, \lnot \formula \,\vert\, \formula \lor \formula \,\vert\,  \formula \until{\interval} \formula'.
\end{equation*}

Towards defining quantitative semantics of the above syntax, we focus on two definitions: (space) robustness and resilience.
The former quantifies the degree of satisfying $\formula$, in which, its value exhibits the distance of $\signal$ from the set of trajectories that either satisfy or violate $\formula$ (i.e., $\signalMapping\left(\signal(t)\right)\geq c)$.
The robustness is formally presented as follows \cite{donze2010robust}: 
\begin{equation}\label{eqn:stl_robustness}
	\stlRobustness{\signalMapping'} = \signalMapping'\left(\signal(t)\right) \quad \text{where} \quad \signalMapping'\left(\signal(t)\right) = \signalMapping\left(\signal(t)\right) - c.
\end{equation}
 \gls{stl}-based resilience characterizes recoverability and durability, in which  recoverability ensures that a signal restores compliance with $\formula$ within a time limit of $\alpha$, while durability guarantees that the signal remains in compliance with $\formula$ for at least a duration of $\beta$.
In this view, \gls{stl}-based resilience is formalized as follows \cite{chen2022stl}:
\begin{equation}\label{eqn:stl_resilience}
	\stlResilience{\alpha}{\beta}( \formula ) = \neg\formula \until{[0,\alpha]}\always{[0,\beta]}\formula.
\end{equation}
To quantify the semantics of the above formulae,  a resilience satisfaction value based on a tuple $(\alpha-\tRecover,\tDurable-\beta)$ is used.
Here, $\tRecover = \min_\tau \left( \{\tau | (\signal,t+\tau) \models \formula\} \cup \{|\signal|-t\} \right)$ defines the time required to recover from violating $\formula$ at $t$.
Then, the durability in terms of satisfying $\formula$ over a period ($t+\tRecover$) is quantified by $\tDurable = \min_\tau \left( \{\tau | (\signal,t+\tRecover+\tau) \models \neg\formula \} \cup \{ |\signal|-t-\tRecover \} \right)$.
Hence, the satisfaction of resilience can be optimized by jointly optimizing $\tRecover$ and $\tDurable$ \cite{chen2022stl}.


\subsection{Epistemic Logic and Kripke Semantics}
\label{sec:kripke}
\begin{figure*}[!t]
\normalsize
\begin{subequations}
\label{eqn:kripke}
\begin{align}
	&& (\kripke,\world) \models \prop
	& \iff
	\prop \in \valuationFunc(w), \\
	&& (\kripke,\world) \models \lnot \formula
	& \iff
	(\kripke,\world) \not\models \formula, \\
	&& (\kripke,\world) \models \formula \lor \formula'
	& \iff
	(\kripke,\world) \models \formula~\text{or}~(\kripke,\world) \models \formula', \\
    && (\kripke,\world) \models \knowledge{\agent} \formula
    & \iff
    (\kripke,\world') \models \formula, \forall \world'\in\worldSet  \text{with}~(\world,\world')\in\relation_{\agent,t}, \\
    &&(\kripke, \world) \models \always \formula 
    & \iff 
(M_{t'}, \world') \models \formula, 
    \forall t' \ge t~\text{ and }~\forall \world' \in \worldSet~\text{ with }~ (\world, \world') \in \relation_{\agent,t'}\\
    && (\kripke, \world) \models \formula \until{} \formula'
    & \iff (\exists t' \ge t, \forall \world' \in \worldSet~\text{ with }~(\world,\world')\in \relation_{\agent, t'}, (M_{t'}, \world') \models \formula') \\
    && \quad & \quad \quad \quad\land (\forall s \in [t,t'), \forall \world'\in \worldSet~\text{ with }~(\world,\world') \in \relation_{\agent,s}, (M_s,\world') \models \formula).
\end{align}
\end{subequations}
\hrulefill
\end{figure*}

In multi-agent systems, the dynamic evolution of each agent's beliefs is as critical as the  truths of the environment they inhabit. 
Classical propositional logic evaluates statements as either true or false, but it does not capture the evolution of agents' beliefs or account for the fact that different agents may have distinct perceptions of the same underlying reality. 
To overcome these limitations, we extend the logical framework by incorporating multi-agent epistemic logic, which enriches the language with epistemic operators, such as $\knowledge{\agent}$, to denote the knowledge or belief of agent $\agent$ among a set~$\agentSet$ of $\AGENT$ agents.

Let the set \(\propSet\) be a non-empty set of atomic propositions.
We define $\mathcal\languageKripke(\propSet)$ as the language defined over the agents that evolves over time consisting of formulae \(\formula\) generated by the following grammar \cite{Kripke}:
\begin{equation*}
    \formula ::= \top \,\vert\, \prop \,\vert\, \lnot \formula \,\vert\, \formula \lor \formula' \,\vert\,  \knowledge{\agent} \formula \, \vert \, \always{} \formula \, \vert \, \formula \until{} \formula',
\end{equation*}
where \(\prop \in \propSet\) represents an atomic proposition. 
Standard logical operations such as the implication operator \(\formula \rightarrow \formula'\) and \(\formula \land \formula'\) are defined in the usual way. In our framework, the environment refers to the actual, objective assignment of truth‐values to all atomic propositions at each moment in time, whereas a possible world represents any (perhaps hypothetical) assignment of truth‐values. We collect all such assignments (actual and hypothetical) in the set $\worldSet$. To assign semantics to the language \(\languageKripke(\propSet)\) while capturing the dynamic evolution of agents' beliefs, we introduce a time-dependent Kripke model defined as
\[
\kripke = \big( \worldSet, \{\relation_{\agent,t}\}_{\agent=1}^\AGENT, \valuationFunc \big),
\]
where
\(\worldSet\) is a non-empty set of possible worlds representing different possible assignments of the environment 
The time-dependent epistemic accessibility relation at time \(t\) for each agent $\agent\in\agentSet$ is denoted by $\relation_{\agent,t} \subseteq \worldSet \times \worldSet$.
Moreover, \(\valuationFunc: \worldSet \to 2^\propSet\) is a valuation function that maps each world \(\world \in \worldSet\) to the set of atomic propositions that are true in \(\world\).
Followed by the above definition, the satisfaction relation $(\kripke,\world) \models \formula$, for $\world \in \worldSet$ and $\formula \in \mathcal\languageKripke(\propSet)$, 
is defined recursively in \eqref{eqn:kripke} while always holding $(\kripke,\world) \models \top$ \cite{Kripke}.

In the existing literature, epistemic logic has not been employed to formally characterize  resilience. 
In contrast, we posit that epistemic logic offers a rigorous framework for analyzing resilience by capturing two essential components: recoverability and durability. 
These properties can be precisely expressed through a mutual knowledge operator that embodies the notion that ``everyone in the set $\agentSet'(\subset\agentSet)$ of agents knows \(\formula\)''. 
Formally, we define this operator as:
\[
\knowledgeCollective{\agentSet'} \formula = \bigwedge\limits_{\agent \in \agentSet'} \knowledge{\agent} \formula.
\]
This formulation encapsulates the mutual knowledge of the group of agents. In this framework, \emph{recoverability} can be understood as the ability of the system to re-establish the mutual knowledge \(\knowledgeCollective{\agentSet'}\formula\) after a disruption, whereas \emph{durability} refers to the persistence of this mutual knowledge in the face of sustained challenges. 
By grounding resilience in this formal epistemic structure, we lay a solid foundation for the systematic analysis and verification of resilient behaviors in multi-agent systems.

\subsection{Network Motifs}

Network motifs are small, recurring subgraph patterns that appear more frequently in a given network than in randomized counterparts \cite{milo2002network}. 
These motifs serve as fundamental building blocks that shape the network’s structural and functional properties. 
Their presence influences overall graph connectivity by introducing alternative paths, enhancing fault tolerance \cite{roy2020motifs}, and reducing network fragmentation \cite{prill2005dynamic,ni2022improve}. 
Certain motifs, such as triangles and feedforward loops, strengthen local connectivity, while others contribute to global cohesion by linking different network regions. 
Conversely, the absence or disruption of key motifs can create structural vulnerabilities, particularly in networks that rely on specific connectivity patterns for efficient information flow.

\begin{figure}[!t]
	\centering
	\includegraphics[width=\linewidth]{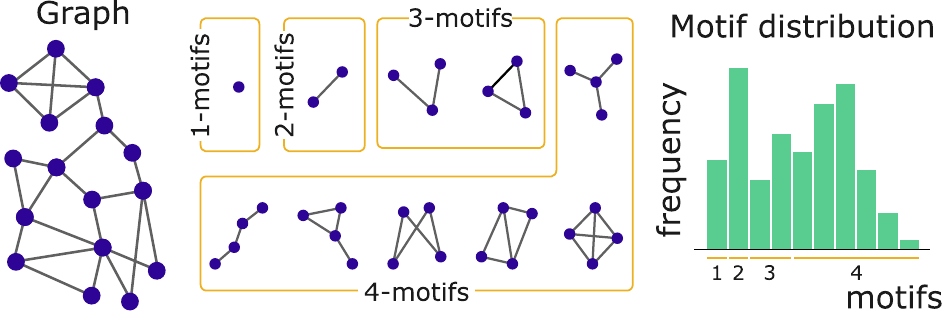}
	\caption{Illustration of up to 4-node motifs of a graph and corresponding  motif distribution.}
	\label{fig:network_motifs}
\end{figure}

Consider a network representation as a graph $ \graph = (\nodeSet, \edgeSet) $, where $ \nodeSet $ is the set of nodes (vertices) and $ \edgeSet \subseteq \nodeSet \times \nodeSet $ is the set of edges (links) between nodes. 
A \emph{network motif} is a small, recurring subgraph pattern $ \subgraph = (\nodeSet_\subgraph, \edgeSet_\subgraph) $ such that $ \nodeSet_\subgraph \subseteq \nodeSet $ and $ \edgeSet_\subgraph \subseteq \edgeSet $, forming a subgraph of $ \graph $ as shown in Fig.~\ref{fig:network_motifs}. 
A motif is considered significant if it appears more frequently in $ \graph $ than in a randomized counterpart $ \graph^* $, which preserves basic structural properties like degree distribution. 
Identifying motifs provides insight into the fundamental building blocks of network connectivity and resilience.

The \emph{absolute count (frequency)} of motif $ \subgraph $ in $ \graph $ is given by \cite{wong2012biological}:
\begin{equation}
	N_\subgraph(\graph) = \sum_{S \subseteq \graph, S \cong \subgraph} 1,
\end{equation}
where $ S \cong \subgraph $ denotes subgraphs $S$ in $ \graph $ that are isomorphic to $ \subgraph $. 
To assess statistical significance, the \emph{motif significance profile} (MSP) is defined as \cite{milo2004superfamilies}:
\begin{equation}
	Z_\subgraph = \frac{N_\subgraph(\graph) - \mathbb{E}[N_\subgraph(\graph^*)]}{\sigma(N_\subgraph(\graph^*)},
\end{equation}
where $ \mathbb{E}[N_\subgraph(\graph^*)] $ and $ \sigma(N_\subgraph(\graph^*)) $ represent the mean and standard deviation of motif occurrences in the randomized graph ensemble $ \graph^* $. 
The \emph{motif distribution} $\left(\motifDistribution=[\motif_\subgraph]_{\subgraph\in\set{\subgraph}}\right)$, describing the relative frequency of different motifs, is given by \cite{prvzulj2007biological}:
\begin{equation}\label{eq:motif-distribution}
	\motif_\subgraph = \frac{N_\subgraph(\graph)}{\sum_{\subgraph' \in \mathcal{\subgraph}} N_{\subgraph'}(\graph)},
\end{equation}
where $ \set{\subgraph} $ is the set of all considered motifs. 
These metrics collectively enable the quantification of motif prevalence and their role in maintaining network connectivity.
The statistical significance of motifs further informs how disruptions propagate and whether the network can self-reconfigure to sustain performance. 
By analyzing motif distributions, one can assess the extent to which small-scale structural patterns influence large-scale network robustness, offering insights into the fundamental mechanisms that govern resilient connectivity.


\subsection{Replication-based Resilient Distributed Learning}\label{Sec4_4}

\begin{figure*}[!t]
    \centering
    \includegraphics[width=.95\textwidth]{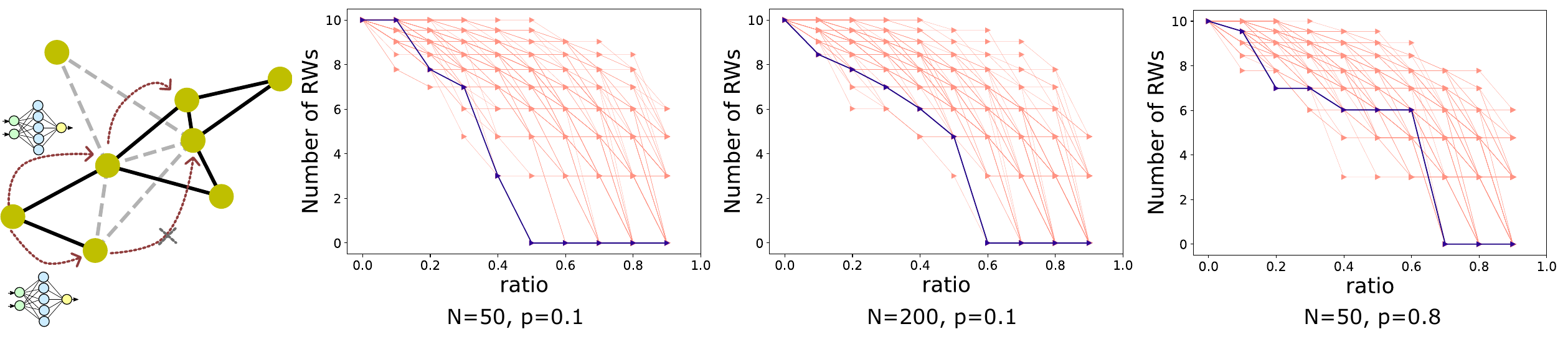}
    \caption{Lack of resilience in random walk distributed learning under link loss, where $N$ is the number of nodes and $p$ denotes the connectivity probability between any two nodes.    }
    \label{fig:sec4_4}
\end{figure*}

Distributed machine learning approaches such as federated learning \cite{chen2021distributed, tian2023distributed} enables the training of neural network models without the collection of data from local devices. Typically, as shown in Fig. \ref{fig:sec4_4},  a decentralized multi-agent communication network,  represented by an undirected graph $\set{G}=(\set{V},\set{E})$, where $\set{V}=\{1,\dots, N\}$ denotes the set of $N$ distributed agents/nodes and $\set{E}=\{\varepsilon_{ij}\}_{i,j \in \set{V}} $ represents the set of communication links between any two adjacent nodes. Let $\set{N}_i$ denote the set of all neighboring nodes connected to node $i$ and we denote the number of nodes in $\set{N}_i$ by $d_i=|\set{N}_i|$.
Each node has its own dataset from the global data distribution, which is denoted by $\set{D}_i=\{\set{X}_i, \set{Y}_i\}$. Distributed model training over the graph~$\set{G}$ is formulated as follows:
\begin{equation}
\label{eq_4_4_1}
\min_{\vect{\theta}} F(\vect{\theta}) := \frac{1}{N}\sum_{i=1}^{N}f_i(\vect{\theta}),
\end{equation}
where $\vect{\theta}$ denotes the model parameters to be learned, and $f_i$ is the loss function associated with the local training data $\set{D}_i$.

There are two  categories of algorithms to solve the problem (\ref{eq_4_4_1}) \cite{ayache2023walk}: (i) a gossip-based distributed learning and (ii) a random walk distributed learning. In the former, agents perform local computations  while sharing their updated parameters to all or a subset of neighboring nodes. While these algorithms are robust to nodes' failure, they often suffer from high computation and communication costs. To alleviate this issue, random walk type of distributed learning can be applied.   However, in practical distributed learning systems, with only \textit{one random walk} across the communication network, the system is \textit{not resilient} to potential links' or nodes' failures, calling for a resilient learning approach, requiring multiple random walks in the graph. 

In Fig. \ref{fig:sec4_4}, we initialize the system with $N_f=10$ and simulate the remaining number of random walks $N_f'$ under different ratio of links' failure, in three different communication network topologies with $p$ denoting the connectivity probability.
The resilience function \cite{vesterby_2022} is defined in the complex parameter space that characterizes the  system,  defined as $10\log_{10}N_f'$ for $N_f'\ge1$ and $0$ for $N_f'=0$. Under this definition, when the number of random walks over the networks becomes no more than $1$, the value of the resilience function is $0$, indicating the system is  not resilient and  distributed learning fails  under any additional perturbations.
The variation pattern across $50$ random realizations is shown in Fig. \ref{fig:sec4_4}, one of which reducing to $0$ is highlighted with blue. 
It can be observed that for small perturbations, the distributed learning system maintains its functionality.
However, under high ratio of nodes' or links' failure, the distributed learning over the network is not resilient due to the substantial loss of random walks. Thus, to ensure resilient distributed learning, the number of random walks should be kept at a well-chosen level, calling for  a decentralized mechanism for  nodes to dynamically recover the number of random walks in the system.

One  solution for enabling resilience in random walk distributed learning systems is through  \textit{replication} \cite{Eggerself2024}. The system information of the random walks and  communication graph topology is abstracted and described through the return time of  random walks at each agent, which is the time interval between receiving the specific random walk model. 
As the process goes on,  agents continuously estimate the number of surviving random walks by abstracting the return time distribution, through which the possible loss in the communication graph is calculated. Then, replication  of the model is conducted when the returning probability is low, to ensure system resilience. Specifically, let $R_i$ describe the first return time of a random walk model to node $i$ after leaving node $i$. At the $t$-th time instance, when a node $i$ receives $k$-th random walk, it measures a sample of $R_i$ by computing $t-L_{i,k}(t)$, where $L_{i,k}(t)$ denotes the last time node $i$ has seen this random walk $k$.
Then it accumulates the information by tracking $L_{i,k}(t)$, i.e., $L_{i,k}(t)=t$.
After some long enough initial time where all random walks of models have visited each node at least once, the nodes can obtain the established empirical \gls{cdf} of the return time established, denoted by $\hat{F}_{R_i}(t)$.
The survival time is defined as \cite{Eggerself2024}
$$
l(t-L_{i,k}(t)):= 1-\hat{F}_{R_i}(t-L_{i,k}(t)),
$$
representing the estimated probability of random walk models returning after time $t$, i.e., $\hat{Pr}(R_i>t-L_{i,k}(t))$, which is an abstraction of the random walk system.
Then whenever one node receives a model, using $R_i$, it can estimate the number of active random walks $N_t$ in the network. To do so, node~$i$~computes
$$
\hat{\beta}_i(t) = \frac{1}{2} + \sum_{l\in\set{L}_i(t)/\{k\}}l(t-L_{i,l}(t)).
$$
It can be proven that $\expect[2\hat{\beta}_i(t)]=N_t$ \cite{Eggerself2024}, thus when it is smaller than a threshold $\epsilon$, i.e., $\hat{\beta}_i(t)<\epsilon$, the node's local model is  replicated with probability $1/N_f$ so that on average, at most one model is replicated at a given time. In this resilient distributed learning system,  there exists an important trade-off between information redundancy and the plasticity/adaptability of the system,  determined by the threshold $\epsilon$ for replication. When $\epsilon$ is large,  nodes can respond quickly to system changes which however leads to higher redundancy and more energy costs. On the other hand, for small values of $\epsilon$,  recovery duration becomes longer while the energy cost can be small. Moreover, the network topology  impacts  resilience, whereby under nodes' or links' losses, a sparse topology leads to higher  loss probability of the random walks, calling for larger $\epsilon$.

\begin{figure*}[t]
\centering
\includegraphics[scale=0.3]{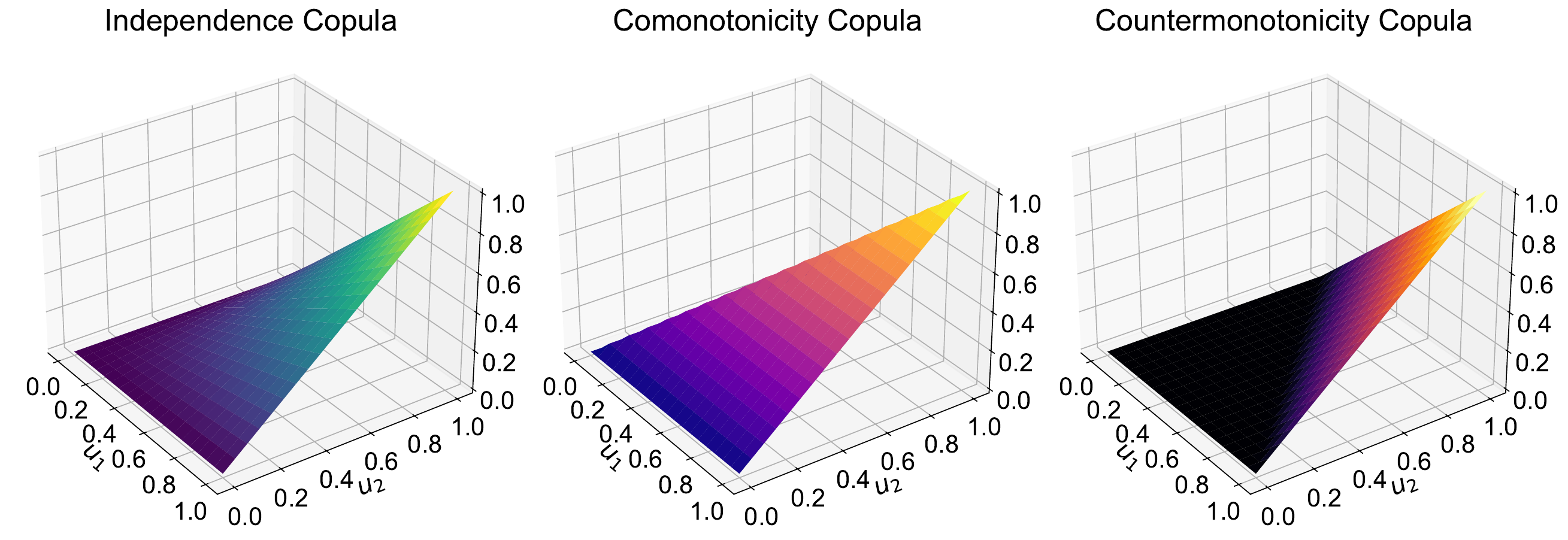}
\caption{Visualization of  three fundamental copulas: independence, comonotonicity, and countermonotonicity.}
\label{fig:fundamental_copulas}
\end{figure*}

\subsection{Copulas}\label{copulas}
\subsubsection{Definition and Properties}
A copula is formally defined as a multivariate \gls{cdf} defined on the unit hypercube $[0, 1]^d$, whose univariate marginal distributions are all uniform on $[0, 1]$ \cite{nelsen2006introduction, joe1997multivariate, mcneil2015quantitative}. More precisely, a function $C: [0,1]^d \rightarrow [0,1]$ is a $d$-dimensional copula if it satisfies the following properties
\begin{enumerate}
    \item \textbf{Grounding:} $C(u_1, \dots, u_d ) = 0$ if $u_i = 0, ~\forall i \in \{1, \dots, d\}$. This property ensures that the joint probability is zero if any of the marginal probabilities is zero.
    \item \textbf{Uniform Marginals:} $C(1, \dots, 1, u_i, 1, \dots, 1) = u_i, ~\forall i \in \{1, \dots, d\}$ and any $u_i \in [0, 1]$. This confirms that if we consider the probability distribution when all variables except the $i$-th one are unrestricted, the resulting marginal distribution for the $i$-th variable is the standard uniform distribution $\mathcal{U}([0, 1])$.
    \item \textbf{$d$-increasing Property (Non-negative Volume):} For any hyperrectangle $B = [a_1, b_1] \times \dots \times [a_d, b_d] \subseteq [0, 1]^d$ with $a_k \leq b_k, ~\forall k$, the $C$-volume $V_C(B)$ must be non-negative. This volume, representing the probability mass assigned by the copula to the hyperrectangle $B$, is calculated as
    \begin{align} \label{eq:d_increasing}
        V_C(B)=\sum_{i_1=1}^2 {\cdots} \sum_{i_d=1}^2 (-1)^{i_1+\dots+i_d} C(u_{1i_1}, \dots, u_{di_d}),
    \end{align}
    where $u_{k1} = a_k$ and $u_{k2} = b_k, ~ \forall k \in \{1, \dots, d\}$ with $V_C(B)\geq 0$. This property is the multivariate generalization of the requirement that $\prob(a \leq X \leq b) = F_X(b) - F_X(a) \geq 0$ for univariate \glspl{cdf}, ensuring that the probability assigned to any region is non-negative.
\end{enumerate}

The  connection between copulas and arbitrary multivariate distributions is established by Sklar's Theorem~\cite{sklar1959fonctions}. Sklar's theorem provides the theoretical foundation for copulas by establishing the relationship between multivariate distributions and their marginals.

\noindent\textbf{Sklar's Theorem.}\cite{sklar1959fonctions} \textit{Let $F$ be a joint $d$-dimensional \gls{cdf} with marginal \glspl{cdf} $F_1, F_2, \dots, F_d$. Then, there exists a $d$-dimensional copula $C$ such that $\forall x_1, \dots, x_d \in \realDomain$, we have
\begin{align}\label{eq:sklar_forward}
F(x_1, x_2, \dots, x_d) = C\Bigl(F_1(x_1), F_2(x_2), \dots, F_d(x_d)\Bigr).
\end{align}
If the marginals $F_1, \dots, F_d$ are continuous, then the copula $C$ is unique. Conversely, if $C$ is a copula and $F_1, \dots, F_d$ are univariate \glspl{cdf}, then the function $F$ defined by \eqref{eq:sklar_forward} is a joint \gls{cdf} with marginals $F_1, \dots, F_d$.}
\vspace{2mm}

Sklar's theorem states that any joint distribution can be decomposed into its marginals and a copula that solely describes the dependence structure. The arguments of the copula, $u_i = F_i(x_i)$, are the random variables transformed by their own \glspl{cdf}. By the probability integral transform, if $X_i$ is a continuous random variable with \gls{cdf} $F_i$, then $U_i = F_i(X_i)$ follows the distribution $\mathcal{U}([0, 1])$. The copula $C$ is precisely the joint \gls{cdf} of these transformed uniform random variables $(U_1, \dots, U_d)$. For continuous marginals, the unique copula can be explicitly expressed by inverting the marginal \glspl{cdf}, using quantile functions $F_i^{-1}$, as follows
\begin{align}\label{eq:sklar_inverse}
C(u_1, u_2, \dots, u_d)\!=\!F\Bigl(F_1^{-1}(u_1), F_2^{-1}(u_2), \dots, F_d^{-1}(u_d)\Bigr),
\end{align}
where $u_i \in [0, 1],~\forall i \in \{1, \dots, d\}$.

Copula theory elegantly frames fundamental dependence concepts. The Fréchet-Hoeffding bounds establish the range of possible dependencies \cite{frechet1951tableaux, hoeffding1940massstabinvariante}. For any bivariate copula $C(u,v)$, these bounds are given by
\begin{multline}
W(u,v) = \max(u + v - 1, 0) \leq C(u,v) \\
\leq \min(u,v) = M(u,v),
\end{multline}
where $W(u,v)$ is the \emph{countermonotonicity copula}, representing perfect negative dependence, and $M(u,v)$ is the \emph{comonotonicity copula}, representing perfect positive dependence. Between these extremes lies the \emph{independence copula}, i.e., $\Pi(u,v) = u \cdot v$, representing stochastic independence. Figure~\ref{fig:fundamental_copulas} visualizes the support of these fundamental bivariate~copulas.

\subsubsection{Copula Construction and Estimation}\label{construction}
\begin{figure*}[t]
\centering
\includegraphics[scale=0.28]{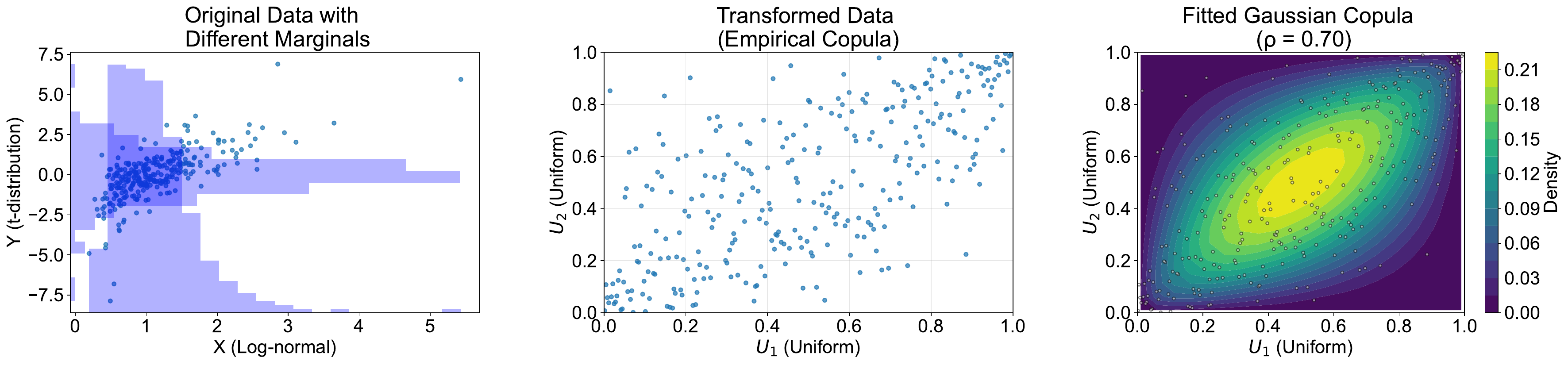}
\caption{Illustration of the copula construction process: original data with different marginals (left), transformed data with uniform marginals (middle), and fitted Gaussian copula (right).}
\label{fig:copula_construction}
\end{figure*}

Applying copulas in practice typically involves a multi-step process, often referred to as semi-parametric estimation  \cite{joe1996estimation, genest1995semiparametric}. First, for each variable $X_i$ in the dataset, its marginal \gls{cdf}, $F_i$, is estimated either parametrically (e.g., fitting a Normal, Gamma, or other distribution) or non-parametrically using the empirical \gls{cdf}. Second, the observed data $(x_{1j}, \dots, x_{dj})$ for $j \in \{1, \dots, n\}$ is transformed into pseudo-observations on the unit hypercube using the estimated marginal \glspl{cdf}, i.e., $(\hat{u}_{1j}, \dots, \hat{u}_{dj}) = (\hat{F}_1(x_{1j}), \dots, \hat{F}_d(x_{dj}))$, which should approximately follow the underlying copula distribution. Finally, a suitable parametric copula family (or potentially a non-parametric estimator) is chosen based on theoretical considerations or exploratory analysis of the pseudo-observations (e.g., scatter plots, dependence measures), and the parameters of the chosen copula model are estimated, often using maximum likelihood estimation on the pseudo-observations. This multi-stage approach allows for great flexibility by separating the modeling of marginal distributions from the modeling of the dependence structure \cite{patton2009copula}. Figure~\ref{fig:copula_construction} illustrates this transformation process schematically for bivariate data.

\begin{figure*}[t]
\centering
\includegraphics[scale=0.28]{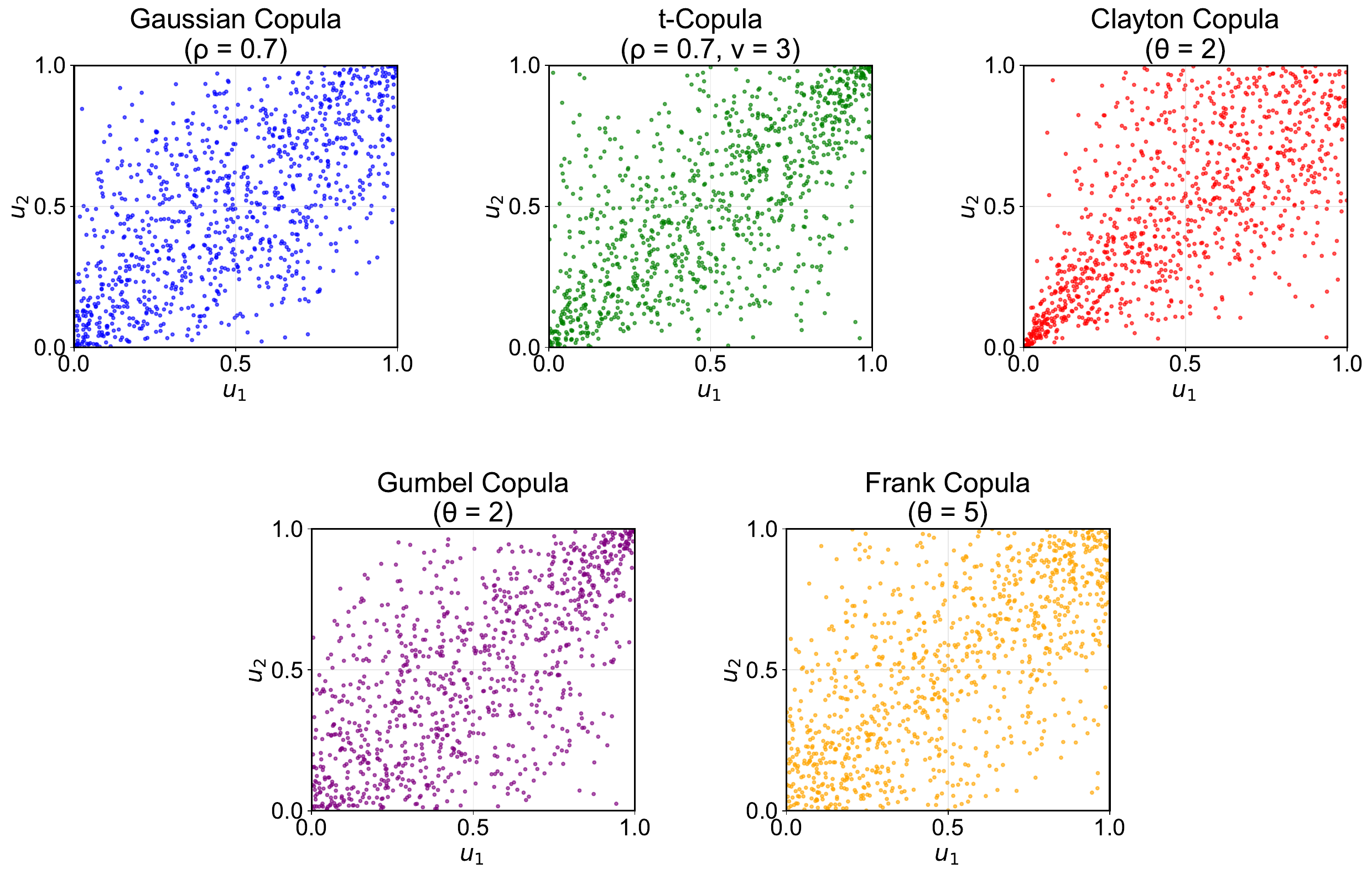}
\caption{Scatterplots of samples drawn from five common copula families: Gaussian, $t$, Clayton, Gumbel, and Frank.}
\label{fig:copula_families}
\end{figure*}

A variety of parametric copula families exist, each designed to capture different types of dependence structures \cite{nelsen2006introduction, joe1997multivariate}. Figure~\ref{fig:copula_families} shows scatter plots of samples drawn from the most common copula families, highlighting their distinct dependence patterns. The choice of copula family significantly impacts the resulting dependence structure, particularly for extreme events. While widespread due to its simplicity and interpretability through the familiar correlation parameter, the Gaussian copula \cite{li2000default} exhibits symmetric dependence but no tail dependence, potentially underestimating joint extreme events. The Student's $t$ copula \cite{demarta2005copula} extends the Gaussian framework by introducing symmetric dependence with both upper and lower tail dependence, whose strength is inversely related to the degrees of freedom parameter $\nu$. Among Archimedean copulas, the Clayton family \cite{clayton1978model} exhibits strong lower tail dependence but no upper tail dependence, with dependence increasing with parameter $\theta$, while the Gumbel family \cite{gumbel1960bivariate} displays the opposite pattern with strong upper tail dependence but no lower tail dependence, also increasing with $\theta$. The Frank copula \cite{frank1979simultaneous} provides symmetric dependence without tail dependence, similar to Gaussian, but distinctively allows for both positive and negative dependence. 

Vine copulas extend the flexibility of standard copula models to higher dimensions by decomposing a multivariate distribution into a cascade of bivariate copulas arranged in a hierarchical tree structure. Introduced by Joe \cite{joe1996estimation} and further developed by Bedford and Cooke \cite{bedford2001probability, bedford2002vines}, vine copulas provide a powerful framework for modeling complex dependencies in high-dimensional data where standard multivariate copulas become restrictive. The mathematical foundation of vine copulas lies in the pair-copula construction, which was operationalized and popularized by \cite{aas2009pair}, decomposing a multivariate density into a product of conditional bivariate copula densities. For a $d$-dimensional random vector $\vect{X} = (X_1, \ldots, X_d)$, the joint density can be expressed as
\begin{align}
&f(x_1, \ldots, x_d) = \prod_{i=1}^{d} f_i(x_i) \times \nonumber \\
& \prod_{j=1}^{d-1} \prod_{k=j+1}^{d} c_{j,k|D_{j,k}}(F_{j|D_{j,k}}(x_j|x_{D_{j,k}}), F_{k|D_{j,k}}(x_k|x_{D_{j,k}})),
\end{align}
where $f_i$ are the marginal densities, $c_{j,k|D_{j,k}}$ are bivariate copula densities, $D_{j,k}$ is a set of conditioning variables, and $F_{j|D_{j,k}}$ represents conditional distribution functions. The construction follows a nested pair-copula approach with three main types: C-vines (centered around key variables), D-vines (based on a sequential chain of dependencies), and R-vines (allowing more general tree structures). This hierarchical construction offers significant modeling advantages, including tailored dependency modeling between specific variable pairs, mixing different copula families within a single model, and accommodating both symmetric and asymmetric dependencies simultaneously. A key benefit is the ability to model conditional dependencies (e.g., how variables $X$ and $Z$ depend on each other, given variable $Y$), providing much more nuanced dependency structures than traditional approaches. For a comprehensive treatment of vine copula theory, estimation, and model selection, see  \cite{czado2019analyzing}.


\subsection{ Persistence Diagrams (PDs)} \label{subsec:Persistence-Diagrams}

This section focuses on  deriving  topological signatures, specifically the \gls{pd}, from point cloud observations. As such, we briefly discuss point clouds followed by a concise description of the computation of \glspl{pd}.

\subsubsection{Point Cloud Representation}
A point cloud \( \mathcal{X} \) is a collection of points embedded in an ambient space, typically Euclidean space, where each point encodes specific information about the underlying structure. From an engineering perspective, sensor technologies such as \gls{lidar}, cameras, and structured light systems can generate such point clouds. For instance, in a grayscale image, each point can be represented as \( (x,y,z) \), where \( (x,y) \) denotes the pixel coordinates, and \( z \) corresponds to the intensity of the \((x,y)\)-th pixel measured in grayscale. To effectively utilize point clouds for higher-order \gls{ml} tasks, it is essential to extract meaningful structural and topological information. One standard approach  is the computation of \glspl{pd}, which capture the underlying topological features of the data.  We next disucss the process of deriving \glspl{pd} from point cloud observations.

\subsubsection{Computation of the Persistence Diagram}

Given a point cloud, the computation of the \gls{pd} relies on a mathematical construct known as a filtration, which is a sequence of nested simplicial complexes. Accordingly, this section is structured into three subsections: (1) Simplicial Complex, (2) Filtration, and (3) Persistence Diagrams from Persistent Homology.

\paragraph{{Simplicial Complex}}

\begin{figure}[!t]
\centering
\includegraphics[width=\linewidth]{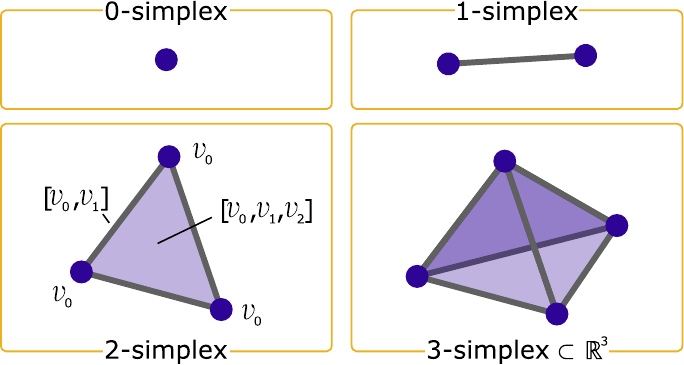}
\caption{Simplices in Euclidean space: $0$-simplex, $1$-simplex, $2$-simplex, and $3$-simplex.}
\label{fig:Simplicial-Complex}
\end{figure}

A simplicial complex extends the notion of a triangle to arbitrary dimensions. Informally, a $0$-simplex represents a point in Euclidean space, a $1$-simplex corresponds to a line segment, a $2$-simplex forms a filled triangle, and a $3$-simplex constitutes a filled tetrahedron as depicted in Fig.~\ref{fig:Simplicial-Complex}. This generalization continues for higher-order simplices, Specifically, a \( k \)-simplex \( \boldsymbol{\sigma} \), is defined as  
\begin{equation}\label{eq:simplex}  
\boldsymbol{\sigma}=\big\{\textstyle\sum_{i=0}^k \theta_i \vertex{i} \ \big | \ \textstyle\sum_{i=0}^k \theta_i{=}1, \theta_i \geq 0, i{=}0,\ldots,k \big\},  
\end{equation}  
where \( \vertex{0}, \ldots, \vertex{k} \) are affinely independent points in a \( d \)-dimensional Euclidean space \( \mathbb{R}^d \), ensuring that \( k \leq d \). Compactly, we denote $\boldsymbol{\sigma} = [\vertex{0},\ldots,\vertex{k}]$. The set \( \mathcal{V}=\{ \vertex{0}, \ldots, \vertex{k} \} \) is referred to as the vertex set. Given the $k$-simplex \(\boldsymbol{\sigma}\), one can compute a $(k-1)$-simplex in  \(\boldsymbol{\sigma}\) for each $0\leq j\leq k$ as $[\vertex{0},\ldots,\vertex{j-1},\vertex{-j},\vertex{j},\ldots,\vertex{k}]$, where the notation $\vertex{-j}$  indicates that $\vertex{j}$ is omitted. Similarly, given any $(k-1)$-simplex, one can compute all $(k-2)$-simplices. This procedure can be continued to compute the all simplices in $\boldsymbol{\sigma}$. For example, simplices of the $2$-simplex in Fig.~\ref{fig:Simplicial-Complex} are $\emptyset$, $[\vertex{0} ]$, $[\vertex{1} ]$, $[\vertex{2}]$, $[\vertex{0},\vertex{1}]$, $[\vertex{1},\vertex{2}]$, $[\vertex{0},\vertex{2}]$, and $[\vertex{0},\vertex{1},\vertex{2}]$. Moreover, an $m$-face of a $k$-simplex $\boldsymbol\sigma$ is an $m$-simplex in $\boldsymbol\sigma$, where $0\leq m\leq k$. For example, $[\vertex{0},\vertex{1}]$ is a $1$-face of $[\vertex{0},\vertex{1},\vertex{2}]$.

A simplicial complex  $\mathcal{K} \subset \mathbb{R}^n$ is a collection of simplices, constructed from simplices of the form  \eqref{eq:simplex}, that satisfies the following properties: (1) every face of a simplex in $\mathcal{K}$ is also included in $\mathcal{K}$ and (2) the intersection of any two simplices in $\mathcal{K}$ is either empty or a face common to both. 

\begin{figure*}[t]
\centering
\includegraphics[width=\linewidth]{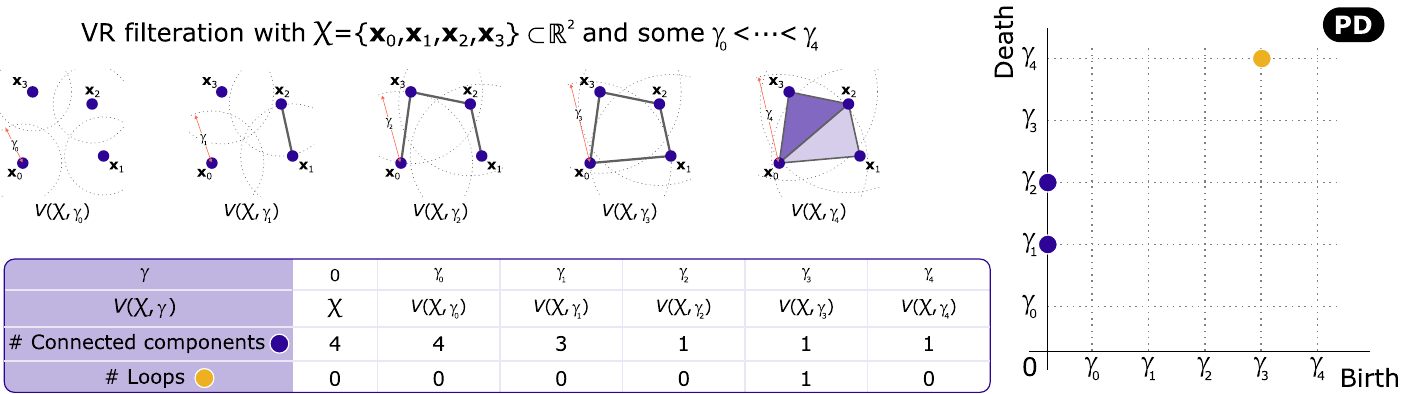}
\caption{An illustration of the VR filtration and the resultant progression of the connected components and loops with the PD.}
\label{fig:pd_materials}
\end{figure*}

\paragraph{\textbf{Filtration}}

A filtration is a sequence of simplicial complexes $\{\mathcal{K}_0, \mathcal{K}_1, \dots\}$ that plays a crucial role in generating the corresponding \gls{pd}. More specifically, a filtration of a simplicial complex $\mathcal{K}$ is a nested sequence of simplicial complexes $\emptyset \subset \mathcal{K}_0 \subset \mathcal{K}_1 \subset \cdots \subset \mathcal{K}_m = \mathcal{K}$. Different filtrations, such as the \gls{vr}, Čech, and Witness~\cite{de2004topological}, can be constructed on top of a given point cloud $\mathcal{X}=\{\PCvertex{0},\ldots,\PCvertex{n}\}$ to analyze its underlying topological structure. Here, we briefly explain only the \gls{vr} filtration to give insights into the subsequent developments. The construction of the \gls{vr} complex from the point cloud $\mathcal{X}$ is governed by a positive scalar $\gamma$ and is denoted $V(\mathcal{X}, \gamma)$. Roughly speaking, a $k$-simplex based on some $(k+1)$ points in $\mathcal{X}$ is included in $V(\mathcal{X}, \gamma)$ if all the $(k+1)$ points are pairwise within a distance $\gamma$. More specifically, for $\gamma > 0$, the \gls{vr} complex $V(\mathcal{X}, \gamma)$ with vertex set $\mathcal{X}$, is defined by the following condition:
\begin{multline}~\label{eq:VR}
  0\leq n_0<\cdots<n_k\leq n,  \ [\PCvertex{n_0}, \ldots, \PCvertex{n_k}] \in V(\mathcal{X},\gamma) 
  \iff  \\
  \ \forall k,\bar k\in \{0,\ldots,k \} \ \mathrm{s.t} \ k<\bar{k},  || \PCvertex{n_k}-\PCvertex{n_{\bar k}}||_2 \leq \gamma.
\end{multline}
It can be shown that for $\gamma_0 < \gamma_1 <\cdots <\gamma_t$, we have
\begin{equation}\label{eq:VR-Filtration}
   V(\mathcal{X},\gamma_0) \subseteq  V(\mathcal{X},\gamma_1) \subseteq \cdots \subseteq V(\mathcal{X},\gamma_t),
\end{equation}
a \gls{vr} filtration that is used to construct the topological signature \gls{pd}. 
Fig.~\ref{fig:pd_materials} shows the idea by using the point cloud $\mathcal{X}=\{\PCvertex{0},\ldots,\PCvertex{3}\}$ and some positive scalars $\gamma_0,\ldots,\gamma_4$.

\paragraph{\textbf{Persistence Diagrams from Persistent Homology}}

Loosely speaking, filtration enables the computation of the evolution of topological features, such as connected components, loops, and voids, as they appear and disappear. For instance, in the example in Fig.~\ref{fig:pd_materials}, one can observe the progression of connected components and loops as the scalar parameter $\gamma$ varies from $\gamma_0$ to $\gamma_1$, then from $\gamma_1$ to $\gamma_2$, followed by $\gamma_2$ to $\gamma_3$, and finally from $\gamma_3$ to $\gamma_4$. This is outlined in the table shown in Fig. \ref{fig:pd_materials}.

Note that when $\gamma = 0$, there are four connected components, each corresponding to an individual vertex of $\mathcal{X}$. This feature has not changed even when $\gamma = \gamma_0$. However, as $\gamma$ increases to $\gamma_1$, the vertex $\PCvertex{3}$ merges with $\PCvertex{2}$, reducing the number of connected components to three. Similarly, when $\gamma$ transitions from $\gamma_1$ to $\gamma_2$, the vertices $\PCvertex{0}$ and $\PCvertex{1}$ become connected to the simplex $[\PCvertex{2},\PCvertex{3}]$, thereby reducing the number of connected components to one. This single connected component remains unchanged for subsequent values of $\gamma$. 
A similar observation reveals that a loop emerges when $\gamma = \gamma_3$ and vanishes when $\gamma = \gamma_4$. From a computational topology perspective, the appearance and disappearance of topological features are formally referred to as \emph{birth} and \emph{death}, respectively. Consequently, the \gls{pd} is defined as a multi-set of points in the upper half-plane, given by $\{(x,y) \in \mathbb{R}^2 \mid x \leq y \}$. Each point in this set represents the birth (corresponding to the $x$-coordinate) and death (corresponding to the $y$-coordinate) of topological features, such as connected components and loops, as the scalar parameter $\gamma$ increases (see the exemplary \gls{pd} in Fig. \ref{fig:pd_materials}).

The mathematical theory used to compute \glspl{pd}, which capture the evolution of topological features across different spatial resolutions $\gamma$ of the filtration, extends beyond just connected components and loops. This theory is known as \gls{ph} and is well-established. However, the detailed mathematical background of \gls{ph} is beyond the scope of this document, and the reader is referred to \cite[\S~IV]{edelsbrunner2010computational},\cite[\S~1]{oudot2015persistence} for an in-depth treatment. It is important to note that \gls{ph} gives rise to persistent homology groups, denoted $H_p$, for each non-negative integer $p$. For instance, the $0$-dimensional \gls{ph} group $H_0$ tracks the evolution of connected components during the filtration process, the $1$-dimensional \gls{ph} group $H_1$ tracks the evolution of loops, and the $p$-dimensional \gls{ph} group $H_p$ tracks the evolution of $p$-dimensional features (e.g., voids for $p=2$).

\subsection{Compositional Active Inference} 
\label{sec:active_inference_theory}

Active inference is a  framework that unifies perception, learning, and action under the principle of variational free energy minimization. Under active inference,  agents maintain a probabilistic internal model of their environment and discrepancies between this model and sensory data (i.e., prediction errors) lead to  belief updates or actions that modify the external state to better conform to the agent’s expectations \cite{FRISTON200670}. This process effectively drives the external world toward alignment with the agent's internal~predictions.

Formally, consider an agent that receives observations \( o \) and seeks to infer the hidden states \( s \) that generate these observations via a generative model \( p(s, o) \). Since the true posterior \( p(s \mid o) \) is often intractable, the agent employs an approximate posterior \( q(s) \). The discrepancy between these two distributions is quantified by the variational free energy~\( F \):

\begin{equation}\label{eq:Free-Energy}
F[q(s)] = \underbrace{\mathbb{E}_{q(s)}\left[\ln \frac{q(s)}{p(s,o)}\right]}_{D_\text{KL}(q(s)\,\|\,p(s|o)) - \ln p(o)},
\end{equation}
where \( D_\text{KL} \) denotes the \gls{kl} divergence. Minimizing \( F \) indirectly minimizes the divergence between \( q(s) \) and the true posterior \( p(s|o) \), effectively performing approximate Bayesian inference.

Nevertheless, agents are not only confined to updating beliefs, but  they also actively select actions to shape sensory outcomes. For a given policy \( \pi \) (a sequence of actions), the expected free energy is given by,
\begin{equation}\label{eq:Expected-Free-Energy}
G(\pi) = \mathbb{E}_{q(s,o \mid \pi)} \left[\ln \frac{q(s \mid \pi)}{p(s,o)} \right].
\end{equation}
The optimal policy is then selected via
\begin{equation}
\pi^{*} = \operatorname*{arg\,min}_\pi\, G(\pi),
\end{equation}
ensuring that actions chosen minimize discrepancies between the agent's internal model and its sensory observations.

To scale active inference to multi-agent settings, we consider a set of autonomous agents $\mathcal N$, each equipped with limited sensing, local computation, and peer-to-peer communication. These agents must collaboratively infer a shared model of their environment, despite having heterogeneous local hidden states reflecting their distinct sensor modalities, perspectives, or internal representations. Each agent $n \in \mathcal{N}$ maintains a local belief $q_n(z)$ over the global hidden state $z$, capturing the common structure underlying their diverse observations. Each agent updates its local belief based solely on its own sensory modality and private hidden state \( s_n \),which encodes a localized or modality-specific view of the environment.

Then, agents fuse these locally informed beliefs into a globally coherent estimate by exchanging their beliefs with neighboring peers and performing iterative geometric mean followed by normalization. Specifically, at each iteration $t$, agent \( n \) updates its belief according to:
\begin{equation}
\label{eq:geom_mean_update}
q_n^{(t+1)}(z) \propto \left[q_n^{(t)}(z) \prod_{j \in \mathcal{V}_n} q_j^{(t)}(z)\right]^{\frac{1}{|\mathcal{V}_n| + 1}},
\end{equation}
where \( \mathcal{V}_n \) denotes the set of neighbors of agent \( n \), and the proportionality symbol indicates normalization to ensure \( \sum_z q_n^{(t+1)}(z) = 1 \). This decentralized update rule aligns the agents' heterogeneous local views into a consistent global belief. Importantly, resilience emerges as a property of the system through this structured composition of diverse beliefs. Individual agents may be unreliable, uncertain, or limited in their isolated perceptions; however, by systematically composing their heterogeneous views, the system as a whole becomes resilient. In other words, resilience arises by a structured composition of diverse and potentially imperfect local hidden states into a coherent global inference.

\begin{figure*}[!t]
    \centering
    \includegraphics[width=.99\linewidth]{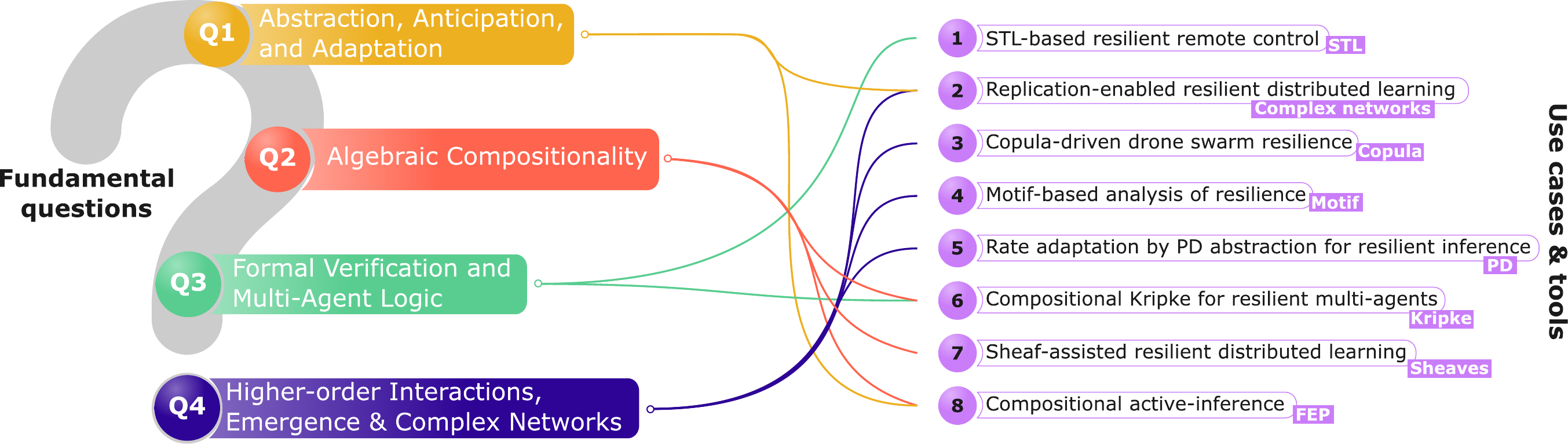}
    \caption{Alignment of use cases with the fundamental research questions (\textsc{Q1}-\textsc{Q4}).}
    \label{fig:Bipartite-Mapping}
\end{figure*}

\subsection{Sheaf Theory in Multi-Agent Learning Systems}
\label{sec:sheaf_theory}

\subsubsection{Representing the Multi-Agent System using Graphs and Cellular Sheaves} 
We consider a decentralized multi-agent communication network represented by an undirected graph \(\set{G}=(\set{V},\set{E})\), where \(\set{V}=\{1,\dots, N\}\) denotes the set of \(N\) distributed agents (or nodes), and \(\set{E}\) represents the set of communication links between adjacent agents. A \textit{cellular sheaf} equips this graph with an additional structure tailored for scenarios involving distributed data or models, such as in federated or distributed learning.

Specifically, a cellular sheaf \( \set{F} \) of \( \realDomain \)-vector spaces over \( \set{G} \) assigns vector spaces (stalks) to the graph's vertices and edges, along with linear maps (restriction maps) that relate them \cite{issaid2025tacklingfeaturesampleheterogeneity}.
\begin{itemize}
    \item \textbf{Vertex Stalks:} For each agent \( i \in \set{V} \), a vector space \( \set{F}(i) = \mathbb{R}^{d_i} \) is assigned. This space, denoted as the \textit{stalk} over agent \( i \), typically represents the space containing the agent's local data representation or model parameters $\vect{\theta}_i$ (i.e., \( \vect{\theta}_i \in \set{F}(i) \)). The dimension \( d_i \) can vary across agents, accommodating heterogeneity in model sizes or feature spaces. The collection of all local models \( \vect{\theta} = (\vect{\theta}_i)_{i \in \set{V}} \) belongs to the total space of $0$-cochains (vertex assignments), defined as \( C^0(\set{F}) := \bigoplus_{i \in \set{V}} \set{F}(i) \), where \( \bigoplus \) is the direct sum of vector spaces. 
    \item \textbf{Edge Stalks:} For each communication link (edge) \( e = (i, j) \in \set{E} \), a vector space \( \set{F}(e) = \mathbb{R}^{d_{ij}} \) is assigned. This edge stalk serves as a common space onto which the information or models from the connected agents \( i \) and \( j \) can be projected for comparison or aggregation. The collection of all edge data resides in the space of $1$-cochains, \( C^1(\set{F}) := \bigoplus_{e \in \set{E}} \set{F}(e) \).
    \item \textbf{Restriction Maps:} For each edge \( e = (i, j) \in \set{E} \) and an incident vertex \( i \), there is a linear transformation \( \set{F}_{i \triangleleft e} : \set{F}(i) \to \set{F}(e) \). This map, called the \textit{restriction map}, projects the information from the vertex stalk \( \set{F}(i) \) onto the edge stalk \( \set{F}(e) \). We often represent this linear map by its matrix \( \vect{P}_{ij} \) (assuming fixed bases), such that \( \set{F}_{i \triangleleft e}(\vect{\theta}_i) = \vect{P}_{ij} \vect{\theta}_i \). The dual maps \( \set{F}^*_{i \triangleleft e} \) correspond to the transpose matrices \( \vect{P}_{ij}^T \).
\end{itemize}
This structure allows for comparing potentially heterogeneous local models \( \vect{\theta}_i \) and \( \vect{\theta}_j \) for connected agents \( i, j \) by examining their projections \( \set{F}_{i \triangleleft e}(\vect{\theta}_i) \) and \( \set{F}_{j \triangleleft e}(\vect{\theta}_j) \) within the common space \( \set{F}(e) \), as shown in Fig. \ref{sheaf1}.

\subsubsection{Measuring Consistency via The Sheaf Laplacian}
A key advantage of the sheaf framework is its ability to quantify the disagreement or lack of consensus among the local models across the network, respecting the potentially heterogeneous nature of the models and their relationships defined by the restriction maps. This is achieved using the \textit{sheaf Laplacian} operator \( L_{\set{F}} \), which acts on the total space \( C^0(\set{F}) \).

The sheaf Laplacian \( L_{\set{F}} \) is a linear operator whose components \( L_{ji} : \set{F}(i) \to \set{F}(j) \) map between vertex stalks. Its action on a collection of local models \( \vect{\theta} \) is given by \( (L_{\set{F}}(\vect{\theta}))_j = \sum_{i \in \set{V}} L_{ji} (\vect{\theta}_i) \), where the operator components \( L_{ji} \) are defined as:
\begin{align} \label{eq:sheaf_laplacian_components}
    L_{ji} &= \begin{cases} 
    \sum_{e \sim i} \set{F}^*_{i \triangleleft e} \circ \set{F}_{i \triangleleft e}, & \text{if } i = j, \\
    -\set{F}^*_{j \triangleleft e} \circ \set{F}_{i \triangleleft e}, & \text{if } e = (i, j) \in \set{E}, \\
    \vect{0} \text{ (zero map)}, & \text{otherwise}.
    \end{cases}
\end{align}
Here, \( e \sim i \) denotes summation over all edges \( e \) incident to vertex \( i \). The condition \( L_{\set{F}} \vect{\theta} = \mathbf{0} \) signifies that the local models \( \vect{\theta} \) are in perfect consensus relative to the structure defined by the sheaf \( \set{F} \).

The degree of disagreement can be measured by the quadratic form associated with the sheaf Laplacian. This form neatly captures the sum of squared differences between the projections of adjacent agents' models onto their connecting edge stalks
\begin{align}
    \label{eq_quadratic_sheaf_laplacian}
     \vect{\theta}^T L_{\set{F}} \vect{\theta} & = \sum_{e = (i,j) \in \set{E}} \| \set{F}_{i \triangleleft e}(\vect{\theta}_i) - \set{F}_{j \triangleleft e}(\vect{\theta}_j) \|^2 \notag\\
     &= \sum_{e = (i,j) \in \set{E}} \| \vect{P}_{ij} \vect{\theta}_i - \vect{P}_{ji} \vect{\theta}_j \|^2,
\end{align}
where \( \| \cdot \|^2 \) denotes the squared Euclidean norm and the second equality uses the matrix representations \( \vect{P}_{ij}, \vect{P}_{ji} \) of the restriction maps \( \set{F}_{i \triangleleft e}, \set{F}_{j \triangleleft e} \), respectively. Minimizing this quadratic form encourages local models to agree in the shared comparison spaces defined by the edge stalks and restriction maps.

\subsubsection{Sheaves for Resilient Multi-agent Learning}
This sheaf-based mathematical foundation enables novel approaches in distributed learning. The correlation structure among agents, encoded in the restriction maps \( \vect{P} \), can itself be learned alongside the model parameters \( \vect{\theta} \). For instance, optimization procedures can alternate between updating \( \vect{\theta} \) (e.g., minimizing local loss plus a consensus term based on \eqref{eq_quadratic_sheaf_laplacian}) and updating \( \vect{P} \) to learn the relationships that best explain observed model similarities or task structures \cite{issaid2025tacklingfeaturesampleheterogeneity}.

By explicitly capturing and learning these inter-agent correlations, the distributed learning process can become more resilient. The system can potentially adapt to adverse events, such as sensor loss or node failure at an agent \(k\), by leveraging the learned relationships (\(\vect{P}_{ik}\), \(\vect{P}_{ki}\)) with its neighbors \(i\) to infer missing information or adjust communication protocols, thereby mitigating performance degradation. This continuous learning and adaptation, grounded in the sheaf structure, offers a pathway towards building more resilient multi-agent learning systems.

\section{Use Cases}\label{sec:use_cases}

This section seeks to  demonstrate the effectiveness and relevance of the various mathematical tools to model, analyze, and optimize system resilience across  several use cases.

\subsection{STL-based Resilient Remote Control}

\subsubsection{Setting}

\begin{figure}
\centering
\begin{subfigure}[t]{.8\linewidth}
\centering
    \includegraphics[width=\linewidth]{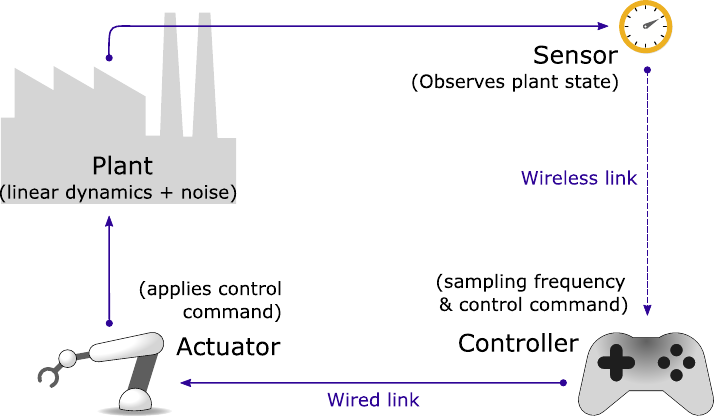}
    \caption{Remote control setup of the \gls{wncs}.}
    \label{fig:uc_remote_control_setting}
\end{subfigure}
\par\medskip 
\begin{subfigure}[t]{\linewidth}
    \includegraphics[width=\linewidth]{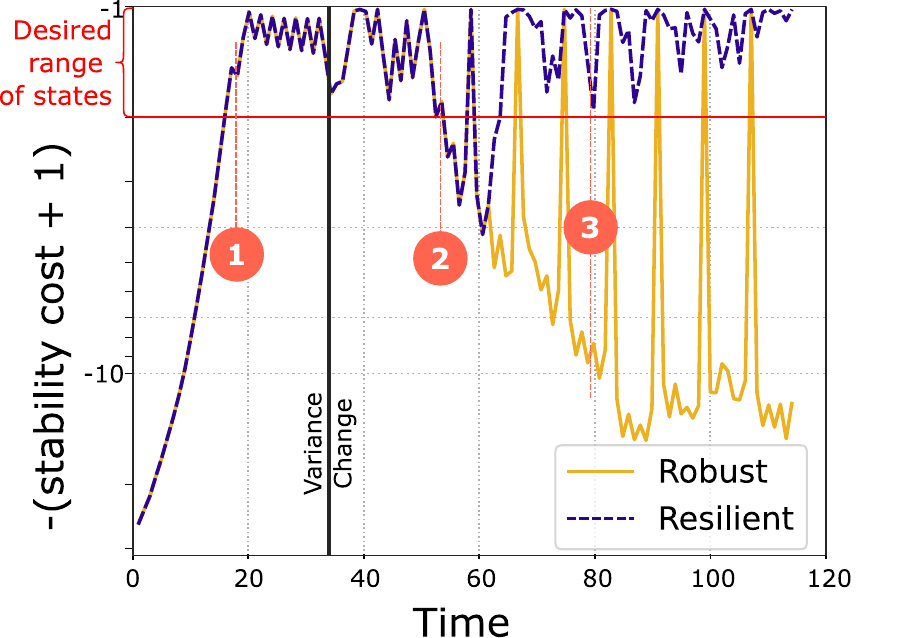}
    \caption{Stability comparison between robust (yellow) and resilient (blue) system designs. }
    \label{fig:uc_remote_control_results}
\end{subfigure}
\caption{Resilient remote control use case.}
\label{fig:uc_remote_control}
\end{figure}

We consider a \gls{wncs} consists of a sensor, a controller, and an actuator interacting with a control plant as illustrated in Fig.~\ref{fig:uc_remote_control_setting}. 
The sensing link is wireless in which, the plant state $\statePlant{k}$ observed by the sensor at control loop $t\in\timeDomain$ 
followed by a time step $k\in\{0,1,\dots,K-1\}$ could be communicated with the controller.
At the controller, its knowledge on the plant state $\stateControl{k}$ is updated based on communications, and a control action $\controlAction{k}$ is computed and applied to the plant through the actuator.

We assume the plant to be a system with noisy linear dynamics 
\begin{equation}\label{eqn:plant_dynamics}
	\statePlant{k+1} =
	\boldsymbol{A}\transpose \statePlant{k} + \boldsymbol{B}\transpose \controlAction{k} +\plantNoise_k,
\end{equation}
where $\boldsymbol{A}$ and $\boldsymbol{B}$ are known static parameters.
The noise $\plantNoise_k \sim \mathcal{N}\left(\zero, \identity\noiseStressor\right)$ follows a Gaussian distribution with mean $\zero$ and variance $ \identity\noiseStressor$ where $ \identity$ is the identity matrix. 
The stressor parameter $\noiseStressor$ controls the plant noise and it is subjected to changes over time.

The sensing link operates based on a communication policy $\comPolicySingle_k$ where $\comPolicySingle_k=1$ denotes the scheduling of sensor-controller communication while $\comPolicySingle_k=0$ otherwise.
Once scheduled, the sensor uses a fixed transmission power $\txpower$ to communicate the plant state to the controller, and thus, $\stateControl{k}=\statePlant{k}$ is held.
In the absence of sensing, the controller updates its estimate of the state following the linear dynamics
\begin{equation}\label{eqn:dynamics_estimated}
	\stateControl{k}=\boldsymbol{A}\transpose \stateControl{k-1} + \boldsymbol{B}\transpose \controlAction{k-1},
\end{equation}
with an estimation error of $\plantNoise_k$.
Under such continuous steps without communication, the error gets accumulated. 
To track the error, we introduce the notion of \gls{aoi} that evolves as follows:
\begin{equation}\label{eqn:aoi}
	\aoi_{k+1} = (1 - \comPolicySingle_k) (1 + \aoi_{k}).
\end{equation}

Due to the noisy dynamics and occasional wireless sensing, the controller employs a \gls{mpc} mechanism over a time horizon $\timeHorizon=\{0,\dots,K-1\}$ to stabilize the system.
Here, the controller determines scheduling decisions $\comPolicy=[\comPolicySingle_k]_{k\in\timeHorizon}$, estimated states $\stateVec=[\stateControl{k}]_{k\in\timeHorizon}$, and control actions $\controlVec=[\controlAction{k}]_{k\in\timeHorizon}$ at each control loop $t$ with the following requirements.   
To ensure  stability, the system state should be maintained around a desired state $\stateTarget$ with a threshold $\stateThreshold$, i.e., $\|\statePlant{k} - \stateTarget\| \leq \stateThreshold$.
To ensure the estimated state $\stateControl{k+\iota}$ at the controller as close to the actual plant state $\statePlant{k+\iota}$ when a successful communication takes place at $k$, we impose a constraint on the estimation error as follows:
\begin{equation}\label{eqn:constraint_estimation_error}
	\var\left( \left\| \statePlant{k+\iota} - \stateControl{k+\iota} \right\| \right) 
	\leq \stressorThreshold,
\end{equation}
where $\stressorThreshold(>0)$ is the maximum allowed estimation error. 
Note that from \eqref{eqn:plant_dynamics}-\eqref{eqn:aoi} we can derive that $\statePlant{k+\iota} - \stateControl{k+\iota} = \sum_{i=0}^{\aoi_{k+\iota}-1} (\boldsymbol{A}\transpose)^i \plantNoise_{k+\iota-1-i}$, in which allowing the estimation error to be tracked by recasting \eqref{eqn:constraint_estimation_error} as follows:
\begin{equation}\label{eqn:constraint_estimation_error_mod}
	\sum_{i=0}^{\aoi_{k}-1} \left\| \left( \boldsymbol{A}\transpose \boldsymbol{A} \right)^i \right\| \noiseStressor
	\leq \stressorThreshold.
\end{equation}

Even under imposing a bound for the error, one cannot guarantee a strict condition of stability. 
In this work, we focus on the events where the stability condition is violated, i.e., $\| \statePlant{k} - \stateTarget \| > \stateThreshold$, and enforce conditions on recovering to the desired operation within a given time duration $\timeRecovery$ and maintaining it over a time period $\beta$ with  guarantees.
Towards this, we resort to  tools from \gls{stl}, more specifically the notion of \gls{stl}-based resilience in \eqref{eqn:stl_resilience}.
Hence, the requirements of recovery and durability are formalized as follows \cite{safaoui2020control,calafiore2006distributionally}:
\begin{equation}\label{eqn:constraint_recovery_and_sustain}
	\max_{\mathbb{P}\in\mathcal{P}}\mathbb{P}\left[ \stateControl{k} 
	\not\models  
	\stlResilience{\alpha}{\beta} \left(\left\| \stateControl{k} - \stateTarget \right\| 
	\leq \stateThreshold \right)\right]\leq \stlThreshold \quad 
	\forall k\in \timeHorizon,
\end{equation}
where 
$\mathcal{P}$ is the set of all probability distributions with variance $\noiseStressor$
and $\stlThreshold(>0)$ is a predefined threshold. 
Note that the uncertainty of $\stateControl{k}$ occurs due to the fact that $\stateControl{k}=\statePlant{k}$ is held for $\comPolicySingle_k=1$ and the dynamics defined under \eqref{eqn:dynamics_estimated}.
This constraint on the \gls{stl}-based resilience, ensures that the conditions on the recovery within $\alpha$ and sustaining the performance over $\beta$ are held with a probability of $(1-\stlThreshold)$ over the uncertainties on the stressor.

The sensing, control  and state estimation within  \gls{wncs} is designed to be energy efficient under the aforementioned stability requirements. 
Towards this, the overall system design rooted in energy  minimization problem is:
\begin{equation} \label{eqn:wncs_optimization}
    \begin{array}{ll}
   \underset{(\controlVec, \comPolicy, \stateVec)}{\text{minimize}} & \dfrac{1}{K} \sum_{k=0}^{K-1} \left( \txpower \comPolicySingle_k  +  \left( \controlAction{k} \right) ^T \boldsymbol{Q} \controlAction{k} \right)\\
    \mbox{subject to} & \eqref{eqn:dynamics_estimated},\eqref{eqn:aoi}, \eqref{eqn:constraint_estimation_error_mod}, \eqref{eqn:constraint_recovery_and_sustain}.
    \end{array}
\end{equation}
Since $\noiseStressor$ used in \eqref{eqn:constraint_estimation_error_mod} is subjected to changes, it is essential to track such variations and incorporate them into \eqref{eqn:wncs_optimization}.
Towards this, the controller aggregates a collection of errors $\left\{ e_{k'} = \left\| \statePlant{k'}- \left( \boldsymbol{A}\transpose \stateControl{k'-1} + \boldsymbol{B}\transpose \controlAction{k'-1} \right) \right\| \right\}_{k'}$ for all $k'$ with $\aoi_{k'}=1$ (when the plant state is successfully received at the controller).
Since these errors should be sampled from $\mathcal{N}\left(\zero, \identity\noiseStressor\right)$, we can conduct Chi-Square test for the variance to validate the null hypothesis $H_0: {\sigma} = \noiseStressor$ against $H_1: {\sigma} \neq \noiseStressor$.
When $H_1$ is held, the belief of the stressor is updated as $\sigma = \sum_{k'} e_{k'} / \sum_{k'} \aoi_{k'}$.

\subsubsection{Results}

For the experiment, we consider a plant with dynamics defined by $\boldsymbol{A}=1.1$ and $\boldsymbol{B}=-0.25$ with the known stressor $\noiseStressor=0.02$.
The desired stability is defined by $\stateTarget=0$, $\stateThreshold=1$, and $\boldsymbol{Q}=1$ while the state estimation error is to be bounded by $\stressorThreshold=1$.
Then, we simulate the stressor to be changed to $\noiseStressor=0.15$ while imposing resilient conditions with $\timeRecovery=20$, $\timeSustain=10$, and $\stlThreshold=0.01$.

With a known stressor $\noiseStressor$ in the initial phase, a robust controller precomputes the optimal sampling frequency to maintain high confidence (i.e., low uncertainty) in its predicted future state over a given time horizon. 
However, this approach assumes that the stressor model remains accurate throughout the operation. 
In contrast, the resilient design, based on active inference, actively monitors variations in the stressor model and employs hypothesis testing (HT) to assess whether an update to its stressor model is necessary. 
Upon detecting deviations, the system refines its internal model and reoptimizes its decision-making strategy accordingly.

The performance in terms of the stability $\|\statePlant{k} - \stateTarget\|$ of the robust and resilient system designs is compared in Fig.~\ref{fig:uc_remote_control_results}.
To illustrate the finer details at the desired stability region, the y-axis is modified and plotted in log scale. 
As shown in Fig.~\ref{fig:uc_remote_control_results}, the robust baseline struggles to maintain stability due to a persistent mismatch between its internal and true stressor models. 
In contrast, the resilient approach detects the mismatch between its internal and true stressor models, and effectively adapts to evolving conditions, mitigating performance degradation. 
It is worth highlighting that  enhancing early detection or prediction of stressor changes can further enhance the faster recovery yielding better resilience.

Note that different design choices with increased $\boldsymbol{Q}$ and decreased $\stressorThreshold$ could enable the robust method to have frequent communication, and thus, ensuring stability over a wider range of stressor changes. 
However, such a resource over provisioning is efficient only when the stressor is at its extreme limits.
Diverting the above excess resource usage towards enabling reconfigurability within the resilient design, offers the system to continuously operate in resource and energy efficient manner highlighting the robust-resilient tradeoffs.  


\subsection{Replication-enabled Resilient Distributed Learning}
\subsubsection{Setting}

Consider a decentralized communication network $\set{G}$, where the network topology is generated randomly using the Erdős-Rényi random graph model, with $N$ nodes in the network and connectivity probability $p$. 
There are $N_f=10$ initial neural network models with ResNet-18 model architectures \cite{he2016deep}, which are transmitted across the communication network through random walk.
We use a cross-entropy loss function. When the $k$-th model arrives at node $i$ in the $t$-th step, the model conducts one stochastic gradient descent (SGD) with the local dataset in node $i$, i.e.,
$$
\vect{\theta}_k^{(t+1)} = \vect{\theta}_k^{(t)} - \alpha g(\vect{\theta}_k^{(t)}, \varsigma_i),
$$
where $\varsigma_i$ is a mini-batch of data samples in node $i$, and $g$ denotes the computed gradients.
After one step of local SGD, node~$i$ randomly chooses a node from the set $\mathcal{N}_i$ to send the neural network model for further updating. 
We further assume that the agents have \gls{iid} data distribution of CIFAR-10 dataset and thus, the sampling distribution of neighboring nodes is a uniform distribution. 
$N_f=10$ models are updated and transferred across the  network.

In the context of distributed random walk based learning,  node, link, or motifs failures can result in poor performance.
Thus, to ensure resilient learning, we utilize the replication procedure as discussed in Section \ref{Sec4_4}, where each node accumulates the latest time it received one model and calculates the empirical \gls{cdf} of returning time.
Based on that, each node can estimate the number of active random walks in the network whenever it receives one model. 
When the estimated number is smaller than a threshold $\epsilon$, the received model is replicated with probability $1/N_f$. 
In our experiments, we set $\epsilon=1.6$.

We randomly generate three different communication network topologies using Erdos\_Renyi random graph model, with $N=50, p=0.1$; $N=50, p=0.8$; and $N=200, p=0.1$, respectively. 
In the random walk process, each agent continuously estimates the \gls{cdf} of models' returning time and replicates the received model under small estimated results.
To show the resilience of the algorithm, we consider that $50\%$ links and $50\%$ nodes fail randomly across the network, in the $1000$-th and $3000$-th step, respectively. 
Such failures lead to the loss of random walks, as indicated in~Fig.~\ref{fig:sec4_4}.
Moreover, we show the replication scaling with the threshold $\epsilon$ under different network topologies, where the replication is indicated through the stable number of models recovered after the first $50\%$ links loss. 

\subsubsection{Results}
\begin{figure}
\centering
\begin{subfigure}[t]{\linewidth}
    \includegraphics[width=\linewidth]{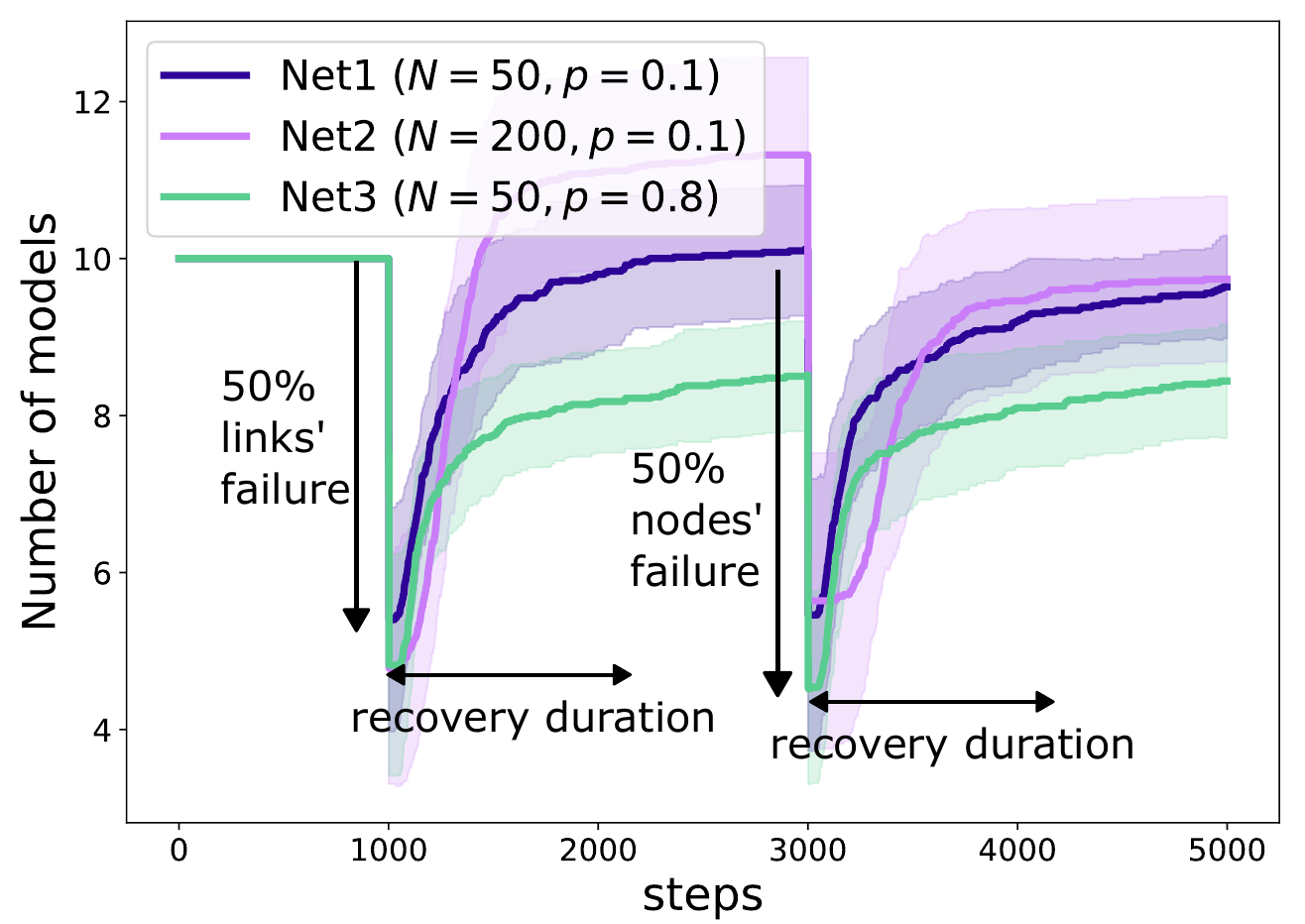}
    \caption{Number of models changing over the steps.}
    \label{fig:uc_rwdl1}
\end{subfigure}
\par\medskip 
\begin{subfigure}[t]{\linewidth}
    \includegraphics[width=\linewidth]{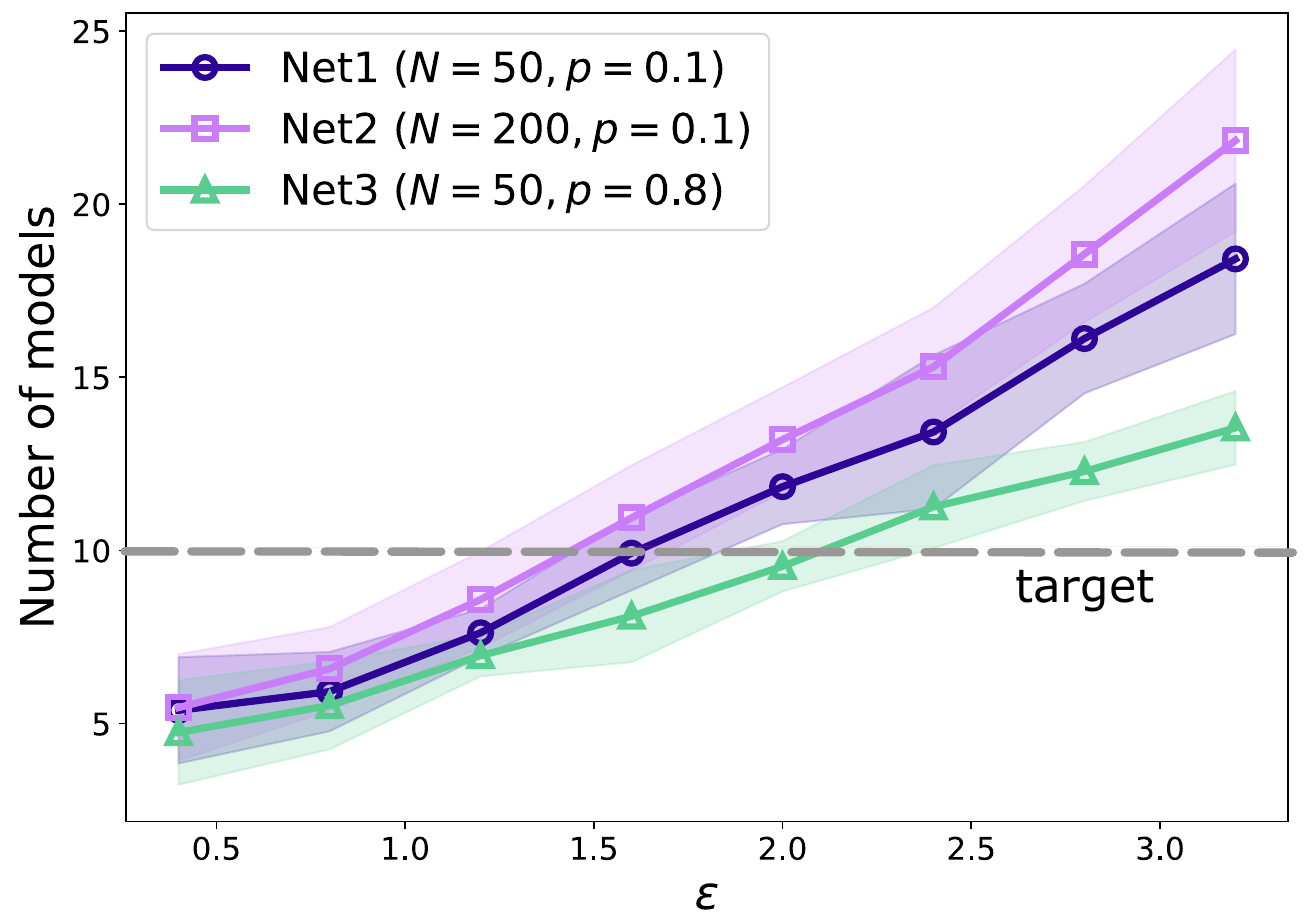}
    \caption{Replication scaling vs. $\epsilon$ under different network topologies.}
    \label{fig:uc_rwd3}
\end{subfigure}
\caption{Replication-enabled resilient distributed learning.}
\label{fig:uc_rwdl}
\end{figure}

We simulate  $50$ runs under random nodes' and links' failures.
The number of models varying with respect to the number of steps  is shown in Fig. \ref{fig:uc_rwdl1}, where the averaged results are shown by the curves and the standard deviations are depicted by shaded areas.
 Fig. \ref{fig:uc_rwdl1} highlights that the number of random walks in the network can recover from the loss resulting from links' or nodes' failures, which becomes stable after a recovery duration time.
Moreover, the same $\epsilon$ shows different performances under different communication network topologies, where the recovery performance in high connectivity of Net2 performs worse than sparser ones. 
This indicates that a larger value of threshold $\epsilon$ is required for denser network topologies. 
Fig. \ref{fig:uc_rwd3} further shows how the replication scales with the threshold $\epsilon$, revealing the intricate trade-off between redundancy and the system's adaptability to deal with unexpected loss events as a function of the replication threshold $\epsilon$. In practical implementations, a higher redundancy of models incurs larger computation cost across nodes, and more computation costs in the model transmission process.
When $\epsilon$ is large, the system recovers to a stable state with high response speed relying on more frequent replication and the resultant resources consumption. On the other hand, given small $\epsilon$, the energy cost becomes smaller while the reconfiguration capacity of the system is highly reduced. As analyzed in Section \ref{subsubsec:Sustainability-versus-Resilience}, the trade-off lies in balancing the adaptability of the system and its energy budget by choosing the optimal threshold $\epsilon$.

\subsection{Copula-driven Drone Swarm Resilience}

\subsubsection{Setting}
A swarm of drones $\set{\droneSet} = \{d_1, d_2, \ldots, d_{\droneCount}\}$ operates in a bounded 2D area $\set{\area} \subset \realDomain^2$ of size $\areaSize \times \areaSize$, where $\droneCount=30$ and $\areaSize = 1000$. Each drone $d_i$ is characterized by its state vector $\vect{\droneState}_i = [\vect{\dronePos}_i, e_i, a_i]$, where $\vect{\dronePos}_i = (x_i, y_i) \in \set{\area}$ is the drone's position, $e_i \in [0,100]$ its energy level, and $a_i \in \mathbb{N}^+$ its age (number of time steps since creation).

Active drones provide circular coverage with a variable radius $r_i$ centered at their position $\vect{\dronePos}_i$. This radius dynamically changes based on the Signal-to-Noise Ratio (SNR) experienced by the drone in its local environment. The radius $r_i \in [r_{\min}, r_{\max}]$, where $r_{\min} = 75$ is the minimum coverage radius (achieved at very low SNR), and $r_{\max} = 225$ is the maximum coverage radius (achieved at high SNR). The total coverage ratio $\coverage(\set{\activeSet}_a^{(t)}) \in [0,1]$ of the swarm at time $t$ is the proportion of $\set{\area}$ covered by at least one drone from the set of active drones $\set{\activeSet}_a^{(t)} \subseteq \set{\droneSet}$, and is defined as
\begin{equation}\label{eq:coverage-ratio}
\coverage(\set{\activeSet}_a^{(t)}) = \frac{|{\vect{x} \in \set{\area} : \exists d_i \in \set{\activeSet}_a^{(t)}, \|\vect{x} - \vect{\dronePos}_i^{(t)}\| \leq r_i^{(t)}}|}{|\set{\area}|}.
\end{equation}
The swarm operates in an environment where drones are subject to stochastic failures. Initially, failures occur independently with probability $\failureProb=0.01$ for each active drone in each time step. However, at time $\transTime=60$, the environment transitions to a regime of spatially correlated failures. The drones are initially positioned in a quasi-uniform grid pattern with small jitter, defined by
\begin{equation}
\begin{cases}
      x_i^{(0)} &= \lfloor i/\sqrt{\droneCount_R} \rfloor \cdot \frac{\areaSize}{\sqrt{\droneCount_R}} + \frac{\areaSize}{2\sqrt{\droneCount_R}} + \epsilon_x,\\
      y_i^{(0)} &= (i \bmod \sqrt{\droneCount_R}) \cdot \frac{\areaSize}{\sqrt{\droneCount_R}} + \frac{\areaSize}{2\sqrt{\droneCount_R}} + \epsilon_y,
    \end{cases}
\end{equation}
where $\droneCount_R$ is the actual number of drones including redundancy, and $\epsilon_x, \epsilon_y \sim \normal(0, \frac{\areaSize}{10 \sqrt{\droneCount_R}})$ are small random perturbations. The initial dependency strength between drones $d_i$ and $d_j$, denoted by $\gamma_{ij} \in [0,1]$, is defined as
\begin{equation}\label{rho0}
\gamma_{ij} = \frac{0.4}{1 + \exp((\|\vect{\dronePos}_i - \vect{\dronePos}_j\| - 200)/50)}, 
\end{equation}
with diagonal elements initialized to 0.2. Finally, the environment contains risk areas $\set{\riskAreas} = \{(\vect{r}_k, w_k)\}$ where $\vect{r}_k \in [0, \areaSize] \times [0, \areaSize]$ is the position of a risk area and $w_k \in [0,1]$ is its intensity.

The objective is to maximize the expected coverage ratio $\mathbb{E}[\coverage(\set{\activeSet}_a)]$ over a finite time horizon $T$, despite drone failures and the transition to correlated failure patterns. The problem can be formulated as 
\begin{align}
\underset{\{\vect{\controlActions}^{(t)}\}_{t=0}^{T-1}}{\text{maximize}} \quad & \mathbb{E}\left[{T^{-1}}\textstyle\sum_{t=1}^{T} \coverage(\set{\activeSet}_a^{(t)})\right] \\
\mbox{subject to} \quad & \vect{\dronePos}_i^{(t+1)} = \vect{\dronePos}_i^{(t)} + \vect{\controlActions}_i^{(t)} \label{eq:pos_update} \\
& e_i^{(t+1)} = e_i^{(t)} - 0.01  \|\vect{\controlActions}_i^{(t)}\| \label{eq:energy_update} \\
& a_i^{(t+1)} = a_i^{(t)} + 1 \label{eq:age_update} \\
&  \vect{\dronePos}_i^{(t)} + \vect{\controlActions}_i^{(t)} \in \set{\area} \label{eq:boundary_constraint} \\
&  \prob(d_i \text{ fails at } (t+1)) \nonumber \\ 
& = f\left(\vect{\droneState}_i^{(t+1)}, \{\vect{\droneState}_j^{(t+1)}\}_{j \neq i, d_j \in \set{\activeSet}_a^{(t)}}, \set{\riskAreas}^{(t+1)}\right)\label{eq:failure_prob}
\end{align}
where $\set{\riskAreas}^{(t+1)}$ is the set of risk areas with their positions and weights, tracking locations of previous failures, and $f$ is the failure probability function. The control actions $\vect{\controlActions}^{(t)} = \{\vect{\controlActions}_1^{(t)}, \dots, \vect{\controlActions}_{|\set{\activeSet}_a^{(t)}|}^{(t)}\}$ are the decision variables at each time step $t$, representing the position adjustment vector for each currently active drone $d_i \in \set{\activeSet}_a^{(t)}$. The drone states $(\vect{\dronePos}_i^{(t)}, e_i^{(t)}, a_i^{(t)})$ evolve according to equations \eqref{eq:pos_update}-\eqref{eq:age_update}.

The robust design uses static redundancy with no adaptation. The system is configured with $\droneCount_R = \droneCount(1+\redundancy)$ drones, where $\redundancy = 0.3$ is the redundancy factor. Rather than simply placing drones statically, the controller solves the following optimization problem
\begin{equation}
\underset{\vect{\dronePos}_i}{\text{maximize}}
 \min_{\substack{\set{S} \subset {1,\ldots,N},\\ |\set{S}|=K}} \coverage(\set{\droneSet} \setminus \set{S}),
\end{equation}
where $K = 5$ is the number of worst-case failures (approximately 15\% of the total drones). This is solved using simulated annealing with a hexagonal grid initialization. Once deployed, the robust controller applies null control actions, i.e., $\vect{\controlActions}_i^{(t)} = \vect{0}, \forall i, t$, relying solely on the optimized initial positioning and redundant drones. 

In contrast to static strategies relying solely on initial placement, the resilient design employs dynamic, adaptive control driven by insights from a vine copula model of drone failure dependencies. This sophisticated statistical model, implemented using an R-vine structure, allows the system to capture and reason about complex, multivariate, and potentially asymmetric correlations in drone failures. The R-vine copula structure is initialized with a proximity-based dependency matrix $\vect{\Gamma} 
 = [\gamma_{ij}]$ using \eqref{rho0}. The copula's dimension is constrained to at most 100 for computational feasibility. The model employs a mix of bivariate copula families, including Gaussian, Clayton, Gumbel, and Frank copulas, selected for their ability to model different types of tail dependencies. This model is a direct application of the principles outlined in Section \ref{copulas}. Following Sklar's Theorem (Eq. \eqref{eq:sklar_forward}), the vine copula decomposes the complex multivariate joint probability of drone failures into two distinct components: (i) the individual or marginal failure probability of each drone, $\mathbb{P}_{\text{fail}}(d_j)$, and (ii) the dependence structure that links these failure events, which is captured by the copula itself. The vine structure is a powerful technique for constructing a high-dimensional copula for the entire swarm from a flexible combination of the simpler bivariate copula families mentioned. The system continuously learns this dependence structure $\gamma_{ij}$ by observing simultaneous failure events, analogous to the copula estimation process described in Section \ref{construction}, allowing it to adapt to the changing failure correlations in the environment.

The controller uses the dependency structure learned by the copula for proactive actions to enhance swarm survivability. Drones actively steer away from neighbours with whom the copula identifies a high statistical dependency $\gamma_{ij}$, especially if those neighbours also exhibit a high predicted individual failure probability $\prob_{\text{fail}}(d_j)$. This repulsion is explicitly calculated using the copula-derived dependency using
\begin{align}
&F_D(d_i) \nonumber \\
&\!=\!\sum_{\substack{d_j \in \set{\activeSet}_a,\\ j \neq i}} \left\{\gamma_{ij}  \mathbb{P}_{\text{fail}}(d_j)  \left(1\!-\!\frac{\|\vect{\dronePos}_i\!-\!\vect{\dronePos}_j\|}{200}\right)  \frac{\vect{\dronePos}_i\!-\!\vect{\dronePos}_j}{\|\vect{\dronePos}_i\!-\!\vect{\dronePos}_j\|} \right. \nonumber \\
&\left.  \times \indict(\|\vect{\dronePos}_i\!-\!\vect{\dronePos}_j\| < 200) \right\}.
\end{align}
Simultaneously, the system seeks to maintain area coverage by attracting active drones towards gaps left by failed ones using a coverage optimization force $F_C(d_i)$. Crucially, this attraction is also modulated by the copula-derived dependency information. The attraction towards a failed drone $d_j$ is weighted by a safety factor $w_{ij}$ that decreases if the statistical dependency $\gamma_{ij}$ between the active drone $d_i$ and the location of the failed drone $d_j$ is high, ensuring safer coverage recovery
\begin{equation}
F_C(d_i) = \sum_{d_j \in \set{\droneSet} \setminus \set{\activeSet}_a} w_{ij} \frac{\vect{\dronePos}_j - \vect{\dronePos}_i}{\|\vect{\dronePos}_j - \vect{\dronePos}_i\|} \indict(150 < \|\vect{\dronePos}_i - \vect{\dronePos}_j\| < 350).
\end{equation}
The combined force calculation balances the competing objectives of avoiding correlated risk and maintaining coverage. The controller adaptively weights the dependency repulsion and coverage attraction forces based on the local dependency environment, prioritizing safety ($\beta = 0.7$) when significant dependencies exist. Hence, the combined force is defined as follows
\begin{equation}
F_{\text{comb}}(d_i)=
\begin{cases}
\beta F_R(d_i)
\\
\quad +(1-\beta) F_C(d_i),&\text{if} \sum\limits_{j \neq i} \gamma_{ij} \prob_{\text{fail}}(d_j) > 0 \\
F_C(d_i), &\text{otherwise}.
\end{cases}
\end{equation}
The resulting control action $\vect{\controlActions}_i^{(t)}$ translates this combined force into a concrete movement vector. Before executing this control action, the controller explicitly performs a crucial safety verification using the copula-derived dependency information $\vect{\Gamma}$. A proposed move is only allowed if it does not place the drone into a configuration deemed excessively risky due to potential failure correlations by ensuring a calculated risk score remains below a threshold. Even actions like filling coverage gaps are guided by dependency awareness, prioritizing moves that minimize statistical entanglement with new neighbours. Furthermore, reactive mechanisms triggered by failures leverage the learned dependency matrix $\vect{\Gamma}$ from the copula model to identify and proactively disperse clusters of drones that are statistically likely to fail together, mitigating cascading failures.

\begin{figure}
\centering
\begin{subfigure}[t]{\linewidth}
\centering
    \includegraphics[width=.9\linewidth]{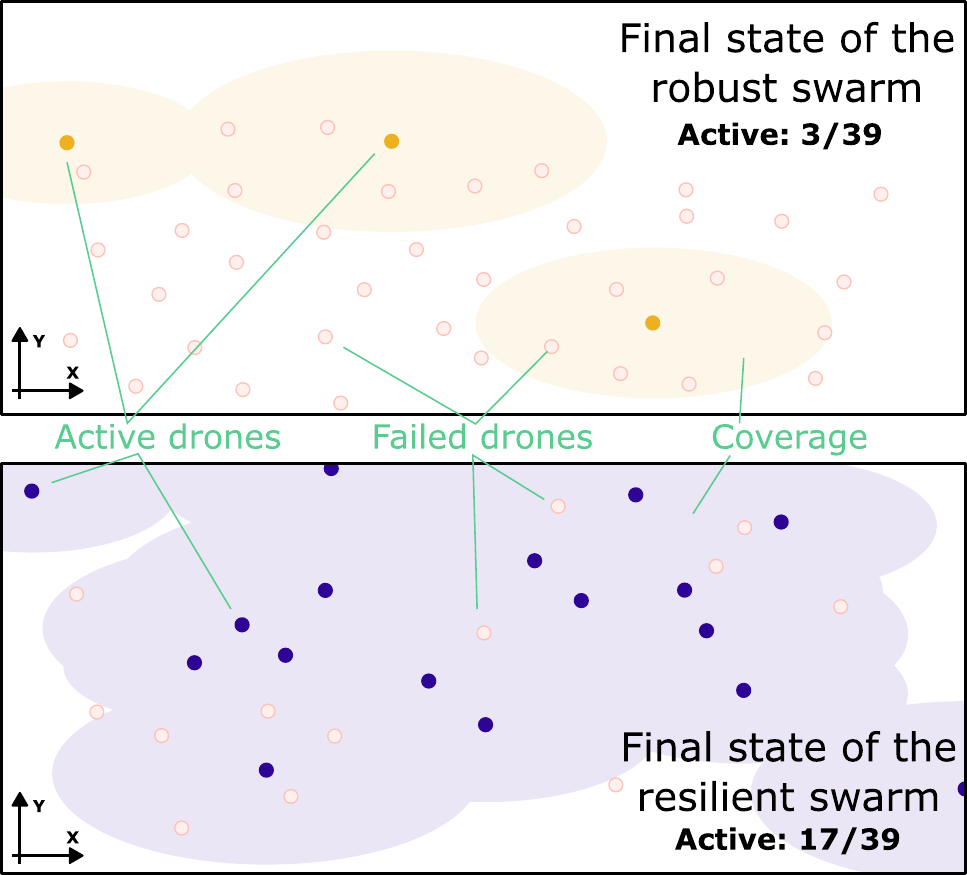}
    \caption{Final swarm states of robust (top) and resilient (bottom) designs showing active and failed drones.}
    \label{fig:swarm_comparison_1}
\end{subfigure}
\par\medskip 
\begin{subfigure}[t]{\linewidth}
    \includegraphics[width=\linewidth]{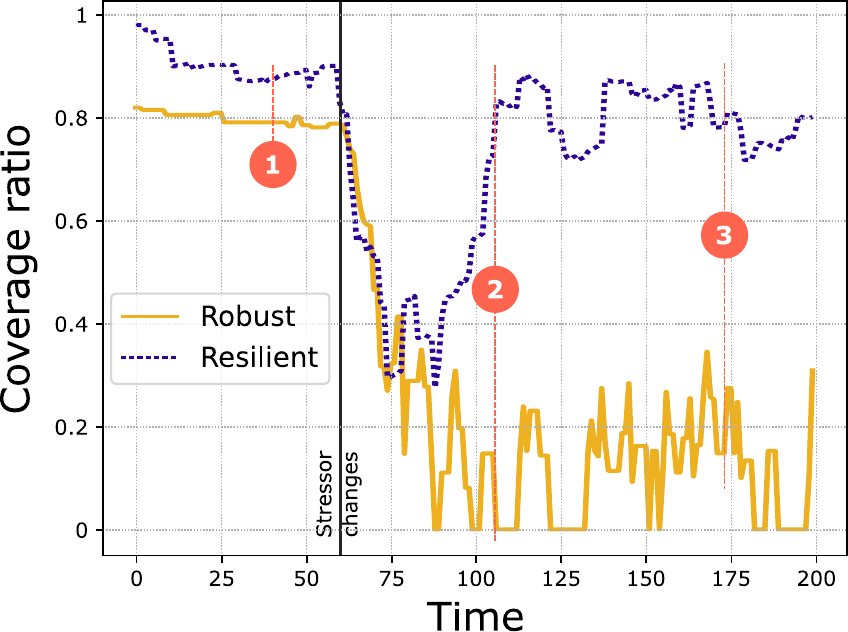}
    \caption{Comparison of robust and resilient Drone Swarm performance showing coverage ratio over time.}
    \label{fig:swarm_comparison}
\end{subfigure}
\caption{Drone swarm resilience use case.}
\label{fig:uc_swarm_comparison}
\end{figure}

The vine copula model  is not static but  dynamically adapts. Initialized using a proximity-based dependency matrix $\vect{\Gamma}$ derived from \eqref{rho0}, it continuously learns by observing which drones fail simultaneously during operation, using this data to update the $\gamma_{ij}$ values. The copula directly influences the failure simulation process by generating correlated random variables, reflecting the learned dependencies. This creates a vital feedback loop: observed failures refine the copula model, and the refined model provides more accurate dependency insights to guide the swarm's adaptive positioning via the control actions $\vect{\controlActions}_i^{(t)}$, significantly enhancing resilience against complex, correlated failure patterns.

\subsubsection{Results}
The performance of the robust and resilient strategies was compared over $200$ time steps, with results visualized in Figure \ref{fig:swarm_comparison}. The time series plot (Figure \ref{fig:swarm_comparison}, left) shows the coverage ratio achieved by each strategy. Initially (before $t=60$), both the robust (blue line) and resilient (orange line) strategies maintain relatively high coverage ratios, with the robust strategy hovering around $0.8$ and the resilient strategy near $0.9$. At $t=60$, the failure model transitions to spatially correlated failures. Both strategies experience a sharp drop in coverage immediately following this change. However, their long-term responses differ significantly. The robust strategy's coverage continues to degrade, failing to recover and fluctuating at very low levels (often below $0.3$) for the remainder of the simulation. In contrast, the resilient strategy, after an initial dip, demonstrates adaptive recovery. Its coverage ratio increases substantially after $t \approx 90$ and stabilizes at a high level (around $0.8$), effectively mitigating the impact of correlated failures.

The final states of the swarms at $t=200$ are shown in Figure \ref{fig:swarm_comparison_1}. The robust swarm (top) exhibits catastrophic failure, with only $3$ out of the initial $39$ drones remaining active. The resulting area coverage is minimal and highly localized. This highlights the vulnerability of the static, redundancy-based approach to correlated failures. The Resilient swarm (bottom), operating with $30$ initial drones, ends with $17$ drones active. Despite experiencing $13$ failures, the adaptive positioning strategy allowed the swarm to maintain a significant number of active drones and achieve substantial area coverage, albeit with a less uniform distribution than the initial state. The variable coverage radii of the active drones are also visible.

The robustness-resilience trade-off is shown in terms of resource allocation. The robust design's static over-provisioning with additional drones proved fragile, while the resilient design invests resources in dynamic reconfiguration, namely the copula model's computational cost and the energy for drone movement. The sustainability-resilience trade-off emerges from this energy cost of adaptation. The resilient swarm's continuous repositioning to restore coverage directly impacts sustainability through increased energy expenditure. Finally, the recoverability-durability trade-off is seen in the resilient controller's logic. An aggressive strategy for rapid recovery compromises long-term durability by moving drones into high-risk areas, whereas a conservative strategy preserves drones at the cost of slower recovery. The controller must therefore balance the immediate need for service (recoverability) against the long-term survival of the system (durability).

\subsection{Motif-based Analysis of Resilience}

\subsubsection{Setting}

Consider a network of a set $\nodeSet$ of $\NODE=50$ \glspl{bs} having a set $\edgeSet$ of \gls{bs}-\gls{bs} wireless links.
As such, the network is modeled as a graph $\graph=(\nodeSet,\edgeSet)$ where all edges in $\edgeSet$ are undirected and nodes are ordered from highest degree to the lowest. 
Additionally, we assume that
an additional set of edges, denoted $\permittedEdges$ with $|\permittedEdges|=10$, which is disjoint from $\mathcal{E}$
can be reconfigured to ensure the desired degree of connectivity 
under potential malicious attacks, if needed.
Hence, the initial configuration of the network is given by $\configMat= [\configuration_{\node,\node'}]_{\node,\node'\in\nodeSet}$ where $\configuration_{\node,\node'}=1$ if $(\node,\node')\in\edgeSet$ while $\configuration_{\node,\node'}=0$ if $(\node,\node')\in\permittedEdges$. 
With such network configuration, we assume that a desired level of connectivity between all the \glspl{bs} is maintained, which is captured by two conditions.
The first condition is on the highest degree of the network to maintain a minimal number of connections, i.e., $\sum_{\node'=2}^{\NODE} \configuration_{1,\node'}\geq\minDegree$ where $\minDegree=10$ is a predefined parameter. 
The second condition is to maintain the graph's 4-node motif distribution $\motifDistribution=[\motif_\subgraph]_{\subgraph\in\set{\subgraph}}\in\realDomain^6$ where $\set{\subgraph}$ represents the set of $4$-node motifs (see~\eqref{eq:motif-distribution}).

\begin{figure}
\centering
\begin{subfigure}[t]{\linewidth}
    \includegraphics[width=\linewidth]{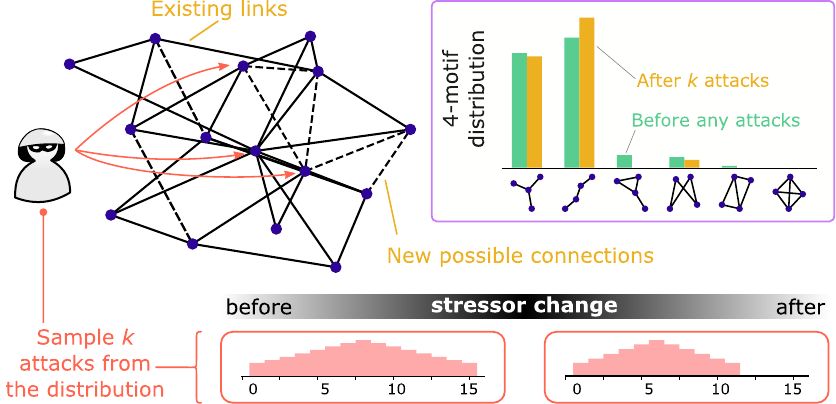}
    \caption{System model illustrating node degree centric attacks and their impact on 4-motif distribution.}
    \label{fig:uc_motif_setting}
\end{subfigure}
\par\medskip 
\begin{subfigure}[t]{\linewidth}
    \includegraphics[width=\linewidth]{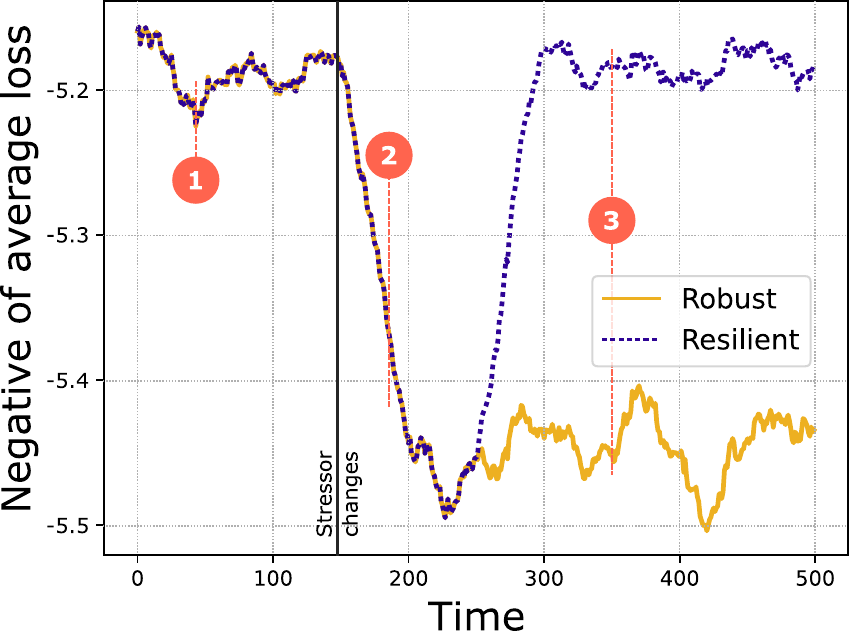}
    \caption{Performance comparison between robust and resilient designs towards network connectivity.}
    \label{fig:uc_motif_results}
\end{subfigure}
\caption{Motif-based resilience analysis use case.}
\label{fig:uc_motif}
\end{figure}

Occasionally, the network is subjected to a sequence of attacks from an external malicious attacker with an intention of disrupting the overall connectivity.  
Therein, a node centrality-based attacks are considered, in which, the attacker targets up to $\targetNodeCount(\ll\NODE)$ \glspl{bs} in the descending order of their degree following an attack distribution $\attackDistribution=[\attackProb_k]_{k\in\targetNodeSet}$ where $\targetNodeSet=\{1,\dots,\targetNodeCount\}$.
Here, $\attackProb_k$ represents the probability of attacking $k$ \glspl{bs} within a selected attack sequence and for the experiments $\targetNodeCount=15$ is considered.
After such $k$ attacks, due to the changes in nodes and edges, the resultant network graph is represented by $\atk{\graph}{k}= ( \atk{\nodeSet}{k}, \atk{\edgeSet}{k} )$, where $\atk{\nodeSet}{k}=\nodeSet\setminus\{1,\ldots,k\}$, and $\atk{\edgeSet}{k}$ is obtained by removing the edges connected to nodes $\{1,\ldots,k\}$ from $\edgeSet$. 
Moreover, we denote by $\atk{\permittedEdges}{k}$ the corresponding set of permissible reconfigurable edges after $k$ attacks, obtained similarly by removing the edges connected to nodes $\{1,\ldots,k\}$ from $\permittedEdges$. 
Over time, the attacker changes its strategy in terms of changing the attack distribution $\attackDistribution$ as illustrated in Fig.~\ref{fig:uc_motif_setting}.

Towards the desired operations, the network needs to reconfigure and maintain the connectivity under such attacks.
In this view, we use \gls{kl} divergence loss to quantify the deviations in the motif distribution from the original setting due to the reconfiguration as follows:
\begin{equation}\label{eqn:kl_divergence}
	\kldiv\big( \motifDistribution(\init{\configMat}), \motifDistribution(\atk{\configMat}{k}) \big)
	=
	\sum_{\subgraph\in\set{\subgraph}} \motif_\subgraph(\init{\configMat}) \log\left(\frac{\motif_\subgraph(\init{\configMat})}{\motif_\subgraph(\atk{\configMat}{k})}\right),
\end{equation}
where $\init{\configMat}=\configMat$ represents the configuration prior to any attacks and $\atk{\configMat}{k}$ denotes the configuration after $k\in\mathcal{K}$ attacks.
Assuming that the system has an estimated knowledge $\hat{\attackDistribution}$ on the attacker's revised strategy, our network design is cast as the following optimization problem:
 \begin{equation} \label{eqn:motif_optimization}
    \begin{array}{ll}
  \underset{\{ \atk{\configMat}{k} \}_{k\in\mathcal{K}}}{\mbox{minimize}} & \mathbb{E}_{\hat{\attackDistribution}} \left[  \kldiv\left( \motifDistribution(\init{\configMat}), \motifDistribution(\atk{\configMat}{k}) \right) \right]\\
    \mbox{subject to}  & \atk{\configuration_{\node,\node'}}{k} = \atk{\configuration_{\node,\node'}}{k-1}, \   k\in\targetNodeSet, \ (\node,\node')\in\atk{\edgeSet}{k}\cup\atk{\permittedEdges}{k} \\
    & \sum_{\node=k+2}^{\NODE} \configuration_{k+1,\node}\geq\minDegree, \  k\in\targetNodeSet \\
   & \atk{\configuration_{\node,\node'}}{k} \in \{0, 1\}, \ k\in\targetNodeSet, \ (\node,\node')\in\atk{\edgeSet}{k}\cup\atk{\permittedEdges}{k}.
    \end{array}
\end{equation}
Here,  additional constraints on configuration impose the need for maintaining original and previously held connectivity between nodes after each attack, while specifying the inability to form connections with the nodes that have been attacked.

\subsubsection{Results}
Under a known stressor model (attack distribution before stressor change as in Fig. \ref{fig:uc_motif_setting}), a robust design determines the optimal network configuration that minimizes the expected loss and employs the configuration designed for the worst-case conditions.
As opposed to this deterministic robust design, the resilient design reevaluates the stressor modality $\hat{\attackDistribution}$ via hypothesis testing whenever a significant performance drop is observed.
Then, using the revised stressor model (attack distribution after stressor change as in Fig. \ref{fig:uc_motif_setting}), the optimal configuration for the system is redefined.

The performance in terms of the average deviations in the motif distributions under different sets of attacks for robust and resilient system designs is shown in Fig.~\ref{fig:uc_motif_results}.
The simulated setting considers that the attacker modifies its attack strategy after $t=150$ in terms of the choice os the maximum number of targeted \glspl{bs} ($\NODE$) and the probability of selecting $k(\leq\NODE)$~\glspl{bs}.

As noted from Fig.~\ref{fig:uc_motif_results}, when the stressor deviates from the initial model to an unaccounted setting, the fixed robust design is incapable of maintaining the performance. 
In contrast, the resilient approach is capable of detecting the stressor change via the significant loss of the performance and estimate the stressor through a distribution estimation framework. 
Afterwards, it adapts the link reconfiguration policy accordingly yielding lower losses (higher performance) compared to the robust design as noted in Fig.~\ref{fig:uc_motif_results} over $t>250$. 
Note that resource over provisioning (adding more \glspl{bs}) and redundancy (allowing more reconfigurable links, i.e., increased $|\permittedEdges|$ ) could improve the performance within the robust design.
In contrast, trading off those with efficient stressor detections and estimations along with faster reconfiguration methods have enabled the resilience in the proposed method.

\subsection{Rate Adaptation by PD Abstraction for Resilient Inference}

\subsubsection{Setting}

Recall that the signatures of the intrinsic geometry of the data, such as topological invariants, serve as abstractions of the underlying data. In this use case, we demonstrate how such an abstractor, with inherent characteristics such as sparsity, enables the emergence of resilience in a \gls{p2p} communication and inference setting despite channel imperfections. In particular, given a point cloud, we leverage its \gls{pd} as an abstraction (See~\S~\ref{subsec:Persistence-Diagrams}), serving as a topological signature of the point cloud.

\begin{figure}
\centering
\begin{subfigure}[t]{\linewidth}
    \includegraphics[width=.8\linewidth]{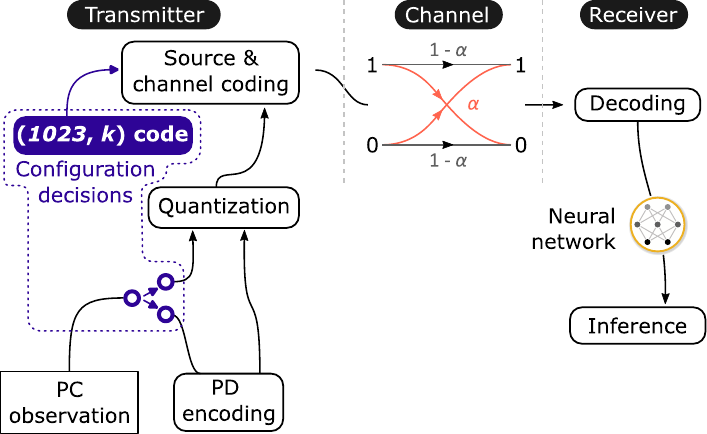}
    \centering
    \caption{\Gls{p2p} communication model.}
    \label{fig:uc_PD_setting}
\end{subfigure}
\par\medskip 
\begin{subfigure}[t]{\linewidth}
    \includegraphics[width=\linewidth]{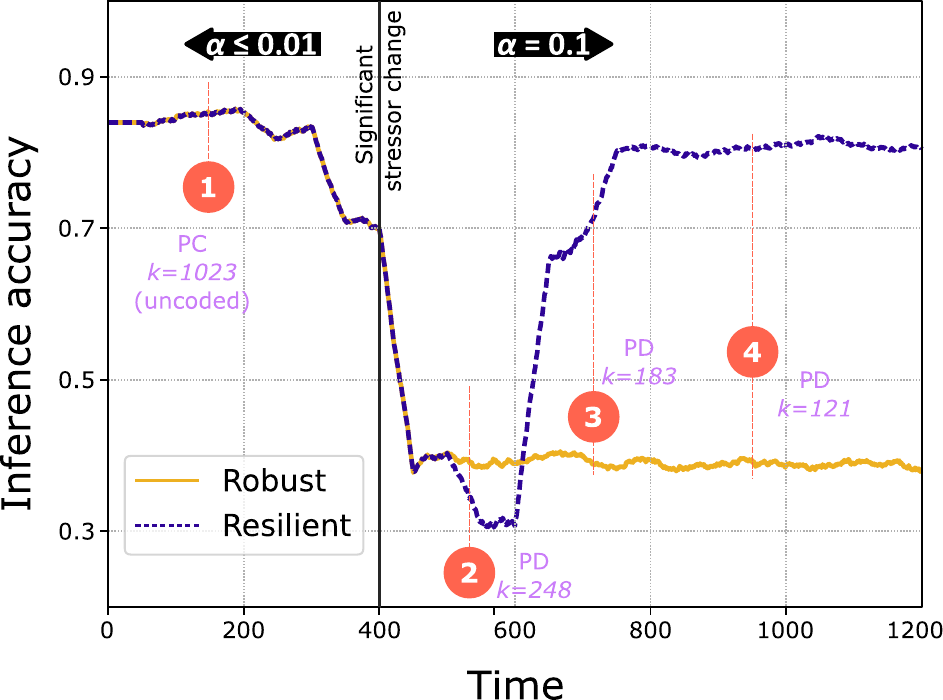}
    \caption{Performance comparison between robust and resilient designs towards inference at the receiver.}
    \label{fig:uc_PD_results}
\end{subfigure}
\caption{Resilient link-level adaptation use case.}
\label{fig:uc_PD}
\end{figure}

We start by considering a \gls{p2p} communication where the channel between the \gls{Tx} and the \gls{Rx} is based on the \gls{bsc} with crossover probability~$\alpha$, see Fig.~\ref{fig:uc_PD_setting}. More specifically, we define our communication channel as $M$ uses of the \gls{bsc}. Thus, the capacity $C$ of the channel is given by $C({\alpha})=M[1-f({\alpha})]$ bits per channel-use, where $f({\alpha})=-\alpha\log_2 \alpha-(1-\alpha)\log_2(1-\alpha)$. Note that the stressor is characterized by $\alpha$ in this example. Roughly speaking, the \gls{Tx} functions as an object sensor, transmitting sensory information over the channel to the \gls{Rx}. The \gls{Rx} then performs object inference on the possibly noisy received data using a trained \gls{ml}~model.

More specifically, the \gls{Tx} senses a sequence of objects $\{O_i\}_{i\in\mathbb{N}}$, where each object $O_i$ is associated with a class label $c_i\in \{1,\ldots,L\}$. Moreover, the object and label pairs $(O_i,c_i)$ are assumed to be \gls{iid} and the distribution over objects and their class labels is assumed to be well-defined. Given an object $O_i$, the \gls{Tx} makes a binary decision on whether to transmit the raw \gls{pc} of the object or its associated \gls{pd}. This decision is represented by a binary variable:  
\[
x \in \{0,1\},
\]
where the transmitted representation $R_i$ of $O_i$ is given by
\[
R_i(x) =
\begin{cases}
    \text{\gls{pc} of } O_i, & \text{if } x = 0, \\
    \text{\gls{pd} of } O_i, & \text{if } x = 1.
\end{cases}
\]
To transmit the representation $R_i(x)$ over a \gls{bsc}, we first apply 2D uniform quantization. The quantized representation is source encoded using Huffman coding, producing $H_i(x)$. It is then channel encoded with a \gls{bch} code, yielding $C_i(x, k)$, where $k$ is the message length of the $(n, k)$ \gls{bch} code. At the \gls{Rx} end, transmitted $C_i(x, k)$ is decoded to obtain $\hat{R}_i(x,k)$, the reconstructed representation of $R_i(x)$. Finally, the \gls{Rx} utilizes $\hat{R}_i(x,k)$ as input to an offline-trained \gls{ml} model, which outputs $\hat{c}_i(x,k)$, representing the inferred class of the transmitted object $O_i$. As such, the inference accuracy denoted $A(x,k)$ is given by 
\begin{equation}\label{eq:Inference-Accuracy-PD-UC}
    A(x,k) = \mathbb{E}\left[\indict\left\{\hat{c}_i(x,k) = c_i \right\}\right],
\end{equation}
where $\indict$ is the indicator function. 
Note that the subscript $i$ in \eqref{eq:Inference-Accuracy-PD-UC} can be dropped without loss of generality due to the \gls{iid} assumption of the sequence of object-label pairs $(O_i,c_i)$. 
Moreover, the expectation in \eqref{eq:Inference-Accuracy-PD-UC} is assumed to be well-defined with respect to the relevant probability distribution of $(O_i,c_i)$ governing the underlying randomness.

The preceding formulation provides a framework for the system's configuration based on two decision variables, namely \( x \) and \( k \). In the presence of minimal or no channel impairments (i.e., \( \alpha \approx 0 \)), the channel capacity is sufficiently large, making it preferable to operate with \( x = 0 \). This allows the full object point cloud representation $R_i(0)$ to be transmitted to the \gls{Rx}, which generally leads to better inference performance. However, when noticeable channel imperfections are present (i.e., \( \alpha \gg 0 \)), it is recommended to reconfigure the system to leverage the sparsity of the \gls{pd} representations. Technically, this reconfiguration is achieved by solving the following feasibility problem:
 \begin{equation} \label{eq:Robust}
    \begin{array}{ll}
    \mbox{find} & x, k\\
    \mbox{subject to} & \mathbb{E}\left\{\ell\left(C(x,k)\right)\right\}\leq M \\
     & k/n\leq C(\alpha)/M \\
    &  A(x,k)\geq \beta \\
    & x\in\{0,1\}\\
    & {k \in \{1,\dots,n\}},
    \end{array}
\end{equation}
where the decision variables are $x$ and $k$. The notation $\ell(\cdot)$ denotes the length of the coded representation (in bits) and $\beta$ is the desired minimum inference accuracy level. 
The first constraint imposes a delay requirement, mandating that, on average, the transmission of the coded representation must be completed within at most $M$ uses of the \gls{bsc}. 
The second constraint restricts the information rate not to exceed the fundamental capacity limit of the channel. 
The third constraint enforces a quality-of-service condition by requiring the inference accuracy to meet a specified minimum threshold. 
The final two constraints reflect limitations imposed by the allowable system configurations.
Although problem \eqref{eq:Robust} appears concise, it is highly challenging due to its non-convexity and the technical difficulties associated with computing the expectation in the constraints. As a result, empirical approximations, such as Monte Carlo simulations for density approximations and table lookups, among others, are necessary to address the problem in practice.

\subsubsection{Results}

Fig.~\ref{fig:uc_PD_results} shows the inference accuracy versus time $t$ in an empirical setup, where the sequence $\{O_i\}_{i\in\mathbb{N}}$ of objects is based on the MNIST~\cite{deng2012mnist}. Each object belongs to one of three classes ($L=3$), where class 1 comprises digits containing two loops, class 2 includes digits with a single loop, and class 3 consists of digits with no loops. 
Given an arbitrary object $O_i$, the corresponding \gls{pc}  is computed as the set of pixel coordinates whose grayscale intensity exceeds $0.7$. The associated \gls{pd} is computed using standard algorithms. For the \gls{bch} code scenario, a $(1023,k,t)$ code is used. Moreover, the number of \gls{bsc} uses $M$ is set to $6035$, and a the minimum classification accuracy threshold $\beta$ is set to $0.7$.

The figure shows that for $t\in[0,250]$, $\alpha$ remains below or equal to $0.01$, resulting in a channel capacity of $M.C(0.01)=5548$ (bits/channel use) or higher. As such, our empirical computations suggest that an initial configuration of $(x,k)=(0,1023)$ is comfortably accommodated so that none of the constraints of problem~\eqref{eq:Robust} is violated. However, the crossover probability $\alpha$ gradually rises, reaching $0.1$ at $t=400$, after which it remains constant. Consequently, at 
$t=400$, the channel capacity experiences an abrupt reduction from $5548$ to $M.C(0.1)=3205$ (bits/channel use). This leads to a sudden degradation in inference accuracy, as the initial configuration \((x, k) = (0, 1023)\) is no longer capable of supporting error-free transmission under the updated channel conditions. In this respect, the resilience mechanism is activated to identify a new configuration $(x, k)$ that is feasible for problem~\eqref{eq:Robust}. It turns out that $x = 0$ is no longer viable if communication delays are not permitted. In particular, even the least redundant \gls{bch} code with $k = 1013$, whose code rate is $0.99$, results in an coded bit rate of approximately $6096\approx 6035/0.99$ (bits/\gls{pc}), which exceeds $M$ violating the first constraint of problem~\eqref{eq:Robust}.
Therefore, the resilience mechanism has configured $x=1$, or in other words, \gls{pd} is used instead of the raw \gls{pc}. The corresponding resilience plot in Fig.~\ref{fig:uc_PD_results} illustrates that employing the feasible configurations $(1, 248)$, $(1, 183)$, and $(1, 121)$ at time instances $t = 500$, $t = 700$, and $t = 850$, respectively, each corresponding to BCH codes with progressively higher redundancy, leads to a gradual improvement in inference accuracy while satisfying the constraints of problem~\eqref{eq:Robust}.

A straightforward computation indicates that the system, under its initial configuration, would have maintained robustness against the stressor variation (i.e., from $\alpha=0.01$ to $\alpha=1$) had it been over-provisioned with $M = 10447$ \glspl{bsc} instead of the initial $M = 6035$. However, rather than resource over-provisioning, the resilience-oriented design must trigger a reconfiguration mechanism, incurring additional computational overhead for stressor change detection and subsequent reoptimization. This illustrates the robustness–resilience trade-off discussed in Section \ref{subsubsec:Robustness-versus-Resilience}. In the resilience-oriented design, the need for reconfiguration to restore inference accuracy  results in additional energy consumption. In other words, recovering from a degradation in inference accuracy comes at the cost of increased energy expenditure, emphasizing the relevance of the sustainability–resilience trade-off discussed in Section \ref{subsubsec:Sustainability-versus-Resilience}.

\subsection{Kripke Semantics for Resilient Multi-agent Systems}

\subsubsection{Setting}
\begin{figure}
\centering
\begin{subfigure}[t]{\linewidth}
\centering
    \includegraphics[width=.7\linewidth]{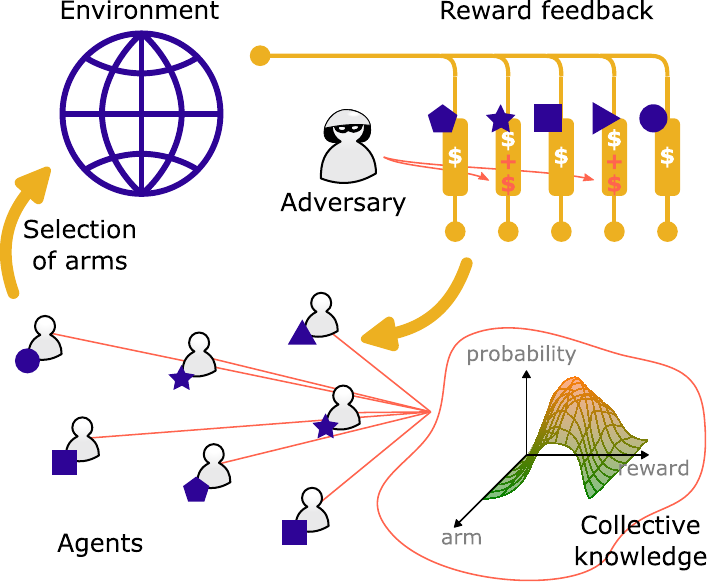}
    \caption{MAMAB setting.}
    \label{fig:UC_logic_setting}
\end{subfigure}
\par\medskip 
\begin{subfigure}[t]{\linewidth}
    \includegraphics[width=\linewidth]{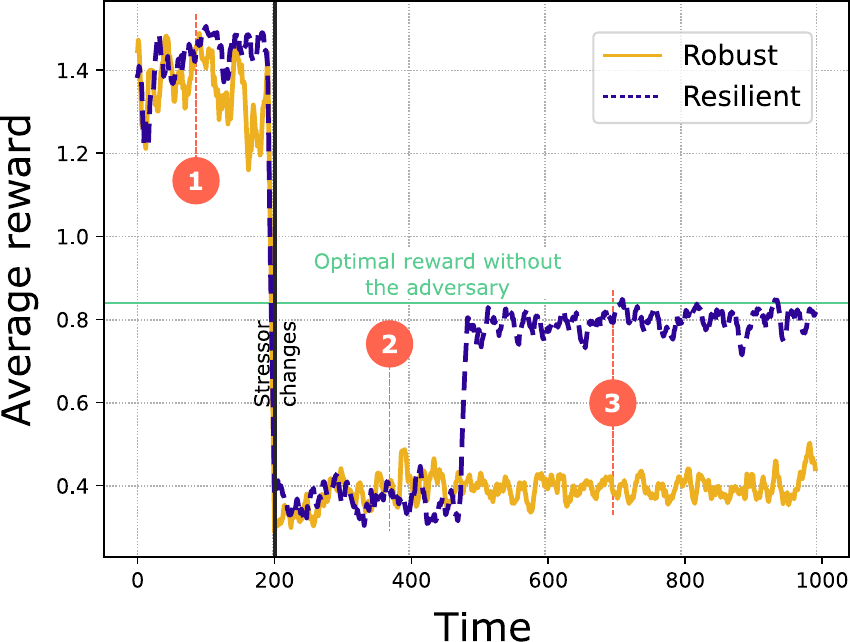}
    \caption{Comparison of the average reward across agents with robust (blue) and resilient (orange) designs under adversarial influence.}
    \label{fig:UC_logic_results}
\end{subfigure}
\caption{Epistemic-based resilient multi-agent systems.}
\label{fig:UC_logic}
\end{figure}

We consider a multi-agent multi-arm bandit (MAMAB) setting, depicted in, Fig.~\ref{fig:UC_logic_setting} involving a set $\mathcal{N}$ of $N$ agents, each selecting actions (arms) $\mathcal{A} = \{1,2,\ldots,A\}$ with unknown stochastic rewards drawn from distributions $P_a$ with mean $\mu_a$. Agents interact with the environment over a discrete and finite time horizon of $T$ rounds, where $T \in \mathbb{N}$ denotes the total number of time steps.  At each time step $t$, agent $\agent$ selects an arm $a_{\agent,t}$ and receives a reward $r_{\agent,t} \sim P_{a_{\agent,t}}$. These reward observations are stored as  agent’s local history of past pulls and outcomes. At round $t$, agent $\agent$ uses this history to compute, for each arm $a$, the empirical mean reward which is the average of all rewards it has observed from arm $a$ over its previous pulls.
Agents aim to maximize their cumulative rewards over a time horizon $T$ via action-selection policies $\pi_{\mathrm{action}}^\agent$ that map each agent’s current empirical mean estimates to arm choices. 
Classical solutions to the multi-arm bandit problem typically rely on statistical algorithms such as Upper Confidence Bound (UCB)~\cite{UCB_MAMAB}, Thompson Sampling~\cite{Thompson_MAMAB}, or epsilon-greedy strategies~\cite{greedy_MAMAB}. These methods balance exploration and exploitation using statistical estimates of arm rewards, confidence bounds, or posterior distributions. However, these strategies inherently assume that reward deviations are purely stochastic and unbiased. This assumption fails under adversarial conditions, where  oracle attacks exploit agents' exploration behavior to inject deceptive reward signals and mislead learning dynamics.

Under oracle attacks, an adversary injects deceptive noise $\epsilon > 0$ into rewards of suboptimal arms $\mathcal{A}_{\text{adv}} \subseteq \mathcal A$ during exploration phase $\mathcal{T}_{\text{explore}} \subset \{1,2, \ldots,T\}$, inflating their empirical means:
\[
\tilde{r}_{\agent,t} =
\begin{cases}
r_{\agent,t} + \epsilon, & \text{if } t \in \mathcal{T}_{\text{explore}} \text{ and } a_{\agent,t} \in \mathcal{A}_{\text{adv}},\\[1ex]
r_{\agent,t}, & \text{otherwise.}
\end{cases}
\]
This causes agents to overestimate suboptimal arms, leading to long-term performance loss.

To address such adversarial scenarios, we equip each agent~$\agent$ with an internal belief model $M_{\agent,t} = (\worldSet, \relation_{\agent,t}, v)$ that forms a `slice' of the global Kripke model $\kripke = \big( \worldSet, \{\relation_{\agent,t}\}_{\agent=1}^\AGENT, \valuationFunc \big)$. As introduced in \S~\ref{sec:kripke}, a Kripke model comprises a set of possible worlds \(\worldSet\), agent‐indexed accessibility relations \(R_{\agent,t}\), and a valuation \(v\).  We also equip each agent~$\agent$ with a set of epistemic actions $\mathcal{E}$, including querying, sharing, exploring, revising, or holding beliefs. These actions allow agents to adapt their internal knowledge in response to observations and contradictions, providing the foundation for resilient decision-making. Each agent's Kripke model is semantically grounded in a set of logical propositions that express beliefs about the environment, particularly concerning the reward properties of arms. For example, let \(\formula_a\) denote a formula asserting that the expected reward of arm $a$ lies within a confidence interval $[l_a,u_a]$, derived from the agent's empirical observations. The agent's belief that this formula holds is expressed as $K_\agent \formula_a$, which reads as: ``agent $\agent$ believes that the expected reward of arm $a$ lies in $[l_a, u_a]$''. These logical propositions are evaluated within the epistemic model and guide the agent's policies. Crucially, agents may also communicate such logical formulae to one another as a mean of epistemic coordination. 
In detail, each agent $\agent$'s decision making is now governed by: 

\begin{itemize}
\item \textbf{Belief-management policy} $\pi_{\text{belief}}^\agent$: maps beliefs and observations to epistemic actions $e_{\agent,t} \in \mathcal{E}$ to update or hold current beliefs.
\item \textbf{Meta-action policy} $\pi_{\text{meta}}^\agent$: takes the recovered beliefs and produces a vector of meta-parameters \(\boldsymbol\theta_\agent\in\Theta\) (for example, an exploration rate \(\epsilon\) or decaying rate) that will configure the downstream action-selection policy. 
\item \textbf{Action-selection policy} $\pi_{\text{action}}^\agent$: selects arms based on current beliefs and meta-parameters.
\end{itemize}

Resilient multi-agent decision-making in the presence of adversarial or uncertain conditions hinges on two critical capabilities: \textit{recoverability} and 	\textit{durability}. Recoverability reflects the agents' ability to detect and promptly correct false or inconsistent beliefs, while durability ensures that once valid beliefs are established, they can be reliably maintained over time. For example, an agent may detect that a previously-held $\formula_a$ is no longer valid when the observed rewards consistently fall outside the believed interval. Specifically, if the received reward repeatedly violates the bounds $[l_a,u_a]$ over time, the agent has evidence that its current belief no longer reflects the environment. Thus, the agent considers its belief invalid and is subject for revision. Let \(\formula'_a\) belong to a set of alternative formulas that characterize the same arm, with \(\formula'_a\) accurately reflecting the current reward behavior. Since these belief formulas represent mutually exclusive confidence models, they cannot be simultaneously valid. Define:
\begin{itemize}
\item \textbf{Recovery time } $t_r$ as the number of time steps required for the agent group \(\mathcal N\) to achieve \(\mathcal{E}_{\mathcal{N}}\,\formula'_a\), in which all agents come to hold and mutually recognize the updated belief that $\formula'_a$ currently holds. 
\item \textbf{Durability time } $t_d$ as the duration for which the updated belief remains valid and commonly known across all agents.
\end{itemize}

The belief-management policy \(\pi_{\text{belief}}^\agent\) is optimized to minimize recovery time $t_r$ and maximize durability $t_d$, thus balancing both objectives. The optimization problem is formally defined as:
\begin{equation}\label{eqn:logic_optimization}
	\begin{array}{ll}
		\underset{ \{\pi_{\text{belief}}^\agent\}_{\agent \in \mathcal{N}} }{\mbox{minimize}} 
		& 
		\lambda \cdot t_{r} \left( \{\pi_{\text{belief}}^\agent\}_{\agent \in \mathcal N}, \mathcal{E}_{\mathcal{N}}\,\formula'_a \right) \\
        & \qquad\qquad- (1-\lambda) \cdot t_{d}\left( \{\pi_{\text{belief}}^\agent\}_{\agent \in \mathcal N}, \mathcal{E}_{\mathcal{N}}\,\formula'_a \right)
		\\
		\hfil\mbox{subject to} 
		& t_{r}\left( \{\pi_{\text{belief}}^\agent\}_{\agent \in \mathcal N}, \mathcal{E}_{\mathcal{N}}\,\formula'_a \right) \leq t_{r}^{\max}, \\
        & t_{d}\left( \{\pi_{\text{belief}}^\agent\}_{\agent \in \mathcal N}, \mathcal{E}_{\mathcal{N}}\,\formula'_a \right) \geq t_{d}^{\min}, 
	\end{array}
\end{equation}
where \(t_{r}^{\max}\) denotes the maximum allowed recovery period, \(t_{d}^{\min}\) represents the minimum required period of belief stability, and \(\lambda \in [0,1]\) denotes a hyperparameter that controls the trade-off between recoverability and durability. Assigning a higher weight to $t_r$ (i.e., increasing $\lambda$) encourages agents to prioritize faster recovery, which means they revise their beliefs more quickly, without waiting long to accumulate supporting evidence, potentially resulting in shorter periods of belief stability (lower $t_d$). Conversely, giving more weight to durability (lower $\lambda$) leads agents to wait longer before updating their beliefs, allowing for greater stability (higher $t_d$) at the cost of slower recovery (higher $t_r$). This presents the recoverability-durability trade-off discussed in \S~\ref{subsubsec:Recovery-versus-durability}.
Upon belief recovery, agents re-optimize their behavior based on their revised understanding and  adapting behavior by solving:

\begin{equation}
    \max_{\pi^\agent_{\mathrm{meta}}}\;\; \mathbb{E} \left[\sum_{t=t_r}^T \Big(r_{\agent,t} \, \Big\vert \, \pi^\agent_{\mathrm{meta}} (M_{\agent,t}), \, M_{\agent,t}\Big)\right].
\end{equation}
Here, \(\pi_{\mathrm{meta}}^\agent\) produces the meta‐parameters that configure the downstream action‐selection policy \(\pi_{\mathrm{action}}^\agent\). Unlike fixed‐horizon planning performed only at the start of an episode, this formulation explicitly supports mid‐horizon re‐optimization triggered by belief revisions, enabling agents to adapt their behavior to evolving epistemic conditions. This approach ensures that the group as a whole remains resilient even if some agents (i) failed to detect the contradiction locally or (ii) revise more slowly. Once \(\mathcal{E}_{\mathcal{N}}\,\formula'_a\) holds within the given time constraints, every agent, regardless of its individual observation history, will align its downstream policies with the recovered belief, thus ensuring system‐wide resilience.

\subsubsection{Results}
Figure~\ref{fig:UC_logic_results} shows the average‐reward trajectories for a purely statistical “robust” design (solid yellow) versus the epistemic “resilient” design (dashed purple). We simulate $5$ agents on a $20$-armed bandit where each arm's reward is drawn from a Gaussian distribution with standard deviation $0.1$ and a true mean sampled uniformly from $[0.1,0.9]$. To induce the oracle attack, during the first $200$ iterations, we add a noise of $1.2$ to the $10$ arms with the lowest true means. 
\begin{itemize}
  \item \textbf{Inflated Exploration Phase (\#1).} For \(t < 200\), agents in both designs pick the best available arms and achieve mean rewards around \(1.4\). These chosen arms are the suboptimal arms with inflated additive noise.
  \item \textbf{Immediate degradation.} At \(t = 200\), the adversary stops its deceptive noise, causing the agents’ reward to collapse to \(\approx0.4\) under both setups.  
  \item \textbf{Belief recovery (\#2).} Resilient agents execute epistemic actions (explore, revise, share) to detect and resolve contradictions. Robust agents, lacking belief‐revision mechanisms, continually refine their corrupted estimates. 
  \item \textbf{Performance Recovery.} Upon belief recovery, agents reconfigure their arm-selection policies, and at \(t\approx 480\), the average reward rapidly rebounds to the no‐adversary optimum of approximately \(0.80\) (green line). Robust agents remain locked into suboptimal arms, sustaining \(\approx0.4\) reward for the remainder of the horizon.  
  \item \textbf{Durable performance (\#3).} From \(t\approx 480\) to \(T=1000\), resilient agents maintain reward \(\approx0.80\), demonstrating that agents' recovered belief remained valid. 
\end{itemize}

These results confirm that formal epistemic logic and collective belief updates enable rapid detection and correction of adversarial perturbations, whereas purely statistical designs lack the mechanisms needed for recovery and remain vulnerable to sustained performance degradation.

\subsection{Sheaf-assisted Resilient Distributed Learning}

\begin{figure*}[!t]
    \centering
    \includegraphics[width=1.0\textwidth]{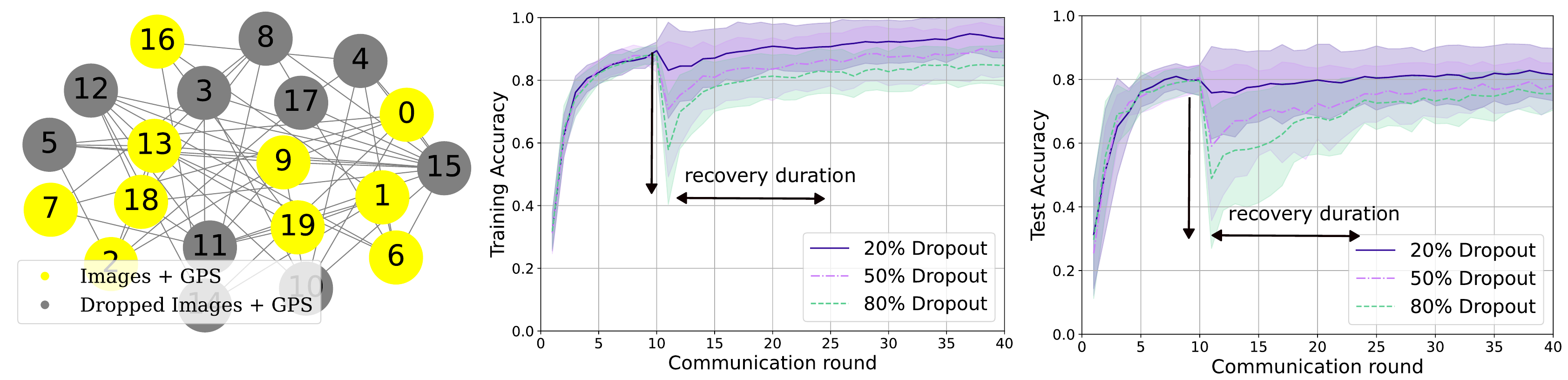}
    \caption{Performance evaluation of the sheaf-assisted resilient distributed learning method.}
    \label{fig:uc_sardl}
\end{figure*}

\subsubsection{Setting}
A distributed mmWave beamforming prediction problem is considered, where each client (BS) has a co-located camera and Global Positioning System (GPS) signal \cite{charan2022towards}. Each BS is equipped with an $U$-element uniform linear array (ULA) and employs a predefined beamforming codebook $\set{F} = \{ \vect{f}_q \}_{q=1}^{Q}$, where each beamforming vector $\vect{f}_q \in \mathbb{C}^{U}$ is designed to direct energy toward a specific spatial direction. 
Let $\vect{h}_c[t] \in \mathbb{C}^{U}$ denote the channel vector between one BS and one drone at subcarrier $c$ and time step $t$. The received signal at the drone can be written as $s_c[t] = \vect{h}_c^H[t] \vect{f}[t] x + n_c[t]$, where, $x \in \mathbb{C}$ is the transmitted symbol with power constraint $\mathbb{E}[|x|^2] = P$, $\vect{f}[t] \in F$ is the selected beamforming vector at time $t$, $n_c[t] \sim \mathcal{CN}(0, \sigma^2)$ is additive Gaussian noise, and $\vect{h}_c^H[t]$ is the conjugate transpose (Hermitian) of the channel vector at time $t$. The objective of the BS is to select the optimal beamforming vector $\vect{f}^*[t]$ that maximizes the SNR at the receiver.
Formally, the set of available sensory information at time $t$ is defined as the set $\{ g[t], I[t] \}$, where $g[t] \in \mathbb{R}^4$ represents the drone's augmented GPS measurements (latitude, longitude, distance, height) and $I[t] \in \mathbb{R}^{W \times H \times C}$ is the RGB image captured by the BS. The goal is to learn a neural network that predicts the optimal beam index $\hat{f}[t]$ in the beamforming codebook $\set{F}$, given the sensory inputs by, minimizing a cross-entropy loss function.

Specifically, $N = 20$ distributed BSs  are assumed, each serving a distinct drone and having access only to its local dataset with modalities $\set{M}_i$ with $|\set{M}_i| = 2$ (GPS and images). 
Based on the DeepSense Drone dataset~\cite{charan2022towards}, each BS predicts the optimal mmWave beam for its drone.
Meanwhile,  BSs cooperate via a predefined communication graph for better performance.
Following~\cite{charan2022towards}, we use a ResNet-50-based feature extractor (pre-trained on ImageNet) for images and a two-layer Multilayer Perceptron (MLP) for GPS data, where the ResNet-50 output is mapped to a beam index of a codebook of size $Q=64$. For multimodal clients, we employ a multi-stream architecture, concatenating and passing features through a shared logistic regression layer. 

As  discussed in Section~\ref{sec:sheaf_theory}, the sheaf-based formulation enables capturing correlation among agents through  \textit{restriction maps} (see Fig. \ref{sheaf1}),  yielding a more resilient  learning process. Specifically, this approach applies the concepts from Section \ref{sec:sheaf_theory} where the parameters of the local neural network at the $i$-th BS reside in a vertex stalk $\set{F}(i)$. The crucial element is the set of learnable linear restriction maps, represented by matrices $\vect{P}_{ij}$, which project the model parameters from two communicating BSs $i$ and $j$ into a shared comparison space (the edge stalk). By learning these maps, the system explicitly models the correlation structure between the agents' tasks. When a sensor fails at one BS, this learning mechanism allows the system to leverage information from its neighbors and adapt to the dynamic state, as the optimization process continuously encourages consistency in the shared projection space by minimizing the term in Eq. \eqref{eq_quadratic_sheaf_laplacian}, thereby improving performance recovery.

To show the resiliency performance, at the $10$-th communication round, we randomly choose different ratios of BSs losing their image modality, as shown on the left side of Fig. \ref{fig:uc_sardl}, i.e., the input of images becomes all zero. 
In practical scenarios, this may arise due to the unexpected camera sensor's failures. In the training process, each agent iteratively updates the model parameters and the linear restriction maps, leading to a resilient  cooperative learning process against potential sensory~losses.

\subsubsection{Results}

Consider different ratios of image modality loss, the training and test accuracy over communication rounds can be shown in Fig. \ref{fig:uc_sardl}. It can be observed that the sensors' loss in different agents results in the performance drop w.r.t. accuracy.  With more nodes facing sensors' failures, the performance becomes more degraded. 
Thanks to continuously learning the restriction maps among agents, agents' correlations can be recaptured quickly, leading to performance recovery and  accuracy increase. The results in Fig. \ref{fig:uc_sardl} reveal the effectiveness and resiliency property of the sheaf-assisted learning in distributed systems.

\subsection{Compositional Multi-agent Active-Inference}

\subsubsection{Setting} 
\begin{figure}
\centering
\begin{subfigure}[t]{\linewidth}
    \includegraphics[width=\linewidth]{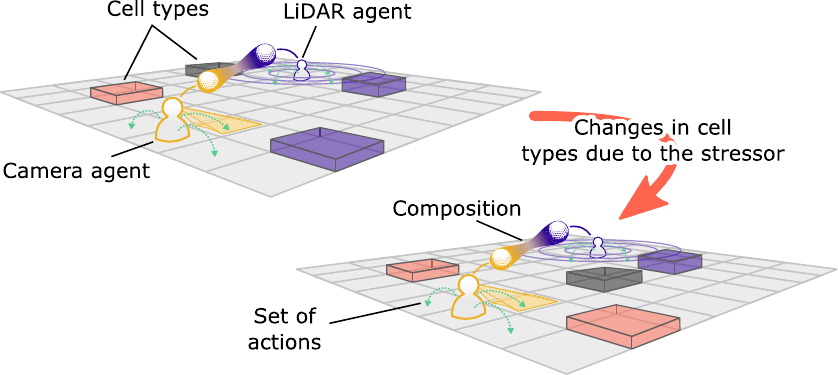}
    \caption{Grid world with different cell types and two agents: Camera and LiDAR. With the stressor change, some cell types are altered.}
    \label{fig:uc_activeInf_setting}
\end{subfigure}
\par\medskip 
\begin{subfigure}[t]{\linewidth}
    \includegraphics[width=\linewidth]{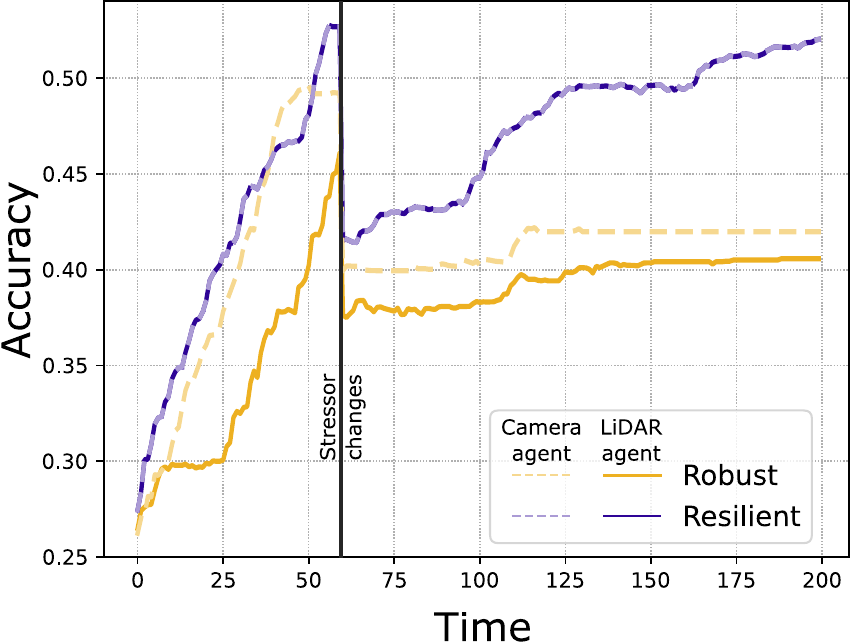}
    \caption{Accuracy of LiDAR and Camera agents with belief composition (resilient) and without (robust). }
    \label{fig:uc_activeInf_results}
\end{subfigure}
\caption{Compositional active inference: resilient vs. robust baseline.}
\label{fig:UC_activeInf}
\end{figure}

In this use case, we consider a multimodal, grid-based exploration scenario involving two autonomous agents equipped with complementary sensors: a LiDAR and a camera, as illustrated in Fig. \ref{fig:uc_activeInf_setting}. 
Let each agent choose an action from the action set defined as
\[
\mathcal{A} = \{\text{N},\; \text{E},\; \text{S},\; \text{W},\; \text{STAY}\},
\]
along with a displacement function
\[
\Delta : \mathcal{A} \to \mathbb{Z}^2, \quad
\Delta(\text{N}) = (0,1),\;
\Delta(\text{E}) = (1,0),\]
\[\Delta(\text{S}) = (0,-1),\;
\Delta(\text{W}) = (-1,0),\;
\Delta(\text{STAY}) = (0,0).
\]
If agent \(n\) is located at grid cell \((x_n, y_n)\) and selects action \(a_n \in \mathcal{A}\), then its new position is computed as
\[
(x_n, y_n) \mapsto (x_n, y_n) + \Delta(a_n).
\]
That is, the actions N, E, S, and W deterministically move the agent one cell in the corresponding cardinal direction, while the STAY action leaves the agent in its current position.

The scenario is motivated by practical applications such as cooperative mapping, indoor robotics, and exploratory tasks where agents must collaboratively construct accurate maps and semantic interpretations of their environment. 
Each agent specializes in a distinct sensing modality, creating a natural necessity for cooperation and belief composition. 
The LiDAR agent excels at determining spatial structures, providing accurate distance measurements to obstacles, yet lacks semantic understanding of what these obstacles represent. 
Conversely, the camera agent can identify and classify obstacles semantically, but it cannot accurately determine the distance or spatial positioning of these obstacles. 
This complementary sensing motivates the two heterogeneous agents to communicate their inferred beliefs, allowing each agent to enhance its understanding by integrating the other’s specialized knowledge.

The discretized $20 \times 20$ grid-world environment under consideration is characterized by a semantic label space $\mathcal L = \{\text{Free},\; \text{Blue},\; \text{Red},\; \text{Black}\}$, representing the four possible cell types. The LiDAR agent maintains local hidden states defined by the discretized distance to the nearest obstacle along four discrete directional bearings (North, East, South, West), quantized into integer bins ranging from $1$ to a maximum detection range of $6$.
Observations for this agent are modeled as true range measurements corrupted by additive Gaussian noise with standard deviation $\sigma=1.0$. 
The camera agent encodes  its local hidden states as semantic labels of the nearest obstacle encountered within a maximum range of $3$ along the same four bearings, classifying each detected obstacle into one of the aforementioned four cell types.
The camera agent's observation model is categorical, with a predefined observation accuracy of $0.6$, indicating a $60\%$ probability of correctly identifying the true semantic class at each observation.

We tackle this problem by formulating it within a compositional active inference framework, as introduced in \S~\ref{sec:active_inference_theory}. This formulation enables agents to maintain heterogeneous local hidden states while still collaborating to infer a shared model of the environment. In standard active inference literature, agents are typically equipped with extrinsic preference distributions that encode desired observations. In contrast, our implementation intentionally omits such extrinsic preferences: each agent treats all possible observations as equally preferable, i.e., their preference distribution is uniform.

Instead, action selection is driven solely by the minimization of multi-step expected free energy as per \eqref{eq:Expected-Free-Energy}, which, in our setup, corresponds to maximizing information gain. This design choice ensures that any observed changes in belief accuracy emerge entirely from the agents’ intrinsic drive to reduce their uncertainty, rather than from any externally imposed reward structure.  
Each agent $n$ maintains a fully factorized belief over the semantic map $z$, represented as:

\[q_n(z) = \Pi_{x,y} q_n(z_{x,y})\] where $z \in \mathcal L^{20 \times 20}$ denote the global latent map composed of discrete semantic labels for each cell in a $20 \times 20$ grid, and $z_{x,y} \in \mathcal L$ represents the latent class of the cell at position $(x,y)$. The distribution $q_n(z_{x,y})$ is  a categorical probability vector maintained independently for each cell, encoding agent $n$’s current belief about the semantic type at that location.
This posterior gets updated based on each agent's own modality and the local hidden state. To compose agents' local beliefs into a coherent global estimate, each agent transmits its belief to its neighbors and then performs a geometric mean update followed by a normalization step. 
This gossip step composes neighboring beliefs to drive the network toward agreement on the global posterior distribution over the global map. 
At iteration $t = 60$, we introduce a significant stressor in the form of a partial environmental change: $30\%$ of randomly selected cells have their labels cyclically permuted (Free $\rightarrow$ Blue $\rightarrow$ Red $\rightarrow$ Black $\rightarrow$ Free). 
This targeted stressor evaluates agents' resilience and adaptability to sudden, partially-overlapping changes in the environment, simulating realistic scenarios where unexpected environmental transformations~occur.

\subsubsection{Results}
Figure~\ref{fig:uc_activeInf_results} shows the temporal evolution of agents' beliefs' accuracy over $200$ steps under two setups: \emph{Composition}, where agents share beliefs through geometric consensus (resilient design), and \emph{No Composition}, where agents operate independently (robust design). 
During the initial phase (iterations $0$–$59$), both agents exhibit progressive improvement in belief accuracy under the independent scenario; however, the composition of beliefs accelerates learning while resulting in higher accuracy levels. Upon introducing the stressor at iteration $60$, a substantial drop in accuracy is observed for both agents under both setups, illustrating the immediate disruptive effect of partial environmental modification. Agents without belief composition experience minimal recovery, stabilizing at  reduced accuracy levels, suggesting limited adaptation. Meanwhile, agents employing compositional inference demonstrate rapid recovery, regaining  a substantial portion of their pre-stress accuracy by iteration $120$.  These results underscore that composing beliefs of individually non-resilient agents can yield a resilient global~behavior.

\section{Conclusions} \label{sec:conclusions}
Despite its critical importance, resilience, a key concept in 6G, has often been overlooked, frequently conflated with robustness and lacking clear metrics. However, resilience could play a crucial role in refining and accelerating current (incremental) 6G visions.  
In contrast to robustness and reliability, resilience assumes  that
disruptions will inevitably happen. Resilience, in terms of elasticity, focuses on the ability to
bounce back to favorable states, while resilience as plasticity involves agents (or networks) that can
flexibly expand their states, hypotheses and course of actions, by transforming through real-time
adaptation and reconfiguration. This constant situational awareness and vigilance of adapting world
models and counterfactually reasoning about potential system failures and best
responses, is a core aspect of resilience. By embedding resilience into networks, ecosystems, social structures, and economic systems, we can enhance adaptability, reduce the risks of failure, and foster long-term sustainability.    Finally, it is our hope that this article will help stimulate more research in network resilience, whose importance will be instrumental in  6G networks and beyond.

\section*{Acknowledgments}

The authors used OpenAI‘s ChatGPT to assist with grammatical corrections and text refinement during manuscript preparation.

\bibliographystyle{ieeetr}  
\bibliography{visionbib}

\end{document}